\documentclass{aa1}

\graphicspath{ {./Figures} }

\usepackage{graphicx}
\usepackage[varg]{txfonts}

\usepackage[colorlinks=true, linkcolor=blue, citecolor=blue, filecolor=blue,
urlcolor=blue]{hyperref}
\usepackage{lscape}          
\bibpunct{(}{)}{;}{a}{}{,} 

\newcommand{\tx}{T_{\rm X}}
\newcommand{\Tx}{$T_{\rm X}$}
\newcommand{\kms}{km\,s$^{-1}$}
\newcommand{\Msun}{M$_{\odot}$}
\newcommand{\Lsun}{L$_{\odot}$}
\newcommand{\Teff}{$T_{\rm eff}$}
\newcommand{\teff}{T_{\rm eff}}

\newcommand{\lx}{L_{\rm X}}

\newcommand{\Lwind}{L_\mathrm{wind}}
\newcommand{\Vwind}{V_\mathrm{wind}}
\newcommand{\dotMw}{\dot{M}_\mathrm{wind}}

\newcommand{\ovii}{\ion{O}{vii}}

\newcommand{\cv}{\ion{C}{v}}

\newcommand{\chan}{\textit{Chandra}}
\newcommand{\xmm}{\textit{XMM-Newton}}
\newcommand{\gaia}{\textit{Gaia}}

\defcitealias{Sandin13}{Paper\,I}
\defcitealias{helleretal.16}{Paper\,II}
\defcitealias{SSW.08}{SSW}
\usepackage[normalem]{ulem}

\begin{document}

\title{Hot bubbles of planetary nebulae with hydrogen-deficient winds}

\subtitle{III. Formation and evolution in comparison with hydrogen-rich bubbles\thanks 
      {This work has made use of data from the European Space Agency (ESA) mission
      {\it Gaia} (\protect\url{https://www.cosmos.esa.int/gaia}), processed by the {\it Gaia}
       Data Processing and Analysis Consortium (DPAC,
       \protect\url{https://www.cosmos.esa.int/web/gaia/dpac/consortium}). Funding for the DPAC
       has been provided by national institutions, in particular the institutions
       participating in the {\it Gaia} Multilateral Agreement. }
       }

\titlerunning{Hot bubbles of planetary nebulae with H-deficient winds III.}

\author{D. Sch\"onberner \and M. Steffen} 

\authorrunning{D. Sch\"onberner \& M. Steffen}

\institute{Leibniz-Institut f\"ur Astrophysik Potsdam, An der Sternwarte 16, 14482 Potsdam,
           Germany\\ \email{msteffen@aip.de, deschoenberner@aip.de}}

\date{Received, \today / Accepted, \today}

\abstract 
{} 
{ We seek to understand the evolution of Wolf-Rayet central stars by comparing the diffuse
  X-ray emission from their wind-blown bubbles 
  with that from their hydrogen-rich counterparts with predictions from hydrodynamical models.  
 }
{ We simulate the dynamical evolution of heat-conducting wind-blown bubbles 
  using our 1D radiation-hydrodynamics code \texttt{NEBEL/CORONA}.
  We use a post-AGB-model of 0.595~\Msun\ but allow for variations of its evolutionary
  timescale and wind power.  We follow the evolution of the circumstellar structures for different 
  post-AGB wind prescriptions: for O-type central stars and for Wolf-Rayet central stars
  where the wind is hydrogen-poor, more dense, and slower.
  We use the \texttt{CHIANTI} software to compute the X-ray properties of bubble models along
  the evolutionary paths. We explicitly allow for non-equilibrium ionisation of key chemical elements.
  A sample of 12 planetary nebulae with diffuse X-ray emission ---seven harbouring an O-type and 
  five a Wolf-Rayet nucleus--- is used to test the bubble models.
 } 
{ The properties of most hydrogen-rich bubbles (X-ray temperature, X-ray luminosity, size) and their
  central stars (photon and wind luminosity) are fairly well represented by bubble models of our
  0.595~\Msun\ AGB remnant.
  The bubble evolution of Wolf-Rayet objects is different, thanks to the high radiation
  cooling of their carbon- and oxygen-rich winds.  The bubble formation is delayed, and the early
  evolution is dominated by condensation instead of evaporation.   Eventually, evaporation begins and
  leads to chemically stratified bubbles.  
  The bubbles of the youngest Wolf-Rayet objects appear chemically uniform,
  and their X-ray properties can be explained by faster-evolving nuclei.
  The bubbles of the evolved Wolf-rayet objects have excessively low characteristic temperatures 
  that cannot be explained by our modelling.
 } 
%
{ The formation of nebulae with O-type nuclei follows mainly a single path, but the formation pathways leading to the Wolf-Rayet-type
  objects appear diverse. Bubbles with a pure Wolf-Rayet composition can
exist for some time after their formation despite the presence of heat
conduction.
}


\maketitle 


\section{Introduction}
\label{sect:intro}

   By means of space-based observations, it became evident that the inner 
   `cavities' of many round or elliptical planetary nebulae are filled up
   with a tenuous but very hot gas that emits preferentially in the soft X-ray
   domain.  The existence of such gas is to be expected: the fast central-star wind collides
   with the slower, denser inner parts of the nebula (the former wind envelope produced during the
   star's previous evolution along the asymptotic giant branch (AGB))
   and becomes shock-heated, producing a `bubble' of very hot gas.  
   Given the typical values of density and velocity of the stellar wind, the wind shock is adiabatic,
   and the shocked gas is expected to reach temperatures of $10^7$~K (=\,10~MK) or more.  
   
   However, all spectral analyses of the X-ray emissions reveal an unexpectedly low mean or
   characteristic X-ray temperature of between about 1~MK and a few MK.  
   Also, the emission measure is much higher than expected.     
   The present status and preliminary results of the extensive \chan\ Planetary Nebula Survey
   (ChanPlaNS) can be found in \citet{kastner.13}  and \citet{freeman.14}. 
   
   \citet[][hereafter SSW]{SSW.08} took up a suggestion by \citet{soker} that
   heat conduction across the hot bubble from the wind shock towards the bubble-nebula interface
   may be responsible for the unexpected bubble properties.
   Indeed, SSW were able to show that, although the dynamics of a model 
   nebula remains virtually unchanged, the bubble structure and its
   characteristic properties, such as 
   X-ray characteristic temperature and luminosity, can well be explained by nebula 
   models that include thermal conduction. Further detailed comparisons between the SSW
   models and observations are given in \citet{ruizetal.13}. 
    
  Another important physical process to reduce the temperature and to increase the density of  
  the X-ray-emitting plasma of wind-blown bubbles of planetary nebulae is mixing between 
  the hot bubble and cooler
  nebular gas across the bubble--nebula interface by Rayleigh-Taylor instabilities. 
  The first `pilot' 2D simulations of \citet[][]{SS.06} cover, however, only a 
  very limited time span ($\la$\,300~yr) and have only simple parameterised boundary
  conditions. They are thus not really suitable for drawing conclusions concerning the extent 
  of mixing and the temporal evolution of the mixing efficiency.
  
           
   More realistic 2D simulations have been presented by 
   \citet{TA.14, TA.16, TA.16b}.\footnote
{We note that both the simulations of \citet{SS.06} and \citet{TA.14} only allow the development 
 of 2D structures in spherical coordinates.
 }
   These are based on the post-AGB evolutionary tracks of \citet{VW.94} and appropriate 
   wind models in a similar manner to the simulations by \citet{villa.02} and
   \citet{peretal.04}, and clearly show that mixing of bubble and nebular matter across the
   bubble--nebula interface generates a region with intermediate temperatures 
   ($ \ga$\,1~MK) and sufficiently high densities for emitting X-rays of the observed properties. 
   However, inclusion of heat conduction increases the amount of cooler matter by
   evaporation of nebular gas at the conduction front 
   to such an extent that the effect of evaporation soon dominates over mixing.     
  
   It is well known that a small fraction of planetary nebulae harbour nuclei with 
   hydrogen-poor but helium- and carbon-enriched surfaces (Wolf-Rayet central stars), 
   which means they also have winds of the same composition that are feeding their hot bubbles.
   However, the formation and evolution of hydrogen-poor central stars is still not fully
    understood, and the existing post-AGB evolutionary models are not applicable, in principle.
   
    Several important questions have to be addressed in this context. (1) It
    is presently unknown at which moment of the post-AGB evolution the
    originally hydrogen-rich stellar wind turns into a hydrogen-poor wind with
    the typical Wolf-Rayet composition. (2) One would like to understand how
    the formation and evolution of hydrogen-poor but carbon- and oxygen-rich
    bubbles is influenced by their high radiation cooling, and (3) how
    important chemical mixing by dynamical instabilities is in comparison to
    evaporation by heat conduction. (4) Finally, it is not clear to what degree
    evaporation and mixing of hydrogen-rich matter into the hydrogen-poor
    bubble, as predicted by the numerical models, actually occur in nature. At
    present, observational evidence for the existence of chemically stratified
    bubbles is still lacking.
  
  Answering these open questions is certainly a very ambitious task.
  A detailed modelling of chemically inhomogeneous stellar--nebular systems in
  combination with appropriate observations will be essential to gain a better understanding of the formation and evolution of hydrogen-poor central stars.
   
  In the first paper of this series \citep[][henceforth Paper\,I]{Sandin13},
  an algorithm for computing heat conduction coefficients for arbitrary
  chemical compositions was developed and tested. It was found that due
  to the high radiation cooling of hydrogen-poor but carbon- and oxygen-rich
  matter, the formation of a wind-blown bubble with heat conduction is
  considerably delayed.

  In a second paper, \citet[][henceforth Paper\,II]{helleretal.16} constructed
  analytical solutions for self-similar hot bubbles, which include thermal
  conduction according to the prescription of \citet{ZP.96} but are modified to
  our needs.  These simple models provide a convenient tool for analysing, for example, the high-resolution X-ray spectrum of the hot bubble around
  BD\,+30\degr 3639.  It could be shown that this particularly young and small
  bubble is fed by the hydrogen-poor and carbon-/oxygen-rich stellar wind and
  most likely does not yet contain evaporated (and/or mixed) hydrogen-rich
  matter.
   
  In the present paper, we report a parameter study addressing some
    of the questions outlined above. To this ed, we constructed a set of
    hydrodynamical sequences consisting of chemically inhomogeneous
    stellar-nebular systems appropriate for comparison with existing X-ray
    observations of planetary nebulae with Wolf-Rayet central stars. This set
    is discussed in comparison with hydrodynamical sequences of chemically homogeneous,
    hydrogen-rich models already presented and discussed in
    \citetalias{SSW.08} and \citet{ruizetal.13}.
  
   The structure of this paper is as follows: 
   We start in Sect.~\ref{sec:model.calc} by providing the details of the new 1D
   radiation-hydrodynamics calculations with hydrogen-poor post-AGB winds,
   including a description of the assumptions
   made. Section~\ref{sec:param.study} continues with a parameter study of
   hydrogen-poor wind-blown heat-conducting bubble models and a
   discussion of the related findings. Section~\ref{sec:PN.xrays} presents a compilation of the
   observed properties of seven O- and five Wolf-Rayet-type central stars and
   their wind-blown bubbles for which diffuse X-ray emissions have been
   observed. These observed properties are compared with the predictions of
   existing (hydrogen-rich) evolutionary simulations in
   Sect.\,\ref{sec:comp.obs.mod_pn}, while Sect.\,\ref{sec:WR-bubbles.observ}
   deals with a careful comparison of the properties of our newly computed
   chemically inhomogeneous models with the observed properties of the bubbles
   around Wolf-Rayet central stars. The results are discussed in
   Sect.~\ref{sec:discussion}, and we end the paper by providing a summary and our
   conclusions (Sect.~\ref{sec:conclusion}).

   The paper is supplemented by two Appendices, one describing how bubble
   temperature and luminosity depend on the chosen X-ray band width
   (Appendix~\ref{appsec:calib.bandwidth}), and the other establishing the
   relations between the post-AGB evolutionary tracks used in this paper and
   the more recent ones of \citet{MMA.06} (Appendix~\ref{app:post.AGB}).
   
\section{Details of the model calculations}
\label{sec:model.calc}
In this section we present a brief overview of our new hydrodynamical
simulations of hydrogen-poor, wind-blown heat-conducting bubbles inside
planetary nebulae and the computations of the bubbles' X-ray emission in terms
of X-ray luminosity and characteristic (or mean) X-ray temperature.

\subsection{General aspects}
\label{subsec:gen.aspects}

We used, as in previous works, the Potsdam \texttt{NEBEL/CORONA} software
package to model the combined evolution of central star and circumstellar
envelope by radiation-hydrodynamical simulations in spherical geometry.  The
details of the \texttt{NEBEL} code can be found in \citet{peretal.98, peretal.04}.
Here we outline only the physical system: A typical model has a radial extent
from $6{\times} 10^{14}$ to $3{\times} 10^{18}$\,cm (0.0002 to 1.0~pc).
Treated in a consistent way, the model contains the freely expanding
central-star wind, the inner reverse shock, the hot bubble of shocked stellar
wind gas, and the nebula proper which is separated from the bubble by a
contact discontinuity (or heat conduction front) and from the surrounding
asymptotic giant-branch (AGB) wind (halo) by an outer, leading shock. The
simulations start at the tip of the AGB and are advanced well into the
white-dwarf regime, thus covering the formation and complete evolution of a
planetary nebula.
  
The \texttt{CORONA} code treats ionisation/re\-combination and heating/cooling
time-dependently for the nine elements H, He, C, N, O, Ne, S, Cl, and Ar
(cf. Table~\ref{tab:abundances}, elements in italics) with up to 12 ionisation
stages. Altogether, non-equilibrium number densities of 76 ions are
evaluated at each time step and at all grid points. The full computational
details and atomic data implemented can be found in \citet{MS.97}.
Heat conduction is included as in SSW, but now the heat conduction formalism
holds for arbitrary chemical composition \citepalias[see][]{Sandin13}. Of
course, heat conduction is only effective within the hot bubble where the
electron density is low and the electron temperature sufficiently high.
  
Given the radial temperature and density structure of the hot bubble, we
applied the well documented \texttt{CHIANTI} code (v6.0.1, \citealt{dere.97};
\citealt{dere.09}) to compute, for selected positions along the evolutionary
sequence, the emergent X-ray spectrum, the X-ray luminosity, $L_{\rm X}$, the
characteristic (or mean) X-ray temperature, $T_{\rm X}$, and the surface
brightness (intensity) profile of the hot bubble model.  For more details
about these calculations, see \citetalias{SSW.08}.
  
We note that this method is inconsistent because the \texttt{CHIANTI} code always
assumes collisional ionisation equilibrium (CIE) while our hydrodynamical
models treat ionisation time-dependently (non-equilibrium ionisation
(NEI)) for the nine elements listed above. Given the low electron densities
of the hot bubbles together with the temperature profile imposed by heat
conduction, significant departures from the ionisation equilibrium can be
expected \citep[e.g.][]{liedahl.99, mewe.99}.
   
We touched this problem already in \citetalias{SSW.08} (cf. Fig.~1
therein) but concluded that the departures from the CIE case are not very
important, especially in the context of the low quality of the existing
X-ray observations, the uncertainty of the distances, and the approximations
used in the implementation of thermal conduction.  The present study,
however, mainly deals with bubbles of hydrogen-poor chemical compositions
where the significance of NEI effects is unknown.  Additionally, the
distances are now well known thanks to \gaia. We therefore
decided to reconsider the NEI case in order to clarify its general
importance for interpreting the X-ray emission of wind-blown bubbles.
   
To this end, we developed an interface which allows the \texttt{CHIANTI}
code to use the individual NEI ionisation fractions provided by our
\texttt{NEBEL/CORONA} code for the nine considered elements. The (standard)
CIE was adopted for the remaining elements.  Since the ions of C, N, O, and
Ne are the most prominent emitters in the X-ray regime of interest here, we
believe that this `hybrid' method suffices to provide realistic results
for comparisons with the observations.

\begin{table}
  \caption{\label{tab:abundances}  
           Chemical composition of the stellar wind (WR) and the nebular gas (PN) 
           used for computing the hydrodynamical models and synthetic X-ray spectra 
           in terms of mass fractions (Cols.~3 and 5) and (logarithmic) 
           number fractions $\epsilon$ (Cols.~4 and 6), arranged by atomic number $Z$.}

  \centering
  \begin{tabular}{r l c r @{\hspace{5mm}} c r }

    \hline \hline\noalign{\smallskip}

$Z$ & El. & \multicolumn{2}{c}{WR} & \multicolumn{2}{c}{PN}\\[1.5pt]
  \cline{3-4}  \cline{5-6} 
  \noalign{\smallskip}              
          &     & Mass & Num.    & Mass & Num.   \\
          
    \hline\noalign{\smallskip}

{ 1}    &{\it H} & 1.990(--02) & 12.00  &  6.841(--01) & 12.00 \\
{ 2}    &{\it He}& 4.160(--01) & 12.72  &  2.979(--01) & 11.04 \\
 3      &  Li    & 5.931(--11) &  2.64  &  5.931(--11) &  1.10 \\
 4      &  Be    & 1.537(--10) &  2.94  &  1.537(--10) &  1.40 \\
 5      &  B     & 2.604(--09) &  4.09  &  2.604(--09) &  2.55 \\
{ 6}    &{\it C} & 4.950(--01) & 12.30  &  6.328(--03) &  8.89 \\
{ 7}    &{\it N} & 1.000(--05) &  7.56  &  2.334(--03) &  8.39 \\
{ 8}    &{\it O} & 5.200(--02) & 11.22  &  4.851(--03) &  8.65 \\
 9      &  F     & 4.682(--07) &  6.10  &  4.682(--07) &  4.56 \\
{ 10}   &{\it Ne}& 1.400(--02) & 10.55  &  1.402(--03) &  8.01 \\
11      &  Na    & 3.336(--05) &  7.87  &  3.336(--05) &  6.33 \\
12      &  Mg    & 6.272(--04) &  9.12  &  6.272(--04) &  7.58 \\
13      &  Al    & 5.405(--05) &  8.01  &  5.405(--05) &  6.47 \\
14      &  Si    & 6.764(--04) &  9.01  &  6.764(--04) &  7.55 \\
15      &  P     & 5.925(--06) &  6.99  &  5.925(--06) &  5.45 \\
{ 16}   &{\it S} & 2.386(--04) &  8.58  &  2.386(--04) &  7.04 \\
{ 17}   &{\it Cl}& 5.028(--06) &  6.86  &  5.028(--06) &  5.32 \\
{ 18}   &{\it Ar}& 7.820(--05) &  8.00  &  7.280(--05) &  6.46 \\
19      &  K     & 3.498(--06) &  6.66  &  3.498(--06) &  5.12 \\
20      &  Ca    & 6.232(--05) &  7.90  &  6.232(--05) &  6.36 \\
21      &  Sc    & 4.513(--08) &  4.71  &  4.513(--08) &  3.17 \\
22      &  Ti    & 3.403(--06) &  6.56  &  3.403(--06) &  5.02 \\
23      &  V     & 3.458(--07) &  5.54  &  3.458(--07) &  4.00 \\
24      &  Cr    & 1.651(--05) &  7.21  &  1.651(--05) &  5.67 \\
25      &  Mn    & 9.153(--06) &  6.93  &  9.153(..06) &  5.39 \\
26      &  Fe    & 1.200(--03) &  9.04  &  1.200(--03) &  7.50 \\
27      &  Co    & 3.327(--06) &  6.46  &  3.327(--06) &  4.92 \\
28      &  Ni    & 7.084(--05) &  7.79  &  7.084(--05) &  6.25 \\
29      &  Cu    & 6.995(--07) &  5.75  &  6.995(--07) &  4.21 \\
30      &  Zn    & 1.767(--06) &  6.14  &  1.767(--06) &  4.60 \\
\hline
  \end{tabular}
  \tablefoot{The sum of the mass fractions is normalised to unity,            
             and the logarithmic number fractions are defined as
             ${\epsilon_{\rm EL} = 12\,+\,\log (N_{\rm El}/N_{\rm H})}$, where $N_{\rm El}$
             is the number density of the element in question.  
             The nine elements considered in our hydrodynamical simulations
             are given in italics (Col. 2). The mean molecular weights
             are 1.33\,/\,0.63 (neutral/fully ionised) for the PN composition
             and 5.92\,/\,1.47 (neutral/fully ionised) for the WR composition.
             The corresponding effective electronic charges for complete
             ionisation are 1.26 (PN) and 4.37 (WR), respectively.}
\end{table}

As in \citetalias{Sandin13}, we allow for a chemical stratification of the
model: The initial circumstellar envelope has always a normal hydrogen-rich
composition (dubbed `PN'), while the central-star wind may be hydrogen-poor
with an element mixture (dubbed `WR') which is typical for the Wolf-Rayet
central stars. The two sets of chemical abundances are listed in
Table~\ref{tab:abundances} in both the mass fractions normalised to unity and
the number fractions normalised to the hydrogen abundance
${ \epsilon_{\rm H} = 12 }$.

The PN mixture is representative of the composition of many planetaries of the
Galactic disk and has already been chosen for the hydrodynamical simulations
of \citet{peretal.04}. The WR composition is based on the
\citet{marco.07} analysis of BD\,+30\degr 3639. All elements not accessible
to observations were supplemented assuming solar mass fractions (see
Table~\ref{tab:abundances}).  Exceptions exist for nitrogen and neon. Since
the WR composition represents the chemistry of the (somehow) exposed
intershell region between the hydrogen- and helium-burning shell of the
progenitor star, we assume that complete hydrogen burning first converts
virtually all C and O nuclei into nitrogen, which is later converted into
neon ($^{22}$Ne) within the helium-burning shell. This scenario implies a
very low nitrogen abundance (we adopt a mass fraction of $10^{-5}$) and
justifies a very high neon abundance for the WR composition
($1.4{\times}10^{-2}$ by mass) which exceeds considerably the original
surface value that is mainly due to \element[][20]{Ne}. We note in this
context that \citet{marco.07} estimated a photospheric neon mass fraction of
${\simeq\!2}$\,\% for BD\,+30\degr 3639, consistent with the 1.4\,\% used
here.

  We emphasise that our WR element mixture is genuinely hydrogen-poor, with
  helium and carbon being the most abundant elements.  This obviously is not
  the case in the study of \citet{TA.18} in which the authors used a chemical
  composition with hydrogen still being the most abundant element for the case
  of BD\,+30\degr 3639.
   
\subsection{The characteristic bubble temperature}
\label{sub:mean.tx}

   Once the temperature and density profiles of a hot bubble are known, 
   one can compute a `characteristic' (or `mean') \hbox{X-ray} temperature 
\begin{equation} \label{eq:tx}
T_\mathrm{X} = \frac{4\pi}{L_\mathrm{X}} \int_{R_1}^{R_2}  r^2 \, T_\mathrm{e}(r) \, 
                \eta_\mathrm{X}(r)\,\mathrm{d}r 
\end{equation}
\noindent
of the bubble plasma. Here, $T_\mathrm{e}(r)$ is the electron temperature between inner, $R_1$,
 and outer radius, $R_2$, of the bubble, and
\begin{equation} \label{eq:lx}
L_\mathrm{X} = 4\pi \int_{R_1}^{R_2}  r^2 \, \eta_\mathrm{X}(r)\,\mathrm{d}r 
\end{equation}
\noindent
is the X-ray luminosity, with
\begin{equation}\label{eq:eta}
\eta_\mathrm{X}(r) = 
         \int_{E_1}^{E_2} \eta(T_\mathrm{e}(r), n_\mathrm{e}(r), \epsilon_i(r), E)\,\mathrm{d}E 
\end{equation}
   being the volume emissivity in the energy band ${E_1 - E_2}$ 
   \citep[cf.][]{SSW.08}.  While $T_\mathrm{X}$ is an emission weighted mean 
   temperature, the true temperature inside a hot bubble is a function of
   distance from the central star and is here ruled by heat conduction.
   It is therefore clear that identical temperature structures can result in different values of
   $T_\mathrm{X}$ because the latter depend also on the emissivity distribution
   (cf. Eq.~\ref{eq:tx}), and therefore on the bubble's chemical composition.
   A thorough investigation of how the X-ray emission of a bubble plasma depends on chemical 
   composition, heat conduction efficiency, and boundary conditions can be found in 
   \citetalias[][]{helleretal.16} on the basis of analytical bubble models.    
   
   Both the X-ray luminosity and characteristic temperature \Tx\ depend on the
   chosen energy window.  The high-energy limit, 2~keV (6.2~\AA), is of no
   problem since above this energy there is no or only very little emission
   for the observed bubble temperatures.  This is different at the low-energy
   end, but there the detector sensitivity becomes low and the extinction
   high.  We used throughout the paper the energy band 0.3--2.0~keV
   (6.2--41.3~\AA) because it also has been used in \citet{ruizetal.13} to
   whose compilation of X-ray data we mostly refer (see
   Appendix~\ref{appsec:calib.bandwidth} for an investigation of the influence
   of the chosen X-ray band width on luminosity and mean temperature).
   
   \citetalias{SSW.08} derived the following important relation for the mean 
   temperature $ \langle T \rangle $ of a heat-conducting bubble:
\begin{equation}   
\label{eq:tx.wind.size.2}  
   \langle T \rangle = \mbox{const}\cdot (L_{\rm wind}/R_2)^{2/7} , 
\end{equation}  
    where ${ L_{\rm wind} = \dot{M}_{\rm wind}\cdot V^2_{\rm wind} /2 }$ is the wind power
    and $ R_2 $ the bubble radius (for the details, see Eqs.~9 through 14 in \citetalias{SSW.08}).  
    In deriving Eq.~(\ref{eq:tx.wind.size.2}) it has been assumed that 
    (i)   the bubble mass is dominated by evaporated matter, 
    (ii)  the bubble is chemically homogeneous, and 
    (iii) the radiation losses are negligible. 
    Of course, a similar kind of relation also must hold for the
    characteristic temperature $ \tx $ as defined by Eq.~(\ref{eq:tx}).
   
    Equation~(\ref{eq:tx.wind.size.2}) shows that the mean temperature of a
    heat conducting bubble is not directly dependent on the wind velocity.
    Rather, the temperature depends exclusively on the actual ratio of stellar
    wind power to bubble size, independently of the bubble mass.  Therefore,
    Eq.~(\ref{eq:tx.wind.size.2}) predicts an evolution of \Tx\ while the
    central star crosses the Hertzsprung-Russell diagram (HRD), and which
    can easily be tested; see discussion in Sect.\,\ref{subsec:obs.xray} in
    the context of Fig.~\ref{fig:tx.r2.pn}).
   
\begin{figure*}
\sidecaption
\mbox{\includegraphics[trim= 0.3cm 13.3cm 0.1cm 0.8cm, width=0.7\linewidth, clip]
{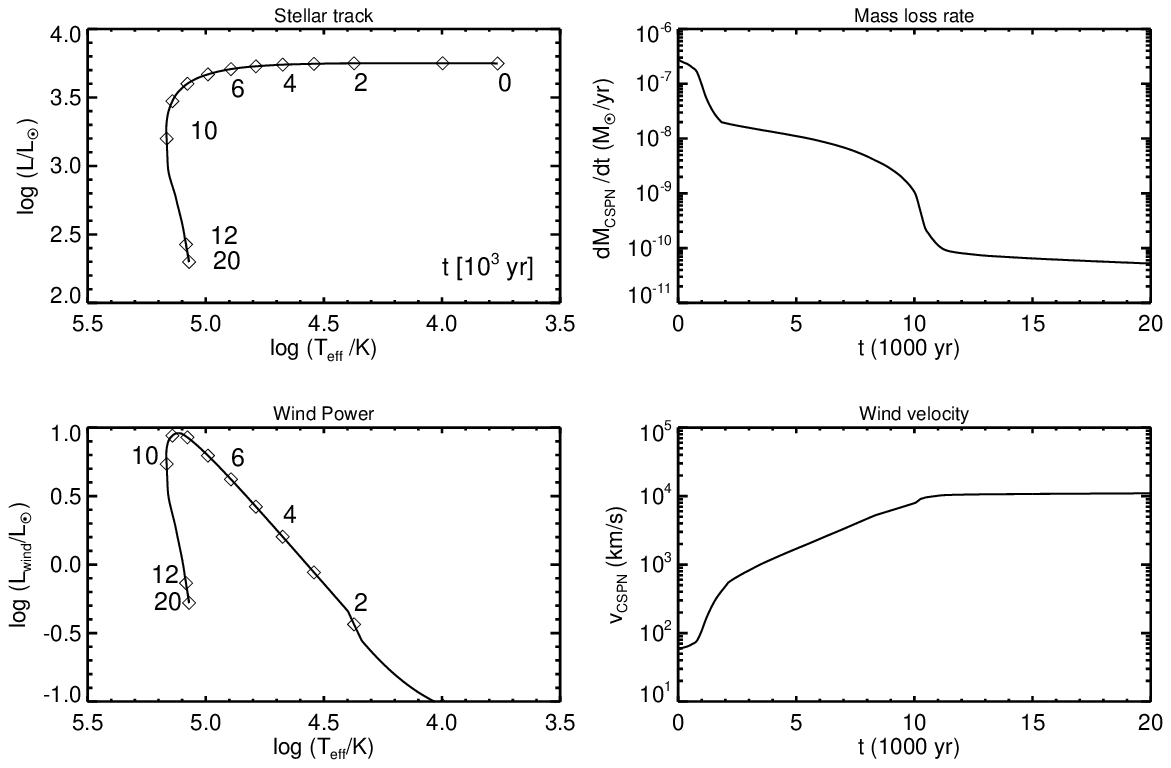}}
\caption{\label{fig:mod.prop}
         Evolution of stellar parameters and wind properties of our standard
         model of a 0.595~\Msun\ AGB remnant.
         \emph{Left column}: Run of stellar bolometric, $ L_{\rm star}/L_\sun $ (\emph{top}), 
         and wind luminosity, $\Lwind$\,=\,$\dotMw\,\Vwind^2/2$ (\emph{bottom}),
         versus stellar effective temperature.  Post-AGB ages
         $ t $ (in units of 1000 years) are indicated along the tracks by diamonds.
         \emph{Right column}:  Corresponding mass-loss rate (\emph{top}) and (terminal) 
         wind velocity versus post-AGB age $ t $ (\emph{bottom}).
         We followed the recommendations of \citet{Pauletal.88} for the central-star wind 
         (${\teff \ge 25\,000~{\rm K}}$ or ${ t \simeq 2000 }$~yr), while we assumed
         a \citep{Reim.75} wind during the transition to the nebula domain
         \citep[cf.\ also][]{peretal.04}.
         }
\end{figure*}
\subsection{Setup of the physical system consisting of a central star and an AGB wind envelope}                                      
\label{subsec:setup}

A realistic modelling of the nebular structures \citep[as done, e.g. in][]
{peretal.04}, is not in the focus of the present work and not
really necessary.  In order to keep the number of model sequences as small as
possible, we employ only a single post-AGB evolutionary model, viz. the
0.595~\Msun\ model and combine it with always the same AGB envelope as
initial outer boundary, both of which have already been used in
\citetalias{SSW.08}.  However, here we vary the strength of the stellar wind,
its chemical composition, and in some cases also the evolutionary speed across
the HRD, in order to mimic various possible evolutionary scenarios. In this
sense we treat wind power and evolutionary timescale as free and independent
parameters.  We emphasise that only the radiation field, the wind
power, and the speed of the stellar evolution enter in the hydrodynamical
simulations, but not the stellar mass.

Figure~\ref{fig:mod.prop} renders the properties of the star-wind model used
in this work in terms of stellar effective temperature and post-AGB time (our
`reference simulation').  The relevant quantity for powering any X-ray
emission is the mechanical luminosity of the stellar wind.  According to the
theory of radiation-driven winds for standard hydrogen-rich chemical
composition in the formulation of \cite{Pauletal.88}, the mass-loss rate and
the wind speed are coupled to the stellar parameters mass, luminosity, and
effective temperature.  Based on this wind prescription, the mechanical energy
transported by the wind increases during the evolution across the HRD, simply
because the slowly decreasing mass-loss rate is over-compensated by the
increasing wind speed (Fig.\,\ref{fig:mod.prop}, bottom right). However, when
the hydrogen-burning shell becomes exhausted, the mass loss rate drops sharply
in line with the stellar luminosity, causing also the mechanical wind power to
drop considerably since the wind speed remains now virtually constant with a
maximum value of about 10\,000 \kms.

Overall, the mechanical power remains rather small and does not exceed 1\%
of the stellar photon luminosity in this particular case
(Fig.\,\ref{fig:mod.prop}, left column).  According to this wind model, the
maximum of the mechanical power occurs close to maximum stellar temperature.
Only very little of the stellar mass is carried away by the wind during the
whole transition to the white-dwarf domain, viz.\
\hbox{$\approx\!3\times\! 10^{-4}$~\Msun}, which is to be compared with the
typical planetary nebula mass of a tenth of a solar mass.

We emphasise that {most} of this mass is already lost with low speed during
the first 1000 years of the transition to the planetary nebula stage. Only
during this phase we have wind speeds as low as a few 100 \kms, i.e.\ low
enough to provide post-shock temperatures of the order of $10^6$ K. These
mass-loss parameters are typical for the `early' wind and are here based on
the \citet{Reim.75} prescriptions. The total stellar mass lost during
the following nebula stage is only about $8\times\!10^{-5}$~\Msun, but this
material has a very high kinetic energy because of its large speed exceeding
1000~\mbox{\kms}, leading to (adiabatic) post-shock temperatures well in
excess of $10^7$~K.
  
The evolutionary timescale of the 0.595~\Msun\ model depicted in the upper
left panel of Fig.~\ref{fig:mod.prop} is fully consistent with the assumed
wind (mass-loss) model; that is, the evolution is driven by envelope consumption
due to both hydrogen-burning at the bottom of the envelope and mass loss from its
surface \citep{SB.93, B.95} in a consistent way.  Mass loss dominates in the
vicinity of the AGB in driving the stellar evolution while later, when
ionisation sets in, mass loss becomes unimportant and nuclear burning takes
over.
      
To simulate the evolution of nebulae with a WR-type central star,
we explicitly allow for different chemistries between the AGB envelope and the
post-AGB wind.  While the chemical composition of the envelope is kept
hydrogen-rich (PN), the post-AGB wind is switched at
${ t=0 }$~yr (cf. Fig.~\ref{fig:mod.prop}) to a hydrogen-poor composition (WR).
With such an initial setup, we implicitly assume that the conversion from a
hydrogen-rich to a hydrogen-poor AGB remnant occurs right at the end of the
AGB evolution when the remnant begins to leave the tip of the AGB.
The contact discontinuity (or the surface of the bubble) is initially
also a chemical discontinuity which will move into the bubble when evaporation
of the hydrogen-rich nebular matter occurs.

The assumption of keeping a hydrogen-rich nebular composition is justified
by the absence of any evidence that the nebulae around [WR] central stars,
or parts of them, have hydrogen-poor compositions \citep[see,
e.g.][]{GS.95, GKA.07}.
      
  It is also possible that the switch of the stellar surface from
  hydrogen-rich to hydrogen-poor occurs at a later point during the
  crossing of the HRD.  Observationally, [WR]-type nuclei show already up as
  rather cool objects ($ \teff \approx 20\,000 $~K, \citealt{hamann.97})
  around very young planetary nebulae.  The formation of these central stars
  must have occurred at or at least close to the tip of the AGB.
  Nevertheless, we set up a test simulation for which the switch from a
  hydrogen-rich to a hydrogen-poor wind occurred 1000~years after departure
  from the AGB (corresponding ${ \teff \simeq 10\,000 }$~K in
  Fig.~\ref{fig:mod.prop}).  It turns out that the resulting bubble
  structures differ only slightly from the case where this switch occurs
  immediately at the tip of the AGB.  Our choice of ${ t = 0 }$~yr as the
  starting point for the hydrogen-deficient wind therefore appears to be a
  reasonable assumption.
      
\section{Parameter study of the formation and evolution of hydrogen-deficient hot bubbles} 
\label{sec:param.study}
      
   In this section, we present and discuss our bubble simulations where a hydrogen-poor wind 
   starts at post-AGB age ${ t = 0 }$~yr and blows into the hydrogen-rich remnant of the AGB wind.
   In a preliminary study by \citet{steffenetal.12}, four simulations have been presented:
   
\begin{enumerate}
\item  The self-consistent reference simulation with the 0.595~\Msun\ post-AGB model and the
       standard \cite{Pauletal.88} mass-loss rates and wind velocities (cf. Fig.~\ref{fig:mod.prop}).
       Both the wind and nebula have the hydrogen-rich PN mixture of Table~\ref{tab:abundances}.
       
\item  The same as in 1. but with the mass-loss rate increased by a factor of 100 (wind velocity
       unchanged).   

\item  The same as in 1. but with a hydrogen-poor wind with the WR composition
       (Table~\ref{tab:abundances}) starting at $ t = 0 $.

\item  The same as in 3. but with the mass-loss rate increased by a factor of 100 (wind velocity
       unchanged).     
\end{enumerate}  

   We dubbed these four simulations `PN1' (1.), `PN100' (2.), `WR1' (3.), and `WR100' (4.),
   where PN and WR stands for the chemical composition of the stellar wind, and the numbers indicate
   the factor by which the \cite{Pauletal.88} mass-loss rate is multiplied. 
   
\begin{table*}
\caption{\label{tab:sequence}
         Parameters of the (new) hydrodynamical simulations used in this work. }
       
        \centering
\begin{tabular}{l c l l c c }  
\hline 
\hline\noalign{\smallskip}
 Sequence  &  Star/\Msun  &  Mass Loss &  Velocity &  Wind Abund. & Neb. Abund. \\
\hline\noalign{\smallskip} 
 PN1\tablefootmark{\,a}&  0.595 & $\dot{M}_{\rm PPKMH}$              & $V_{\rm PPKMH}$   & PN & PN \\
 PN100                &  0.595 & $\dot{M}_{\rm PPKMH}{\times} 100$  & $V_{\rm PPKMH}$   & PN & PN \\
PN100x5.5            &  0.595 & $\dot{M}_{\rm PPKMH}{\times} 100$  & $V_{\rm PPKMH}$   & PN & PN \\
 WR1                  &  0.595 & $\dot{M}_{\rm PPKMH}$              & $V_{\rm PPKMH}$   & WR & PN \\
 WR3V05           &  0.595 & $\dot{M}_{\rm PPKMH}\,{\times} 3$   & $V_{\rm PPKMH}$/2 & WR & PN \\ 
 WR10V05          &  0.595 & $\dot{M}_{\rm PPKMH}\,{\times} 10$  & $V_{\rm PPKMH}$/2 & WR & PN \\ 
 WR10V05x2.0      &  0.595 & $\dot{M}_{\rm PPKMH}\,{\times} 10$  & $V_{\rm PPKMH}$/2 & WR & PN \\
 WR10V05x0.25     &  0.595 & $\dot{M}_{\rm PPKMH}\,{\times} 10$  & $V_{\rm PPKMH}$/2 & WR & PN \\ 
 WR100                &  0.595 & $\dot{M}_{\rm PPKMH}{\times} 100$  & $V_{\rm PPKMH}$   & WR & PN \\
 WR100x5.5            &  0.595 & $\dot{M}_{\rm PPKMH}{\times} 100$  & $V_{\rm PPKMH}$   & WR & PN \\
 WR100V05             &  0.595 & $\dot{M}_{\rm PPKMH}{\times} 100$  & $V_{\rm PPKMH}$/2 & WR & PN \\  
 WR100V05x5.5         &  0.595 & $\dot{M}_{\rm PPKMH}{\times} 100$  & $V_{\rm PPKMH}$/2 & WR & PN \\ 
 
\hline
\end{tabular}
\tablefoot{The sequences are labelled by the chemical composition of the
  post-AGB wind (`PN' or `WR', Col.~5), followed by the mass-loss rate
  enhancement factor relative to the \citet{Pauletal.88} prescription
  (cf. Col.~3), and the corresponding wind velocity factor (cf. Col.~4).  The
  additions `\ldots x2.0' or `\ldots x5.5' mean that the evolution of the
  0.595~\Msun\ remnant is accelerated by a factor of 2 or 5.5,
  `\dots x0.25' means a decelerated evolution by a factor of four.
  \tablefoottext{a}{Sequence 6a-HC2 of \citetalias{SSW.08} but recalculated
    with the updated \texttt{NEBEL} code.}
          } 
\end{table*}  
Increasing the \citet{Pauletal.88} mass-loss rate by multiplying the wind
density by an appropriate factor while keeping the wind velocity unchanged
is, however, not sufficient to model the wind power of [WR] central stars.
We showed in \citetalias{Sandin13} that their wind speeds are only about
half as high as for their hydrogen-rich counterparts (see also
Fig.~\ref{fig:wind.model.2} and the discussion in Sect.~\ref{sec:PN.xrays}.)
In order to account for this, we also computed bubble sequences where the
WR-wind speed is reduced by a factor of two with respect to the
\citeauthor{Pauletal.88} formalism. We dubbed these simulations `WR\#V05',
where \# indicates the mass loss rate enhancement factor.
    
Since the timescale for the evolution of [WR] central stars across the
HRD is unknown, we also computed hydrodynamical simulations with modified
evolutionary timescales.  For instance, `WR100V05x5.5' means that the
evolution of the 0.595~\Msun\ central-star with a WR100V05 wind model is
`accelerated'; that is, all time labels in Fig.\,\ref{fig:mod.prop} are
divided by a factor 5.5.  The wind parameters remain bound to either the
stellar luminosity (mass-loss rate) or effective temperature (wind speed)
and evolve therefore faster as well.  This particular acceleration has been
applied to our 0.595~\Msun\ model in \citetalias{helleretal.16} in order to
mimic the evolution of BD\,+30\degr 3639's bubble.
   
In the following, we consider the evolutionary speed and the wind properties
as free parameters, adjustable if demanded by the observations. All
simulations used or computed for the present study are listed in
Table~\ref{tab:sequence}.

\subsection{Bubble formation}
\label{subsec:form.bubble}  

Although the formation and evolution of wind-blown bubbles is a well-known
physical phenomenon \citep[see e.g.][for an overview]{mellema.98},
we repeat here the essentials for the following reasons:
\begin{itemize}
\item  Quantitative numbers have rarely been given because existing hydrodynamical
       simulations were not focused on the details of bubble formation and evolution. 
       
     \item The case of WR compositions has so far only been addressed by
       \citet{ML.02}, but only for massive stars.

     \item In contrast to the cited literature, our \texttt{CORONA} code
       treats line cooling in a fully time-dependent way. 

     \item Our grid of hydrodynamical bubble sequences with heat
       conduction, time-dependent wind and stellar properties, and
       consistent chemistry-dependent radiative cooling, provides
       new quantitative insight into the physics of bubble formation and evolution.
\end{itemize}   

Whether and when a hot bubble forms and persists depends, according to
\citet{KMcK.92a, KMcK.92}, on three timescales:
the crossing time of the free wind\footnote
{The time needed for a wind parcel to reach the wind shock.}, 
the age of the bubble, and the radiation-cooling time of the bubble matter.  
With $ R_1 $ being the position of the wind shock (the inner bubble boundary) and $ R_2 $ the
position of the contact discontinuity (the bubble surface), the following three cases
are possible:
    
\begin{enumerate}
\item   If the cooling time is much shorter than the crossing time of the wind, no real hot bubble 
        exists because the thermal energy of the shocked wind matter is effectively
        radiated away.  The bubble is (geometrically) thin (${ R_1 \simeq R_2 }$) and is said to be
        `radiative' or `pressure-dominated'.  
            
\item   If the cooling time is larger than the crossing time but still shorter than the bubble
        age, a hot bubble with ${ R_1 < R_2 }$ may form, but radiation cooling plays still        
        some role.  The bubble is said to be `partially radiative'.
               
\item   If the cooling time gets larger than the age of the bubble, cooling is unimportant,
        and the bubble is said to be `adiabatic' or `energy-dominated' (${ R_1 \ll R_2 }$).     
\end{enumerate}   
   
   The conditions concerning the hot-bubble formation change while the whole system evolves.  
   At the beginning, the wind density is high and the velocity low, and therefore  the 
   crossing time is long and the cooling time very short.  A hot bubble cannot exist (case 1).
    As the system expands and gets older, the crossing time decreases because of the rapidly
    increasing wind speed.  At the same time, radiation cooling gets less effective because
    of the generally decreasing densities, and the cooling time exceeds the wind's crossing time
    at some stage.  
    A hot bubble can form: the wind shock at $ R_1 $ detaches from $ R_2 $ and moves against the ram
    pressure of the stellar wind (case 2).   
    Further expansion of the system leads eventually to case 3 where radiation cooling becomes almost
    negligible, and the hot bubble can persist independently of the crossing time.  
       
    The moment of hot-bubble formation depends, for given chemistry, not only
    on the values of mass-loss rate and wind velocity (i.e. the wind density)
    and how both evolve with time, but also on the variation of the radiation
    cooling with plasma temperature.  The main line coolants, next to hydrogen
    and helium, are highly ionised species of carbon and oxygen
    \citep[cf.][]{CT.69}.  For a solar-like composition (here PN of
    Table~\ref{tab:abundances}), the cooling function increases rapidly to a
    bump at about 0.01~MK due to (collisional) ionisation of hydrogen and
    then increases more gradually towards maxima at about 0.1~MK due to
    \element[+1][][]{He} and \element[+2][][]{C}, \element[+3][][]{C}, and at
    about 0.2~MK due to \element[+3][][]{O}, \element[+4][][]{O}. Then the
    cooling function decreases with plasma temperature until it becomes
    dominated by free-free emission for $ T \ga 10 $~MK.
   
    The extreme hydrogen-poor, carbon- and oxygen-rich WR composition leads to
    a completely different shape of the cooling function.  The first peak is
    at 0.1~MK due to \element[+1][][]{He} and \element[+2][][]{C}, followed by
    a minor bump at about 1~MK caused by \element[+4][][]{C} and
    \element[+5][][]{C}.  Both kinds of cooling functions are illustrated in
    Fig.~1 of \citet{ML.02}, demonstrating that the cooling efficiency of the
    WR composition is orders of magnitudes higher than for solar-like
    composition. Therefore, we expect a much higher influence of the
    radiation-cooling on bubble formation and evolution than for
    hydrogen-rich bubbles (\citetalias{SSW.08}, Fig.~4 therein).
   
\begin{figure}
\includegraphics[width=0.72\linewidth, angle= -90, clip]{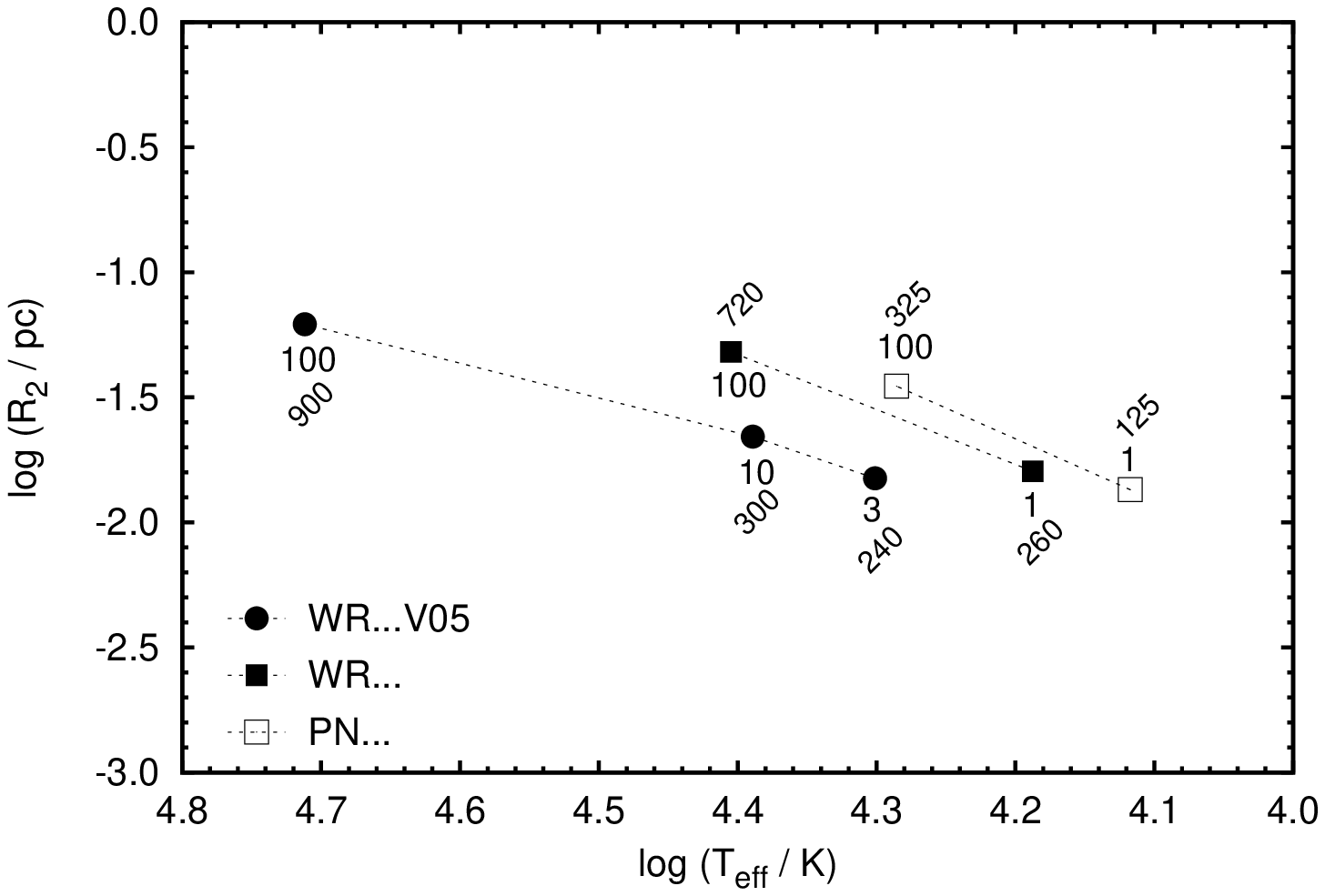}
\vskip -2mm     
\caption{\label{fig:bubb.form}  
          Positions of hot-bubble formation in a $ \teff $-$ R_2 $~diagram for three different 
          `families' of the simulations from Table~\ref{tab:sequence} (see legend). 
          The numbers along the symbols are the usual mass-loss rate factors of the simulations, 
          while the tilted numbers indicate the wind velocity (in \kms) at the moment of hot-bubble
          formation.  The dotted lines connect members of the same family. 
          } 
\end{figure}

    Figure~\ref{fig:bubb.form} shows the moment of the formation of a hot
    bubble in the $ \teff $-$ R_2 $~diagram for three different `families'
    of the hydrodynamical simulations listed in Table~\ref{tab:sequence}.  The
    formation of a hot bubble occurs when $ R_1 $ detaches from $ R_2 $,
    This figure demonstrates how severely the formation of a
    hot bubble depends on the wind properties like mass-loss rate, wind
    speed, and chemistry.
    If the evolutionary timescale is changed, the moment of hot-bubble
    formation will change accordingly. For instance, a faster evolving
    central star results in a smaller and denser bubble with shorter
    radiative timescales, and hence the bubble formation is delayed (not
    shown).

Comparing the two sets of simulations, PN1/PN100
and WR1/WR100, which have the same wind velocity but two different mass-loss
rates, we find that higher wind densities always lead to
later bubble formation.
Secondly, the bubbles of the WR1/WR100 sequences form at higher wind speeds
(or stellar temperatures) because of the higher efficiency of the radiation
cooling of the shocked hydrogen-poor WR wind matter. The difference between
the bubble simulations with the standard velocity law (WR\#) and the ones where
the wind velocity is halved (WR\#V05) is remarkable, too. For
given mass-loss rate, a lower wind velocity increases the wind density,
thereby lowering the cooling time and increasing the crossing time. Both
effects together lead to a considerable delay of hot-bubble formation.
   
As demonstrated in Fig.\,\ref{fig:bubb.form}, the formation of a hot bubble with
X-ray emission around WR-type central stars of very high wind power may be
delayed to well after the formation of the planetary  nebula proper,
corresponding to  stellar temperatures up to 50\,000~K or more where
the wind speeds are already quite high \mbox{($ \la\! 1000 $~\kms)}. In
contrast, hot bubbles around hydrogen-rich central stars form already early
at low effective temperatures and wind speeds (${ \simeq\! 100 }$~\kms),
prior to the creation of the typical planetary nebula structures.
  
We note in passing that the wind speed at hot-bubble formation found for
our PN1 sequence, 125~\kms, is nearly identical to the value estimated by
\citet{KB.90} for a 0.6~\Msun\ remnant with the \citet{Pauletal.88}
mass-loss formalism.
Our simulations are also consistent with the work of \citet{ML.02} who noted
that hydrogen-poor but carbon-rich bubbles form further away from the star at
higher wind velocities and lower wind densities than bubbles with normal
chemical composition.
  
\subsection{Bubble evolution}  
\label{sub:bubble.heatcon}

Once a hot bubble has formed, the further evolution depends on whether heat
conduction occurs within the bubble or not. Figure~4 in
\citetalias[][]{SSW.08} illustrates how heat conduction changes the structure
of the bubble.  Without heat conduction, the bubble is very hot
(${ \ga\! 10 }$~MK) and very diluted, and radiation cooling is negligible.
The typical temperature and density profile set up by thermal conduction
across the bubble results in a relatively cool and dense region at the
bubble's surface where radiation cooling may not be negligible anymore.  It
depends on the balance between the energy input by the stellar wind and the
loss by radiation whether the bubble is stable and can expand further towards
the evaporating stage.  If radiative losses exceed the energy input,
evaporation stops and may even turn into `condensation' out of the bubble
\citep{BBF.90} with a possible reduction of the bubble mass if condensation
exceeds the mass input by the stellar wind.  \citetalias{SSW.08} showed that
(i) radiation-cooling of hydrogen-rich heat-conducting bubbles is unimportant
and (ii) the evaporated mass soon dominates the whole bubble mass.  Only
because of evaporation are the models able to explain the observed
X-ray luminosities of the bubbles around O-type central stars.

In contrast, the mass evolution of bubbles with a hydrogen-poor WR chemical
composition is strongly dependent on radiative line cooling, for two reasons:
\begin{enumerate}
\item   The much higher line-cooling  efficiency of the WR matter, already discussed above.    
        Most relevant are the carbon ions \element[+4][][]{C} and \element[+5][][]{C} at about 1~MK
        and the oxygen ions \element[+6][][]{O} and \element[+7][][]{O} at about 2~MK because
        of their high abundances in a hydrogen-poor WR plasma (cf. Table~\ref{tab:abundances}).

\item   Because of the reduced thermal conduction efficiency, the temperature increases more
        steeply inwards than in hydrogen-rich bubbles.  Hence, the matter is more concentrated 
        towards the bubbles' low-temperature and denser surface layers where cooling is generally 
        most efficient.  
\end{enumerate}      

\begin{figure}
\center
\vskip -5mm
\includegraphics[width= 0.72\linewidth, angle= -90]{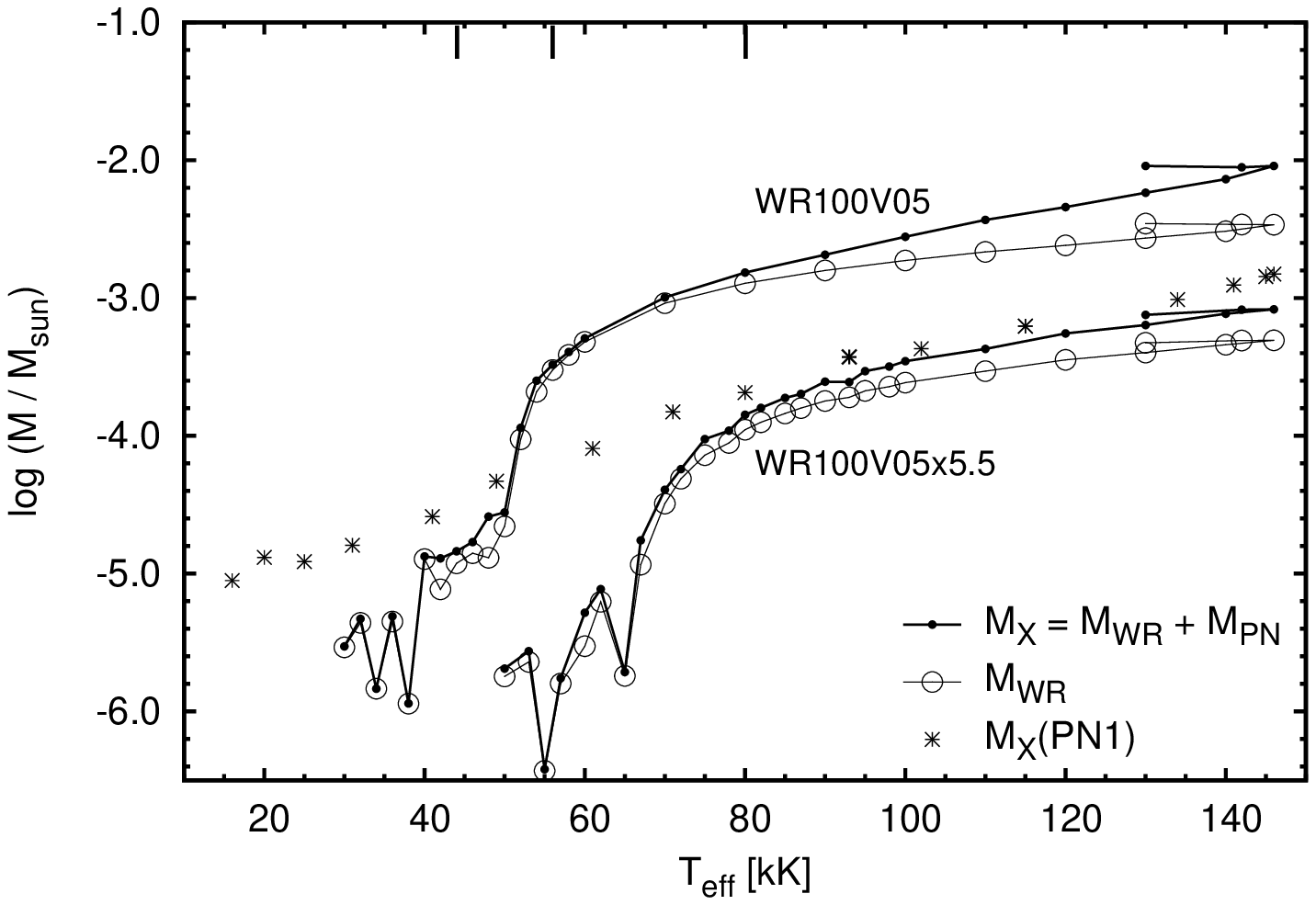}
\vskip -2mm
\caption{\label{fig:WR100V05.WR10V05}
         Evolution of different parts of the bubble mass with stellar temperature $ \teff $
         for the simulations WR100V05 and WR100V05x5.5. The total
         X-ray-emitting bubble mass is ${ M_{\rm X} = M_{\rm WR} + M_{\rm PN} }$,  
         where $ M_{\rm WR} $ is the mass contributed by the stellar wind and $ M_{\rm PN} $ that of 
         the evaporated nebular matter. The symbols mark the models for which the data were evaluated.  
         The evolution of $ M_{\rm X}(\rm PN1) $  of the hydrogen-rich PN1
         reference simulation is shown for comparison (asterisks).  
         The three bubble models of the WR100V05 simulation whose temperature profiles are displayed in 
          Fig.~\ref{fig:WR100V05.suite.bubbles} are marked along the top abscissa.    
         }
\end{figure}
   
In the following, we describe the evolution of two representative
simulations, WR100V05 and WR100V05x5.5, as depicted in
Fig.\,\ref{fig:WR100V05.WR10V05}. These models were found to be most relevant for
interpreting the observations (cf. Sect.~\ref{sec:WR-bubbles.observ}).
   
The hot bubble of the WR100V05 simulation forms out of the radiatively
controlled stage (case 1) at ${ \teff \simeq 50\,000 }$~K. At first, the hot
bubble is only fed by the (hydrogen-poor) stellar wind.  Evaporation starts
later, approximately between 60\,000 and 70\,000~K, but at a very low pace
because the plasma temperature close to the bubble's surface is around 2~MK
where cooling by \element[+6][][]{O} and \element[+7][][]{O} is very
efficient.  Beyond ${ \teff = 70\,000 }$~K, the WR100V05 bubbles get so hot
that radiation cooling decreases and the cooling time soon exceeds the bubble
age (case 3).  Evaporation of hydrogen-rich nebular matter dominates over the
mass fed by the stellar wind to such an extent that the hydrogen-rich matter
makes already up for 62\% of the total bubble mass at maximum stellar
temperature.
  
\begin{figure} [t!]
\centering
\mbox{\includegraphics[bb=14 00 580 353, width= 0.995\linewidth, clip]
{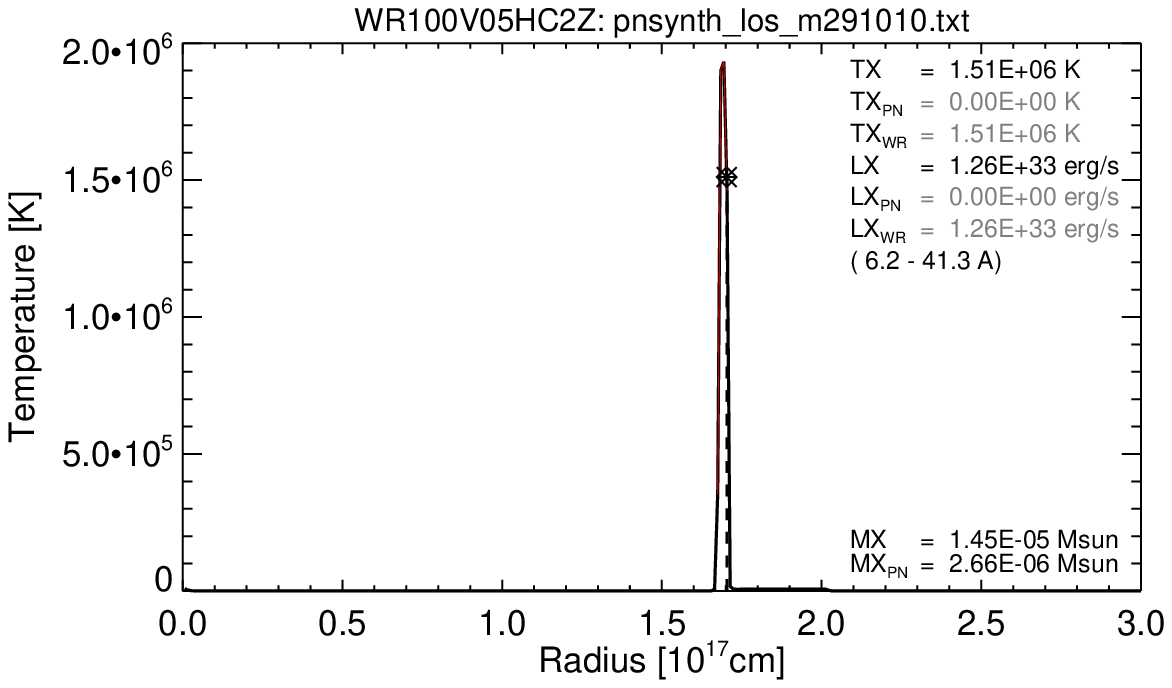}}
\vskip -4mm
\mbox{\includegraphics[bb=14 00 580 353, width= 0.995\linewidth, clip]
{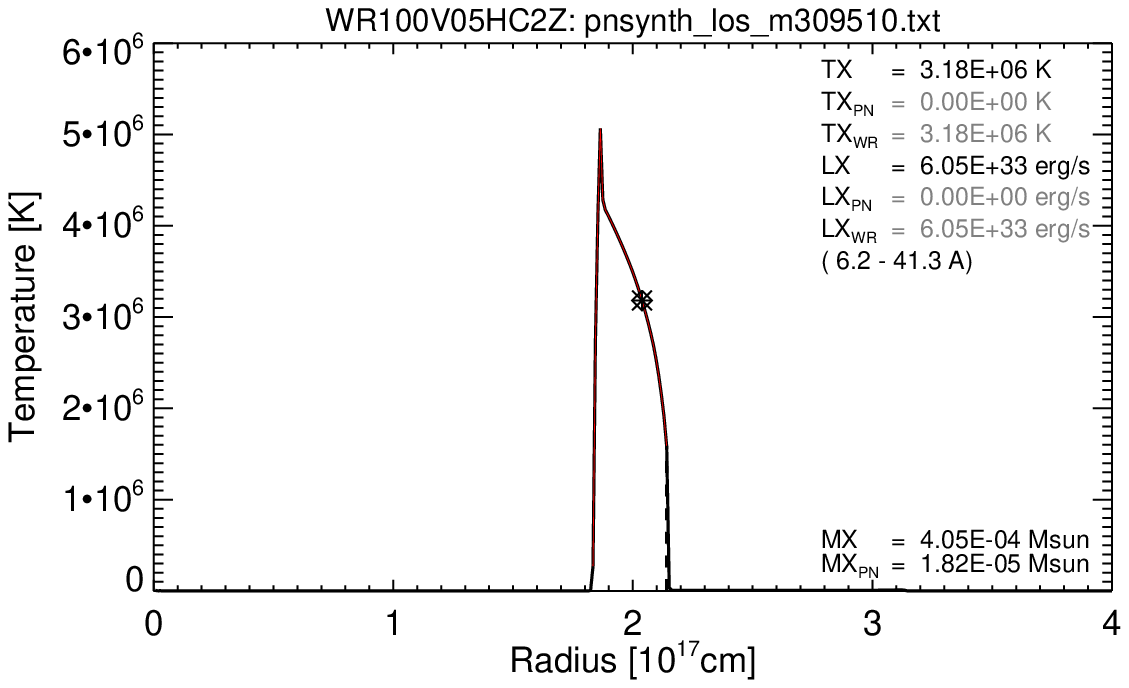}}
\vskip -4mm
\mbox{\includegraphics[bb=14 00 580 353, width= 0.995\linewidth, clip]
{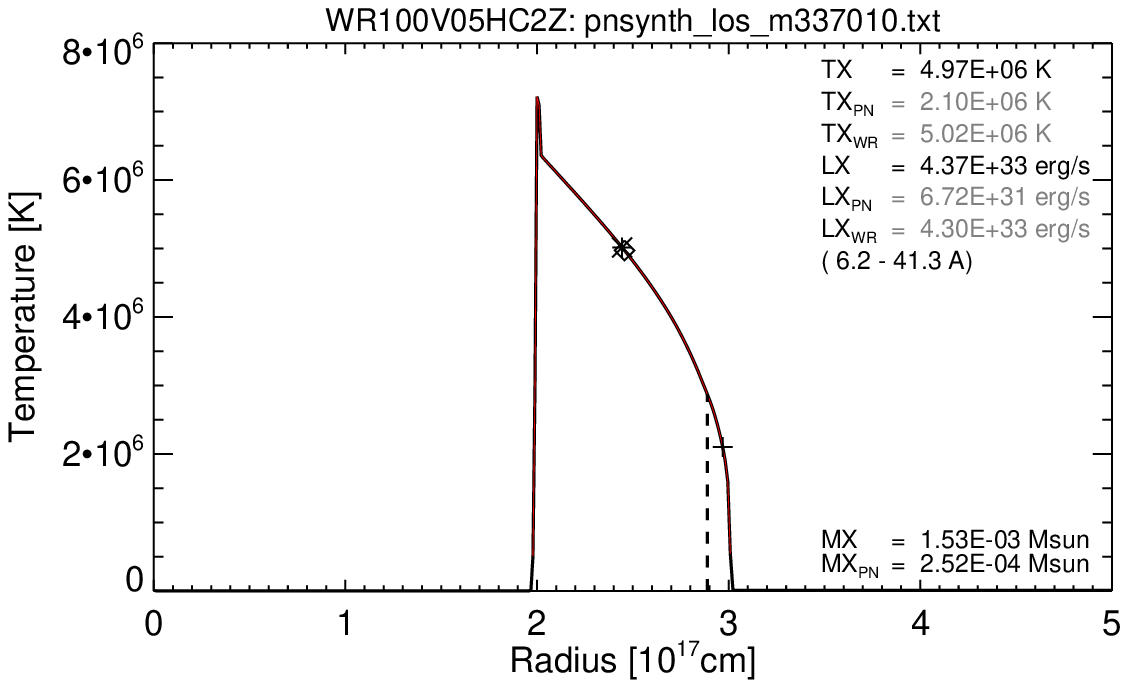}}
\vskip -5mm
\caption{\label{fig:WR100V05.suite.bubbles}
         Radial temperature profiles of three typical bubbles of the WR100V05
         simulation. From top to bottom, the models represent case 1 ($ \teff = 44\,000 $~K), 
         case 2 ($ \teff = 56\,000 $~K), and case 3 ($ \teff = 80\,000 $~K,
         with beginning evaporation).  The central star is at ${r=0 }$~cm.
         The vertical dashed lines mark the transition from the hydrogen-poor WR (left) to the
         hydrogen-rich PN chemistry (right).
         The bubble parameters like characteristic X-ray-temperatures, X-ray luminosities, 
         and masses  (for the entire bubble and for the respective subregions with the WR or PN
         composition) are given in the legends.  The symbols on the bubble's temperature profile
         mark different characteristic temperatures: of the entire bubble (diamond), the 
         hydrogen-poor part only (star), and the hydrogen-rich evaporated part only (plus). 
       }
\end{figure}

The temperature profiles of three selected bubbles from the WR100V05
simulation are shown in Fig.~\ref{fig:WR100V05.suite.bubbles}.  The top panel
contains a still radiative bubble (${ R_1 \simeq R_2 }$) with a peak
temperature well below 2~Mk where we have strong cooling by
\element[+6][][]{O} and \element[+7][][]{O}. The middle panel shows a moment
after bubble formation where the bubble's dense layers close the surface have
${ T \la 2 }$~MK, and evaporation is still suppressed by strong radiation
cooling (case 2).  Finally, the bubble temperature becomes very hot also at
the surface, so that evaporation is now in progress (bottom panel).  The WR-PN
transition started to move from the bubble-nebula interface into the bubble's
interior (dashed vertical line).
   
The WR-PN abundance transition is defined as the radius where the helium to
hydrogen ratio $n({\rm He})/(n({\rm H})=0.255$, corresponding to a chemical
composition obtained by mixing equal amounts (by mass) of PN and WR matter.
Of course, the transition from hydrogen-poor to hydrogen-rich matter is not
sharp due to numerical diffusion. Nevertheless, we assume a discontinuous
chemical profile when computing synthetic X-ray properties with \texttt{CHIANTI}.

Figure~\ref{fig:WR100V05.WR10V05} also shows the mass evolution of the
hydrogen-rich PN1 bubbles (asterisks) for comparison.  Although bubble
formation and evaporation begin much earlier, the total bubble masses remain
nearly a factor of ten lower than those of the WR100V05 bubbles.  The high
mass-loss rate inherent to the WR100V05 models more than compensates for the
late bubble formation and the even later start of evaporation.
   
The bubble evolution of the `accelerated' WR100V05x5.5 simulation is
different.  As expected, the higher wind densities for given $ \teff $ lead to
a somewhat delayed bubble formation (${ \teff \simeq 65\,000 }$)~K and beginning
of evaporation ($ \teff \simeq 80\,000 $~K).  Because of the fast evolution,
the bubble mass remains correspondingly smaller, especially also the
evaporated hydrogen-rich mass (see Fig.~\ref{fig:WR100V05.WR10V05}).
At the end of our simulation, the hydrogen-rich bubble part makes up
for about 40\% of the total mass.
      
Based on these hydrodynamical simulations of hydrogen-poor bubbles we come to the following conclusions:
\begin{itemize}
\item   After a bubble with WR chemical composition has formed, it can remain hydrogen-free 
        for some time thanks to efficient radiation cooling that does not
        allow appreciable evaporation.   
        For the particular case of our WR100V05 simulation, 
        the chemically homogeneous stage persists for about 1500~years.
        The existence of chemically homogeneous hydrogen-poor bubbles is therefore
       not necessarily an indication that heat conduction is suppressed (by whatever means).
       
     \item  However, once evaporation has started the WR bubble becomes quickly
        chemically stratified by evaporated nebular matter, and rather soon
        the hydrogen-rich mass fraction may even exceed the WR matter fraction provided by the
        stellar wind. The hot bubbles of old PNe with a WR-type central star are therefore
        expected to show signatures of a chemical stratification.
        
\end{itemize}   
   
\begin{figure*}
\vskip -2mm
\hspace*{-3mm}
\mbox{\includegraphics[bb=4 28 580 750,clip=true,width=0.34\textwidth]
  {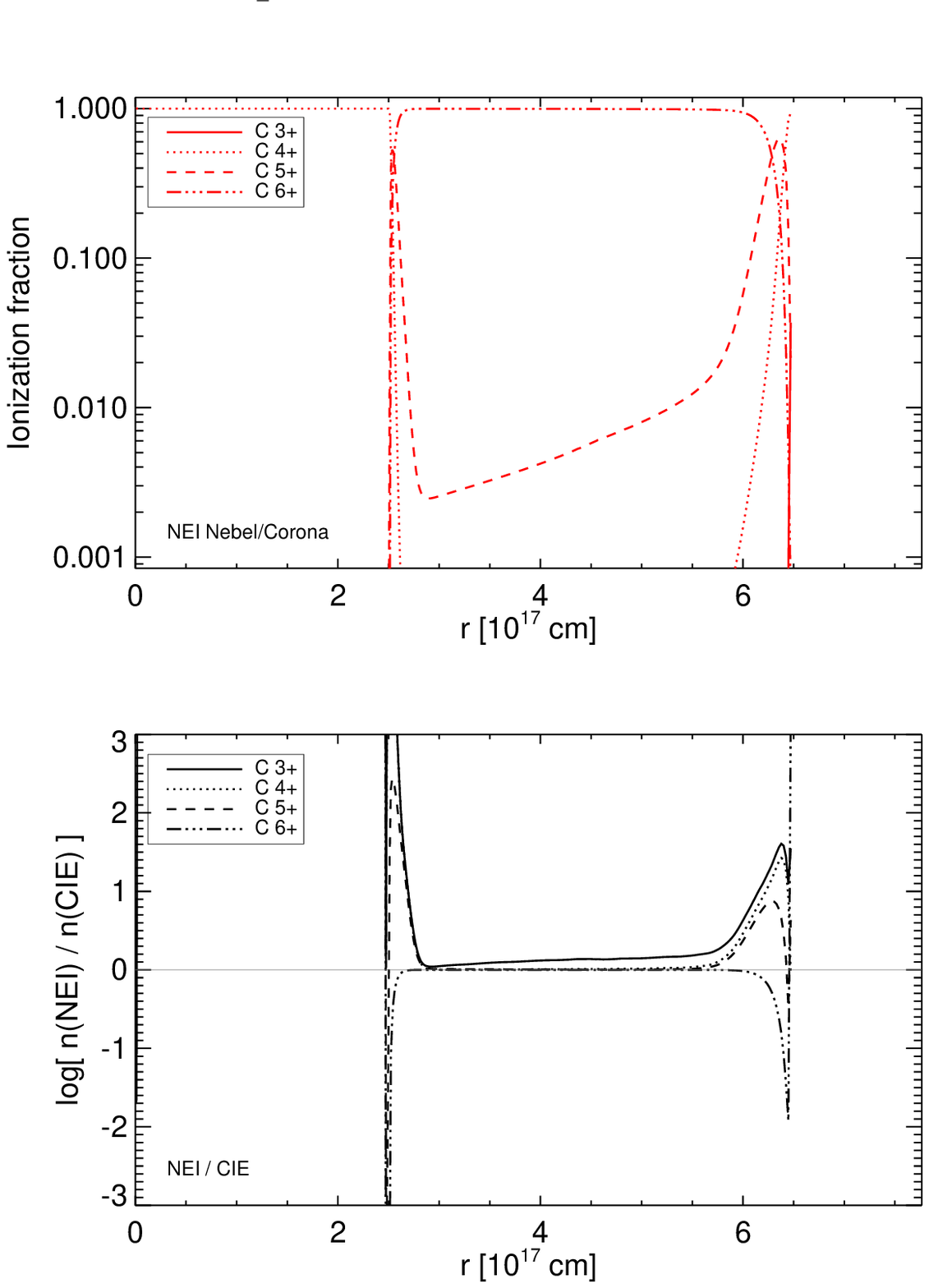}}
\hspace*{-3mm}
\mbox{\includegraphics[bb=4 28 580 750,clip=true,width=0.34\textwidth]
  {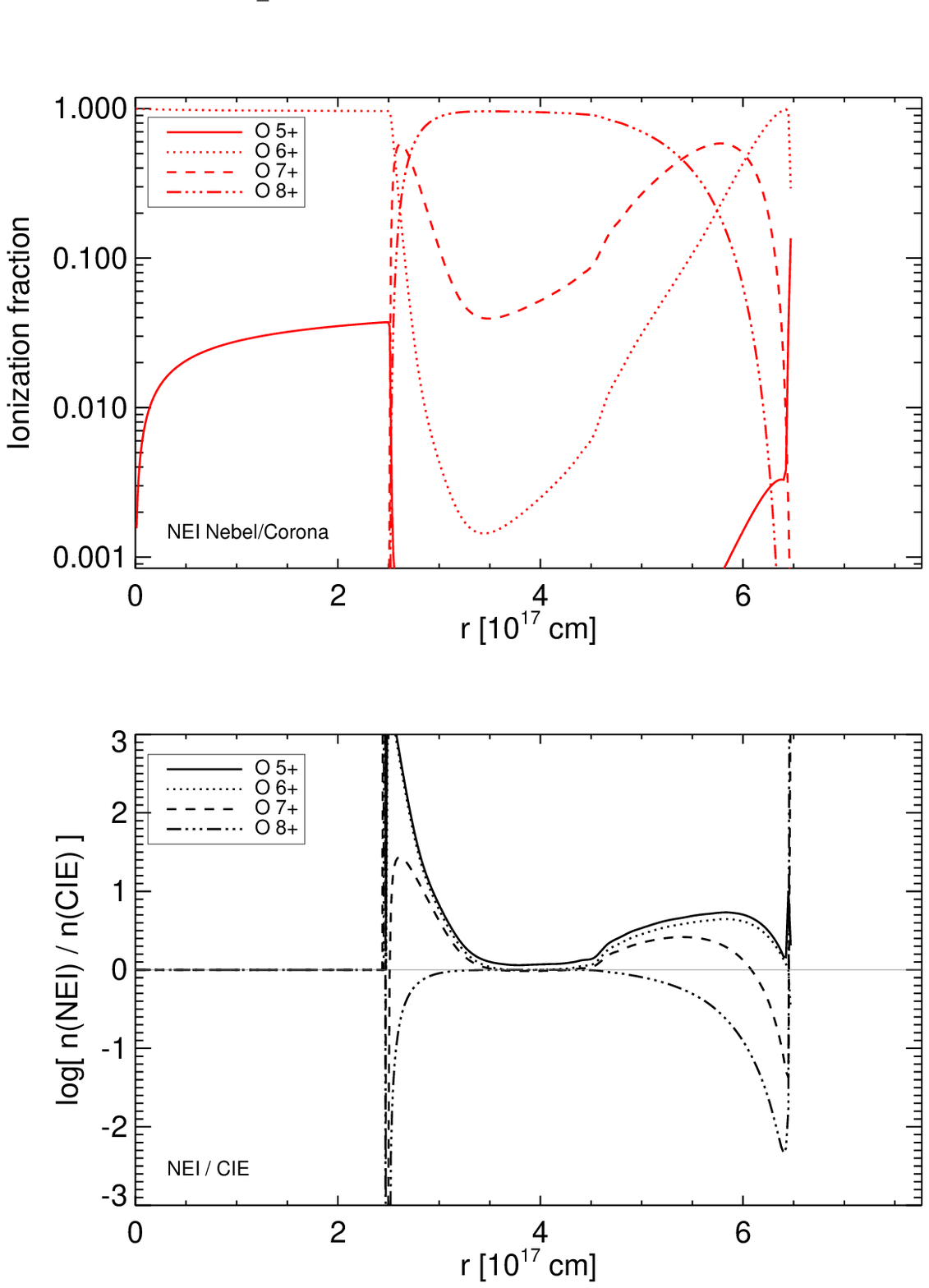}}
\hspace*{-3mm}
\mbox{\includegraphics[bb=4 28 580 750,clip=true,width=0.34\textwidth]
  {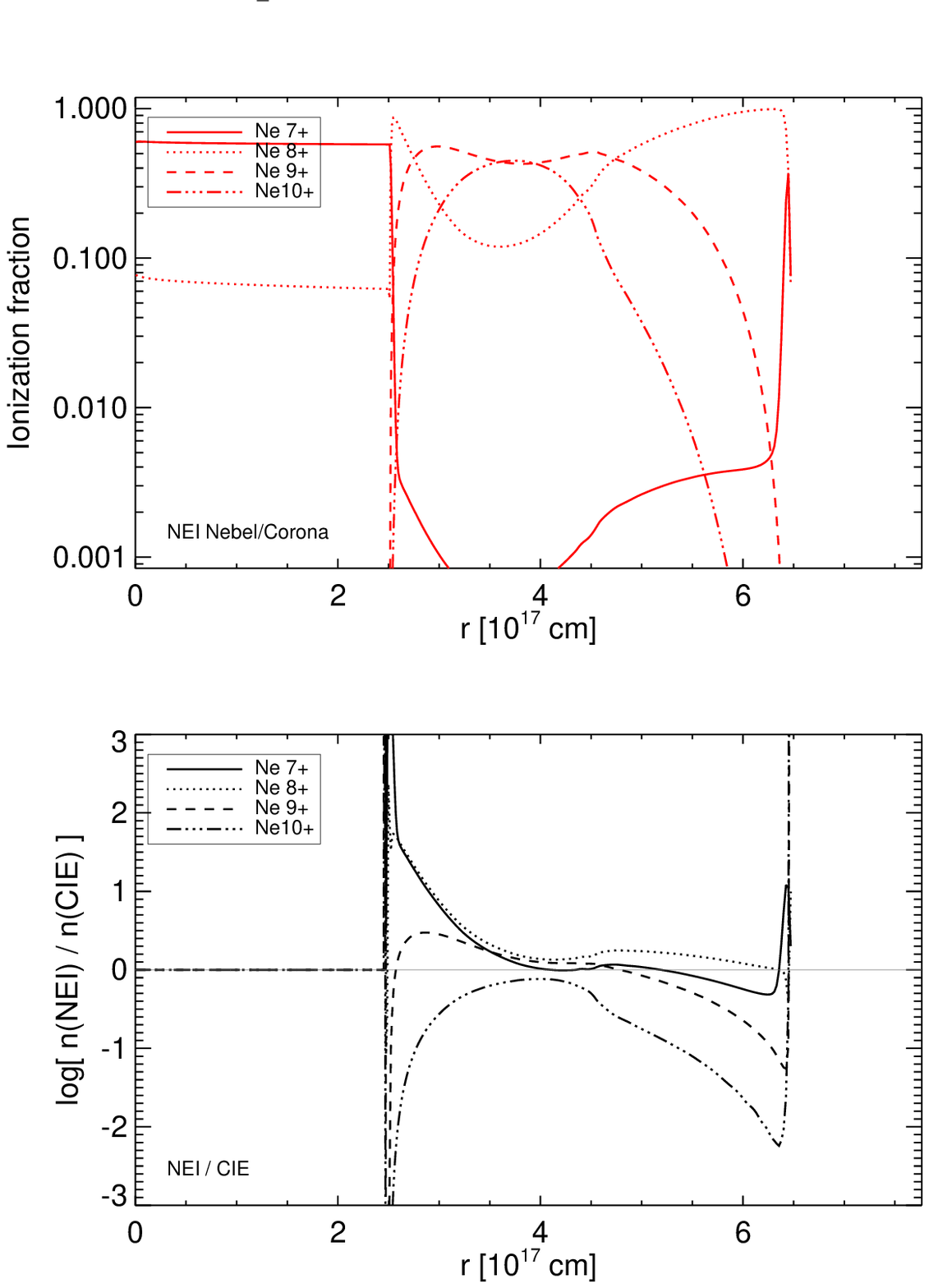}}
\vskip 0mm
\caption{\label{fig:ioneq}
  Ionisation structure of the WR100V05 model displayed in Fig.~\ref{fig:bubble.structure}.
  The \emph{upper panels} show the radial profiles of the NEI ionisation
  fractions for the last four ionisation stages of carbon (\emph{left}), oxygen
  (\emph{middle}), and neon (\emph{right}) as computed with \texttt{NEBEL/CORONA}.
  The NEI to CIE logarithmic number density ratios are plotted in
  the respective \emph{bottom panels}. The wind shock is at   ${R_1 = 2.5\times 10^{17}}$~cm,
  the bubble's surface (conduction front) at  ${R_2 = 6.5\times 10^{17}}$~cm.
  }
\end{figure*}    

\subsection{The ionisation inside a wind-blown bubble}
\label{subsec:ion.stage.bubble}

\begin{figure}[t!]
\hspace*{4.5mm}
\mbox{
\includegraphics[trim= 1cm 0.0cm 1cm 2.2cm, width= 0.93\linewidth, clip]
{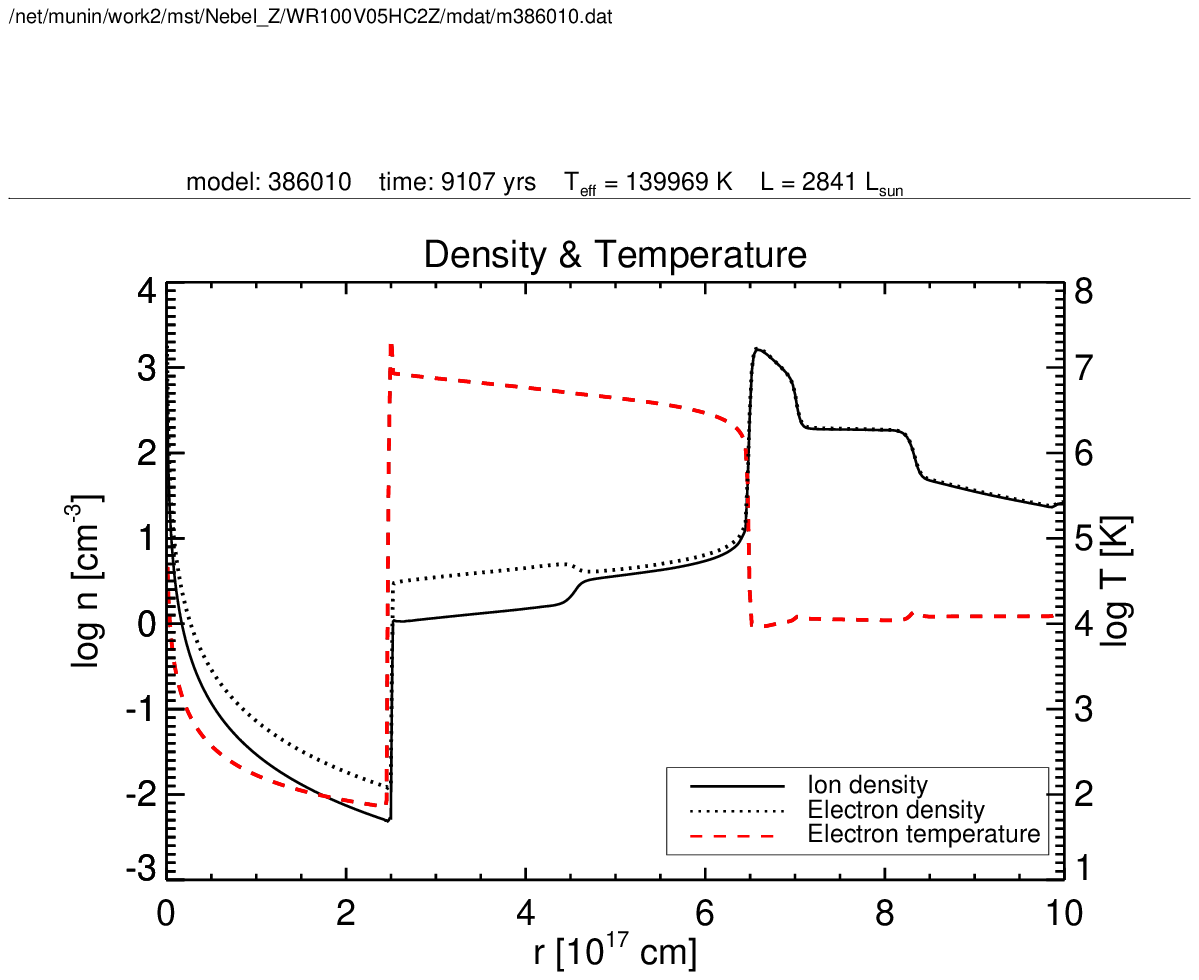}}
\vskip 2mm
\mbox{
\includegraphics[trim= 0cm 0.5cm 0cm 2.0cm, width= 0.97\linewidth, clip]
{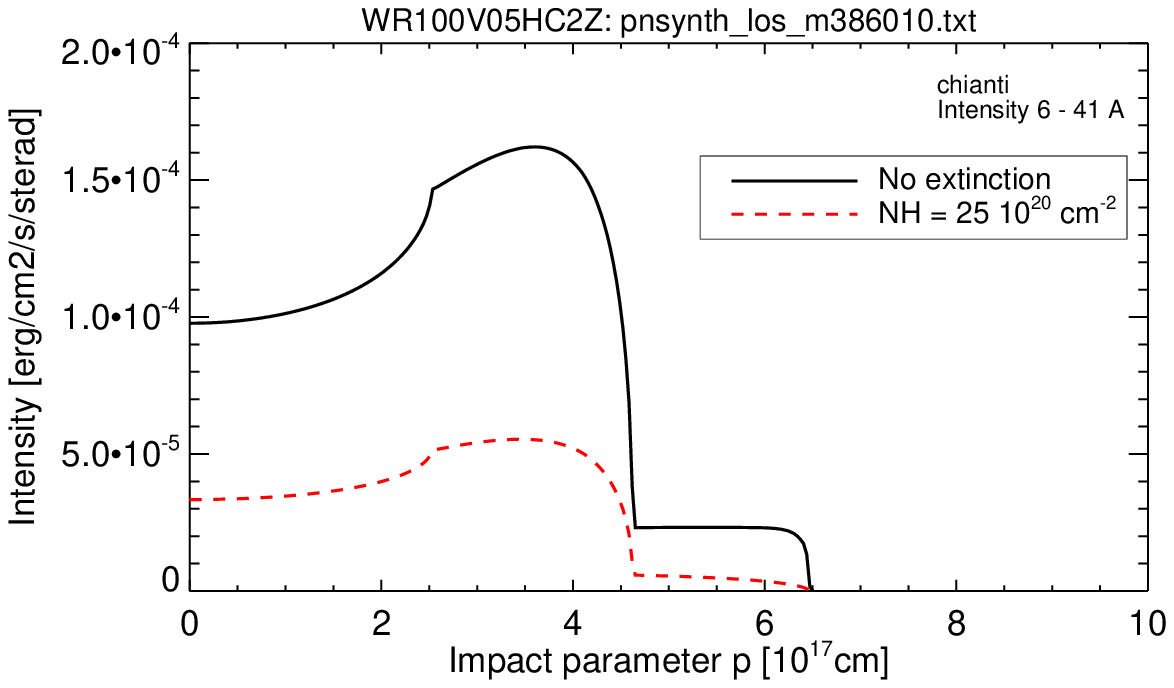}}
\vskip 2mm
\mbox{
\includegraphics[trim= 0cm 0.5cm 0cm 2.0cm, width= 0.97\linewidth, clip]
{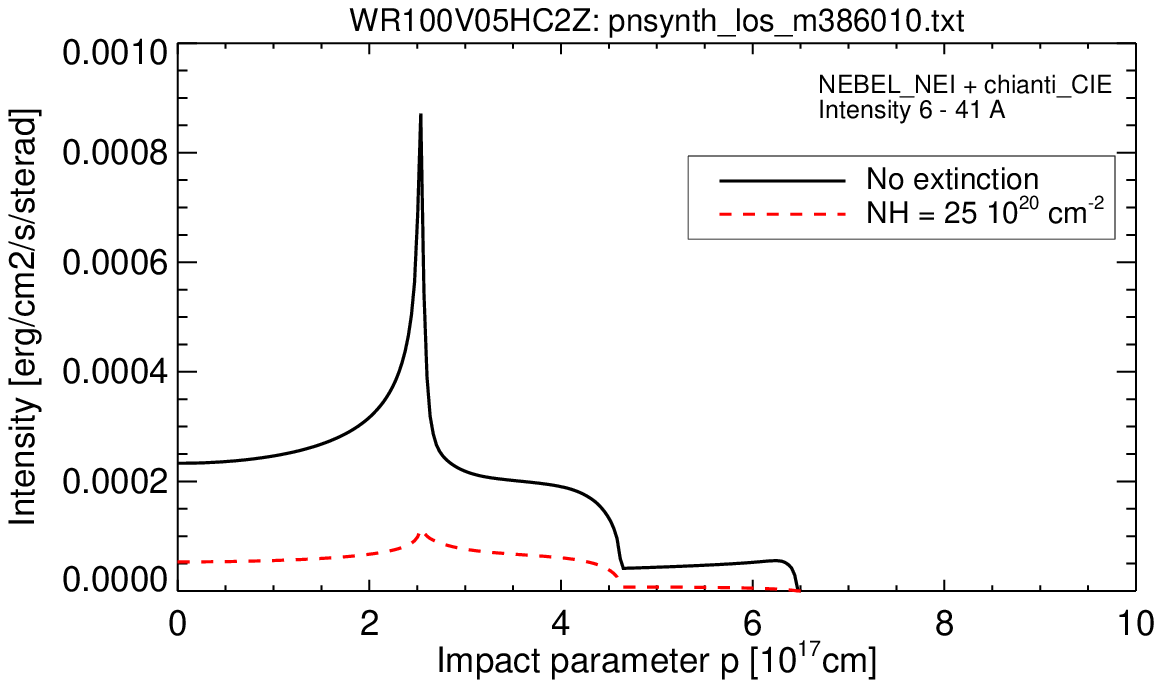}}
\caption{\label{fig:bubble.structure}   
         Hot-bubble structure of the WR100V05 simulation at 
         ${t = 9107}$ yr, ${\teff = 139\,968}$ K, ${L_\star = 2841} $ \Lsun. 
         {\em Top}: radial profile of electron temperature (dashed, right ordinate),
         electron density (dotted), and ion density (solid).  
         The star is at ${r=0}$, the wind's reverse shock at ${R_1 = 2.5\times 10^{17}}$~cm.
         The transition from the wind-shocked hydrogen-poor WR matter to the evaporated nebular
         PN matter is at ${r = 4.6\times\! 10^{17}}$~cm ($T \simeq\! 5.0$~MK), 
         the heat conduction front (bubble-nebula interface) at ${R_2= 6.5\times\! 10^{17}}$~cm, 
         and the high-density region between 6.5$\times 10^{17}$ and 
         8.3$\times 10^{17}$ cm is the fully ionised planetary nebula proper with its double-shell 
         structure and an electron temperature of about $1.2\!\times\! 10^4$~K. 
         {\em Middle}: X-ray surface brightness distribution (6--41~\AA) for the same model,
         assuming CIE, before (solid) and after (dashed)
         absorption by a hydrogen column density of ${ N_{\rm H} = 0.25\times 10^{22}\rm\ cm^{-2} }$.  
         The masses contained in the two bubble parts are 
         $ 3.05\times\!10^{-3} $~\Msun\ of WR matter, supplied by the stellar wind, and
         $ 4.24\times\!10^{-3} $~\Msun\ of evaporated PN matter.  
         {\em Bottom}: same as middle panel, but assuming NEI of nine key elements. 
         }
\end{figure}

The implicit assumption commonly used for the interpretation of nebular X-ray
emission is that the hot plasma within a wind-blown bubble is in CIE; that is, ionisation due to electron collisions is in
equilibrium with radiative recombinations. In this case, the ionisation
structure depends only on the temperature structure and adjusts
instantaneously to any spatial and temporal variations. It can easily be seen,
however, that the physical conditions within a tenuous hot bubble are far from
favourable for the establishment of CIE \citep[see, e.g.][]{liedahl.99}.
First of all, we are dealing with a very rarefied entity with correspondingly
long ionisation/recombination timescales. Furthermore, the electron
temperature jumps by orders of magnitude when the wind matter crosses the
inner wind shock, which makes it difficult for the ionisation to adjust quickly
to the local temperature.  Also, the flow of matter inside the bubble
encounters a radially decreasing (or increasing) plasma temperature
(cf. Fig.~\ref{fig:bubble.structure}) that can drive ionisation away from
CIE. Phases of very strong departures from equilibrium ionisation are expected
(i) during the early bubble evolution where the mean bubble temperature
increases quite rapidly, and (ii) when the stellar luminosity and the wind
power drop while the star settles onto the white-dwarf sequence.
   
Indeed, our hydrodynamics simulations clearly show that NEI is the prevailing situation within a wind-blown bubble. This
concerns especially the early radiative stage but more or less all following
evolutionary phases as well. As an example, we show in Fig.~\ref{fig:ioneq}
the ionisation fractions of the important elements carbon, oxygen, and neon
for the NEI case (top) and the corresponding NEI to CIE number density ratio
(bottom). We selected the same model of the WR100V05 bubble sequence that is
also depicted in Fig.~\ref{fig:bubble.structure} below. It belongs to a late
evolutionary phase where (i) the mean bubble temperature does not change
rapidly, and (ii) the evaporation of PN matter is already quite
efficient. Therefore this model allows us to disentangle the influence of chemistry
and flow of matter on the ionisation in the hot bubble. Inspection of the
figure reveals two regions that are especially prone to non-equilibrium
ionisation:
\begin{itemize}
\item  downstream of the wind shock (post-shock region) where the 
       rather cool and only photo-ionised wind matter must adjust to the now very high electron 
       temperature, and
\item  behind the conduction front where the evaporated matter flows inwards, facing a steadily
       increasing electron temperature.
\end{itemize}   
The only possibility to achieve or come close to the CIE conditions is given
when the matter has left these two more extreme regions.  But we see also that
there are differences from element to element.  Carbon has the lowest number
of ionisation stages of the elements considered here and reaches full
ionisation under CIE conditions in most parts of the bubble. Departures only
occur in the post-shock region and behind the conduction front where
\element[+5]{C} dominates (cf. left panel of Fig.~\ref{fig:ioneq}), but large
regions in the bubble interior are close to CIE.
  
In the case of oxygen with its two additional ionisation stages (middle panel
of Fig.~\ref{fig:ioneq}), the region where CIE is a good approximation shrinks
considerably with respect to carbon. The whole bubble part containing
evaporated matter ($r > 4.5\times 10^{17}$\,cm) is in non-equilibrium, and
only the inner region containing hydrogen-poor wind matter is practically in
ionisation equilibrium. Presumably, the relatively high electron density, in
conjunction with the high oxygen abundance, favours CIE.

The deviations from CIE are even more extreme for neon: only in a very small
region centered at ($r \approx 4\times 10^{17}$\,cm) are all ionisation stages
reasonably close to the equilibrium values predicted by \texttt{CHIANTI}
(Fig.~\ref{fig:ioneq}, lower left panel).
    
We conclude that the ionisation of wind-blown PN bubbles is, in general, far
from CIE. The differences between the CIE and NEI cases depend on the element
and are generally smaller in the central regions of high electron density
and/or low flow velocity. The general trend is that the NEI ionisation always
lags behind the CIE predictions. As will be shown below, the X-ray
emission depends critically on the details of the ionisation structure inside
the bubble. Since all previous interpretations of the diffuse X-rays from
wind-blown bubbles assumed ionisation equilibrium, we provide both
CIE and NEI results in the present work.
             
\subsection{The structure and X-ray emission of chemically stratified heat-conducting bubbles}
\label{subsec:bubb.neb}

   Here we present an overview of the typical structure of a heat-conducting, wind-blown bubble
   with stratified chemical composition. 
   As an example, we render in Fig.~\ref{fig:bubble.structure} the full radial
   structure of a complete hydrodynamical model (top) and its X-ray surface brightness
   as function of the impact parameter (middle and bottom), taken from the WR100V05 sequence 
   close to the maximum of the stellar temperature at ${\teff \simeq 146\,000}$~K.  
   
   The top panel encompasses the freely expanding wind, the hot bubble proper
   consisting of shocked wind and evaporated nebular matter, and the nebular structure. 
   Because the bubble is isobaric, the different element mixtures
   produce jumps of the ion and electron densities at the
   composition transition.

   Contrary to the particle densities, the run of the electron temperature is smooth across the
   composition change (red dashed line in the top panel of Fig.~\ref{fig:bubble.structure}).  
   However, there is a very small change of the gradient at the position of the composition change.
   The temperature profile is a bit steeper in the
   hydrogen-poor part of the bubble because of the lower heat-conduction efficiency of the 
   hydrogen-poor WR composition.  A detailed discussion how the heat-conduction efficiency and 
   the bubble's temperature structure depends on the chosen element composition is given in 
   Papers~I and II.
    
   The synthetic X-ray brightness distributions shown in the middle and bottom
   panels of Fig.~\ref{fig:bubble.structure} clearly reflect the two different
   chemistries: The outer, hydrogen-rich part is significantly fainter
   than the inner part containing the WR matter. The reason is the much higher
   X-ray emissivity of the WR plasma with its high mean electronic charge of
   ${ Z\simeq 4 }$. The shape of the brightness profile and its absolute
   magnitude depend sensitively on the treatment of ionisation. Assuming
   non-equilibrium ionisation (on the basis of our hybrid approach outlined in
   Sect.\,\ref{subsec:gen.aspects}) we see a geometrically extremely thin
   `spike' of very high X-ray intensity immediately behind the
   wind shock at ${R_1 = 2.5\times 10^{17}}$~cm (bottom). This is the signature
   of the post-shock transition region where the shock-heated gas is still
   far from being in ionisation equilibrium. Closer inspection reveals that
   line emission of \cv\ at 34.97 and 40.27~\AA, originating from a
   \element[+5]{C} `pocket' just behind the reverse wind shock, is
   primarily responsible for this emission spike, with minor
   contributions of \ovii\ lines. Unsurprisingly, the intensity spike
   disappears when CIE is assumed (middle panel).
  
   Observationally, the (intrinsic) intensity distribution is distorted by
   extinction because the low-energy X-rays from the cool outer bubble regions
   are more affected by absorption than the high-energy X-rays from the hot
   inner parts. The same holds for the post-shock intensity spike whose
   amplitude is significantly reduced after extinction (dashed line in bottom
   panel of Fig.\,\ref{fig:bubble.structure}) because the lower-ionised
   species emit preferentially at lower energies.
    
    The dominant X-ray emission from the hotter WR bubble region has a profound influence on the 
    mean bubble temperature.  The definition of Eq.~(\ref{eq:tx}) leads to a formally much higher value 
    of \Tx\ than for a chemically homogeneous bubble.
   The calibration of the mean bubble temperature against line ratios breaks down because both 
   depend differently on the radial position of the chemical discontinuity within the bubble:  
   \Tx\ is an emissivity weighted mean over the entire bubble whereas the spectrum is always 
   dominated by the hotter, inner hydrogen-poor bubble part. We therefore determined separately the mean 
    temperatures of both bubble parts and found, assuming NEI (CIE) for the bubble shown in 
    Fig.~\ref{fig:bubble.structure}, $\tx({\rm WR}) = 6.70\;(5.86)$~MK and 
    $\tx({\rm PN}) = 2.68\;(2.87)$~MK, respectively. The characteristic temperature of the whole
    bubble is $\tx = 5.68\ (5.18)$~MK, i.e. close to the mean temperature of the hydrogen-poor WR matter.
    
   Observationally, only the mean temperature of the entire bubble can, in principle, be determined
   from the bubble spectrum, provided the chemical stratification is known a priori,
   which is usually not the case.
   A thorough discussion of how the spectral appearance of a bubble depends on the position of
   the WR-PN chemical transition can be found in \citetalias{helleretal.16} (Figs.~16 and 17).
   
   According to our models, the image of an evolved WR bubble with ongoing 
   evaporation should consist of an X-ray-bright central region confined by 
   a ring of very low emission (cf. Fig.~\ref{fig:bubble.structure}, lower panels).
   The intrinsically bright but thin post-shock region will be difficult to detect
   given the limited spatial resolution and comparatively poor photon statistics of
   existing observations. This is especially true for the hydrogen-rich bubbles with their much
   lower carbon abundance.
             
\section{Planetary nebulae with diffuse X-ray emission}     
\label{sec:PN.xrays}

  In this section we compile data of planetary nebulae with diffuse X-ray
  emission. With the advent of reliable distances provided by the \gaia\
  satellite, we found it worthwhile to reevaluate the object parameters
  to put the following investigation and the conclusions of earlier publications
  (\citetalias{SSW.08}, \citealt{ruizetal.13}) onto a firm basis. 

\begin{table*}
\caption{\label{tab:data}
          Compilation of stellar and X-ray properties of the objects discussed in this paper. 
         }       
\centering 
\tabcolsep=3.7pt
\begin{tabular}{r l c c c c c c l c c c c c l}
\hline \hline\noalign{\smallskip} 
 No. & ~~ Object  & Sp. T.  & $ d $   & $ \teff $    & $ \log L_{\rm star} $    
         & $ \log\dot{M}_{\rm wind} $  & $ \Vwind $   & ~~Refs. & $ \log \Lwind  $ 
         & $ \log \lx $                & $ R_2 $      & $ \tx $ &  Refs.         \\[2pt]       
     & &  & [kpc]  & [kK]  & $\rm [L_\sun]$ & [\Msun/yr]  & [km/s] &  & $\rm [L_\sun]$  
          & $\rm [L_\sun]$  &  [pc]  &  [MK]  &  \\[2pt]                   
(1) & ~\quad (2) & (3) & (4) & (5) & (6) & (7)  & (8) & ~~~(9) & (10) & (11) & (12) & (13) & (14)
  \\[1.0pt]
\hline\noalign{\smallskip}  
 1 & \object{BD\,+30\degr 3639}
                  & [WC9] & 1.61  & 46  & 3.96 & $ -5.40 $ & ~~730 & (3, 4, 5)    & 2.24                          
                  & $-0.96 $ & 0.015 & $ 1.8\!\pm\! 0.1 $ & (1) \\
 2 &                          
\object{IC\,418}  & O7fp  & 1.37      & 36    & 3.84 & $ -7.06 $ & ~~650 & (6, 7, 8, 9) &  0.48
                          & $ -3.55 $ & 0.012 & {\it 3}  &  \\   
 3 &
\object{IC\,4593} & O7fp  & 2.63      & 40    & 3.77 & $ -7.60 $ & ~~830 & (17)         &  0.15
                          & $ -2.97 $ & 0.025 & {\it 1.7} & (17) \\                                                              
 4 &
\object{NGC\,40}  & [WC8] & 1.79     & 71    & 3.88  & $ -5.61 $ & 1000  & (3, 10)      &  2.31
                          & $-1.71 $ & 0.158 & {\it 1}  &     \\                                                      
 5 &                          
\object{NGC\,2392}& O6f & 1.82       & 44    & 3.87  & $ -7.52 $ & ~~370 &(6, 8, 9, 11)&$ -0.47$~~~   
                          & $-2.02 $ & 0.062 & $ 2.0^{+0.1}_{-0.3} $ &  (18)   \\                                                                                        
 6 &         
\object{NGC\,3242}& O(H)  & 1.33     & 79    & 3.73  & $ -8.27$   & 2350 & (6, 8, 9)  &  0.39
                          & $-2.47 $ & 0.046 & $ 2.2\!\pm\! 0.1 $ &  \\    
 7&
\object{NGC\,5189}& [WO1] & 1.47     &165~~  & 3.58  & $ -6.67 $  & 2500 & (15)       &  2.04
                          & $-1.26 $ & 0.313  & $ 1.6\!\pm\! 0.1 $ & (16) \\                               
 8 &                          
\object{NGC\,5315}\tablefootmark{\,a} 
                  & [WO4] & 2.50     & 75    & 3.69  & $ -5.83 $  & 2300 & (3, 5, 12) &  2.81
                          & $-1.16 $ & 0.012 & $ 2.7\!\pm\! 0.3 $ &     \\ 
 9 &                         
\object{NGC\,6543}& wels  & 1.37     & 64    & 3.57  & $ -7.05 $  & 1420 & (11, 13)   &  1.17                           
                          &$ -1.85 $ & 0.036 & $ 1.7\!\pm\! 0.1 $ &  \\ 
 10 &                        
\object{NGC\,6826}& O6fp  & 1.30     & 47    & 3.56  & $ -7.44 $  & 1200 & (6, 8, 9)  &  0.64
                          &$ -3.29 $ & 0.032 & {\it 2.3}  &  \\                                             
 11 &                        
\object{NGC\,7009}& O(H)  & 1.23     & 82    & 3.51  & $ -8.65 $  &  2770 & (6, 14)   &   0.15
                          &$ -2.08 $ & 0.070 & $ 1.8\!\pm\! 0.2 $ &  \\    
 12 &                                             
\object{NGC\,7026}\tablefootmark{\,b}  & 
                    [WO3] & 3.23     & 126~~ & 4.80  & $ -5.83 $  & 2000 & (14)       &    2.69
                          &$ -0.38 $ & 0.191 & $ 1.1^{+0.5}_{-0.2} $  & (2) \\[2pt]
\hline   
        
\end{tabular}   
\tablefoot{Given next to object and spectral type are distance (Col.~4),
         stellar effective temperature (Col.~5), stellar bolometric luminosity (Col.~6),
         stellar mass-loss rate (Col.~7), terminal stellar wind velocity (Col.~8), references 
         of adopted stellar parameters (Col.~9), stellar-wind luminosity (Col.~10), 
         bubble X-ray luminosity (Col.~11), hot-bubble radius (Col.~12), 
         characteristic bubble temperature (Col.13), and reference thereof (Col.~14).
         Distances based on the \emph{Gaia} Data Release 3 (DR3), except for NGC~5315. 
         Care was taken to ensure that all distance-dependent data are consistent with the
         distances given in Col.~4.
         The X-ray luminosities and X-ray temperatures refer to the energy range 0.3--2.0~keV
         (6.2--41.3~\AA) and are from \citet{ruizetal.13} if not otherwise noted; 
         \Tx\ values derived from spectra of low quality are in italics.\\                                
\tablefoottext{a}{Distance based on an assumed stellar luminosity of 4900~\Lsun; see text. \\  } 
\tablefoottext{b}{A new  analysis performed by Todt (2019, priv. comm.) came up with a distance
                  of 1~kpc for an extinction of $ E(B-V) = 0.85 $ and an assumed stellar 
                  luminosity of 6000~\Lsun.  This distance is not compatible with the
                  extinction-distance relation derived by \citet{SW.84}, according to which the
                  distance should be $ \ga\! 2 $~kpc.  Therefore, we kept the \gaia\ DR3 parallax and note
                  that its error is 10\% only (see text for more details).       
                 }                        
          }           
\tablebib{(1) $ \tx $ value from \citetalias{helleretal.16}; 
          (2) \Tx\ from \citet{gruendl.06}; (3) \citet{marco.07}; (4) \citet{CMS.06}; (5) 
          \citet{FCAP.93}; (6) \citet{MKH.92}; (7) \citet{MG.09}; (8) \citet{KUP.06}; 
          (9) \citet{PHM.04}; (10) \citet{toalaetal.19}; (11) \citet{HB.11}; (12) \citet{todt.15};
          (13) \citet{georgiev.08}; (14) Todt (2019, priv. comm.); (15) \citet{keller.14}; 
          (16) \citet{toala.19}; (17) \citet{toala.20}; (18) \citet{guerrero.05}.
          }          
\vspace*{2mm}                     
\end{table*}

Table~\ref{tab:data} presents a compilation of the relevant parameters of the
12 planetary nebulae from which diffuse X-ray emission has been observed and
analysed in the past.  Seven have normal, hydrogen-rich O-type central stars
(including NGC\,6543 with spectral type `wels'), and five have hydrogen-poor
central stars of various spectral types, all of which we refer to as of
[WR]-spectral type for simplicity.
Since our aim also is to relate stellar wind parameters with X-ray emissions
of wind-blown bubbles, only objects for which both X-ray spectra and stellar
wind data are available are included in the sample of Table~\ref{tab:data} .
All of the listed objects are contained in the \gaia\ Data Release 3 (DR3) and
have parallax errors well below 10\% in most cases.\footnote
   {We use the measured parallaxes, $\pi$, without the (very small) zero
       point corrections to compute distances as $d=1/\pi$.}

   There are two objects for which the \gaia\ parallaxes appear to be
   questionable, at least: NGC~5315 and NGC~7026, both with a nucleus of [WR]
   spectral type. The \gaia\ distance to NGC~5315 of only 0.96~kpc leads to a
   (bolometric) stellar luminosity of about 670~\Lsun, much too low for a
   post-AGB object with $ \teff = 75\,000 $~K.  We therefore decided to keep the
   distance recommendation of \citet{marco.07} of 2.5~kpc that corresponds to a
   reasonable luminosity of 4900~\Lsun\ (Table~\ref{tab:data}).  Our distrust
   is supported by the high renormalised unit weight error (RUWE) of the
   \gaia\ measurement, RUWE = 2.63, indicating a problematic result.
   
   For the [WR] object NGC~7026, a \emph{Gaia} DR3 distance of 3.23~kpc (RUWE
   = 1.05) is reported, which in turn leads to a stellar luminosity of
   62\,000~\Lsun, an unreasonably high value for a post-AGB object, well above
   the Chandrasekhar-limit of about 50\,000~\Lsun\ for an electron-degenerate
   carbon-oxygen stellar core.  However, the mass-loss rate, wind luminosity, and X-ray
   luminosity are high but still reasonable (cf. Cols.~7, 10, and 11
   in Table~\ref{tab:data}).  Since we are unable to resolve this discrepancy
   with our limited information, we adopted here the rather accurate \gaia\
   distance.  We note that the (bolometric) luminosity of NGC~7026's nucleus
   is not needed for the present study.

   Because of the generally rather small distance uncertainties provided by
   the \gaia\ measurements, we would have liked to get likewise accurate
   values for the stellar bolometric and wind luminosities.  To this end, we
   investigated and compared more recent work based on sophisticated stellar
   atmosphere or photoionisation modelling methods.  The relevant papers are
   also listed in Table~\ref{tab:data}, and appropriate averages of the
   parameters were derived for cases for which more than one paper was
   avaliable.  We used only data from publications where next to the
   luminosities and mass-loss rates also the assumed distances were provided,
   which is not always the case.  The distance-dependent quantities were
   rescaled to our adopted distances accordingly, i.e. the luminosities with
   $ d^2 $ and the mass-loss rates with $ d^{ 1.5} $
   \citep[cf.][]{leuetal.96}.
   
   To our surprise, the determination of a reliable stellar luminosity
   appeared to be more difficult than believed: for given distance, the
   bolometric stellar luminosity, $ L_{\rm star} $, may still differ by a
   factor of two between different authors!  We believe that, next to
   systematic errors introduced by the quality of the modelling, an
   underestimated source of error is the extinction towards the object in
   question.

   Altogether, the `intrinsic' uncertainty of the bolometric luminosities
   listed in Table~\ref{tab:data} may amount to about 30\% (0.11 dex).
   If we consider this as a typical intrinsic luminosity uncertainty for all
   sample objects, the contribution of the distance uncertainty
   ($ \la\! 10\% $) is comparatively small for objects with good \emph{Gaia}
   distances.
   
   The mass-loss rates (and therefore also the wind luminosities) are the most
   uncertain parameters: we found that the lowest and highest mass-loss rate
   of a particular object can differ by a factor of up to three.  The typical
   error of the mean mass-loss rate in Table~\ref{tab:data} is estimated to
   be about 60\% (0.20 dex). As above, the low distance uncertainties do not
   significantly contribute to the mass-loss error.

   However, we believe that the real mass-loss uncertainties are much larger
   by systematic effects. The rates listed in Table~\ref{tab:data} are based
   on the assumption of a homogeneous wind flow.  However, detailed line
   analyses showed that the winds may have inhomogeneous structures. This kind
   of `clumping' can be approximated by the so-called `volume-filling
   factor approach'.  Although it is difficult to determine the filling
   factor $ f $ for individual cases, \citet{HM.99} estimated that a value of
   0.1 would be reasonable.  Because the true mass-loss rate scales as
   $ \dot{M}(f)=f^{1/2}\,\dot{M}(f$=$1) $, clumping always reduces the mass-loss
   rate.  For instance, a value of ${ f = 0.1 }$ reduces the mass-loss rate by
   a factor of three ($ -0.5 $ dex).
    
   These uncertainties of the mass-loss rates translate directly into the
   respective errors of the wind powers, $ \Lwind = 0.5\, \dotMw\, \Vwind^2 $.
   The wind velocities can be measured from the absorption troughs of
   optically thick wind lines or the widths of emission lines and are much
   less uncertain.  Although the velocity errors enter in quadrature, their
   contribution to the total error budget of the wind power is virtually
   negligible.
   
   An important quantity in the context of this work is the bubble radius,
   $ R_2 $, which is, by definition, identical with the inner nebula boundary.
   We measured the bubble sizes from the existing X-ray images, and in cases
   of elongated bubbles we took the geometrical average of their minor and
   major semi-axes.  The derived bubble sizes are listed in
   Table~\ref{tab:data}, too.  They agree, after distance adjustments,
   reasonably well with those presented in \cite{ruizetal.13}.\footnote
    {\citet{kastner.08} used higher bubble radii taken from the catalogue
    of \citet{cahn.92}.  However, this catalogue only lists the outer nebular radii!
    }
    
    Table~\ref{tab:data} also contains the relevant X-ray data, i.e. observed
    X-ray luminosities $ \lx $ and the characteristic X-ray temperatures
    $ \tx $.  Both are taken from the compilation of \citet[][Table~2
    therein]{ruizetal.13}, rescaled to our distances, and refer to the energy
    range 0.3--2.0~keV (6.2--41.3~\AA). The influence of the chosen X-ray
    band width on the values of $ \lx $ and $ \tx $ is discussed in
    Appendix~\ref{appsec:calib.bandwidth}.

    In general, the accuracy of the derived X-ray properties is limited by
    (i) the unknown column density of absorbing intervening matter, and (ii)
    the quality of the X-ray observations themselves (low-number photon
    statistics with total photon counts as low as about 30 in the
    quoted energy range). Though a reasonable determination of the X-ray
    luminosity is still possible (on the 20\% level, extinction not
    considered), the determination of a meaningful X-ray temperature
    appears questionable.
    We therefore treat mean bubble temperatures based on spectra with photon counts
    of less than 100 with caution and give them in italics in
    Table~\ref{tab:data} (Col.~13).  This applies to IC~418, IC~4593, NGC~40,
    and NGC~6826.

\section{Confronting our standard  hydrogen-rich post-AGB simulations with the observations}
\label{sec:comp.obs.mod_pn}
   
In this section we compare the stellar and bubble parameters of the sample
objects listed in Table~\ref{tab:data} with the predictions of our existing
post-AGB models which have already been used in \citetalias{SSW.08} and
\cite{ruizetal.13}. However, we investigate here for the first time the
effects of non-equilibrium ionisation on the characteristic properties of
the X-ray emitting plasma, in comparison to the usual assumption of
collisional ionisation equilibrium.

   We follow the paradigm of single star evolution where
   extensive (spherical) mass loss along the AGB leads to the depletion of the stellar envelope,
   forcing eventually the star to leave the AGB.
   We note that the evolutionary calculations are valid for hydrogen-rich objects only.  
     
   Nevertheless, we include the [WR] objects in the comparison with the
   predictions of our hydrogen-rich post-AGB models. This will allows us to
   interpret any differences in the parameters of both type of objects
   (PN/WR) in physical terms, and will provide us with preliminary
   constraints on the evolution of the [WR] central stars that can guide us
   in setting up the parameters of new hydrodynamics simulations
   specifically designed to reproduce the observed hydrogen-poor hot bubbles
   (see Sect.\,\ref{sec:WR-bubbles.observ}).
   
\subsection{Stellar luminosities and wind powers}
\label{subsec:stellar.wind}

\begin{figure}
\center
\includegraphics[trim= 1.0cm 0cm 0.3cm 0.5cm, width= 0.99\linewidth, clip]
                {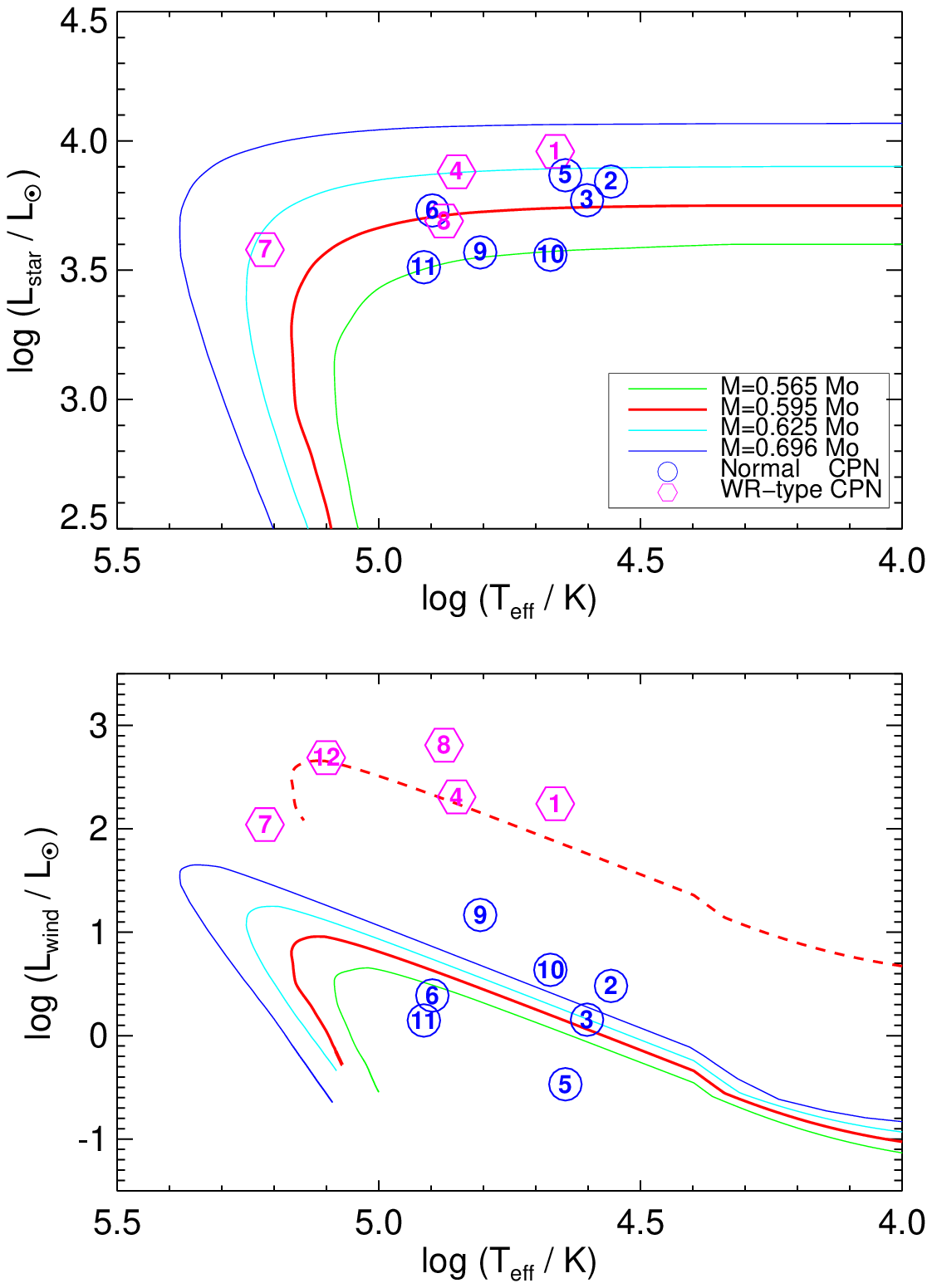} 
\vskip -1mm
\caption{\label{fig:wind.model}
  Evolution of stellar energy output of four AGB models with remnant masses
  of 0.565, 0.595, 0.625, and 0.696~\Msun\ (already used in
  \citetalias{SSW.08} and \citealt{ruizetal.13}) versus
  their effective temperatures. The `reference' track of
  Fig,~\ref{fig:mod.prop}, 0.595~\Msun, is highlighted by a thick red line.
  The positions of the objects from Table~\ref{tab:data} are labelled with the
  object's number within either a circle (O-type central star) or a hexagon
  ([WR]-type central star). \emph{Top}: stellar bolometric photon luminosity;
  NGC~7026 (no.~12) falls outside the plotted luminosity range. \emph{Bottom}:
  wind luminosity of the same AGB-remnants; NGC~7026 is now included. Solid tracks refer
  to the `standard' mass-loss model $\dot{M}_{\rm PPKMH}$ \citep[][see
  Sect.~\ref{sec:model.calc}]{Pauletal.88}.
  The dashed track corresponds to a 0.595~\Msun\ model with the wind power
  increased by a factor of 50 (see text).}        
\end{figure}   

\begin{figure}
\vskip 2mm
\includegraphics[trim= 0.5cm 0cm 0.3cm 0.5cm, width= 0.99\linewidth, clip]
                {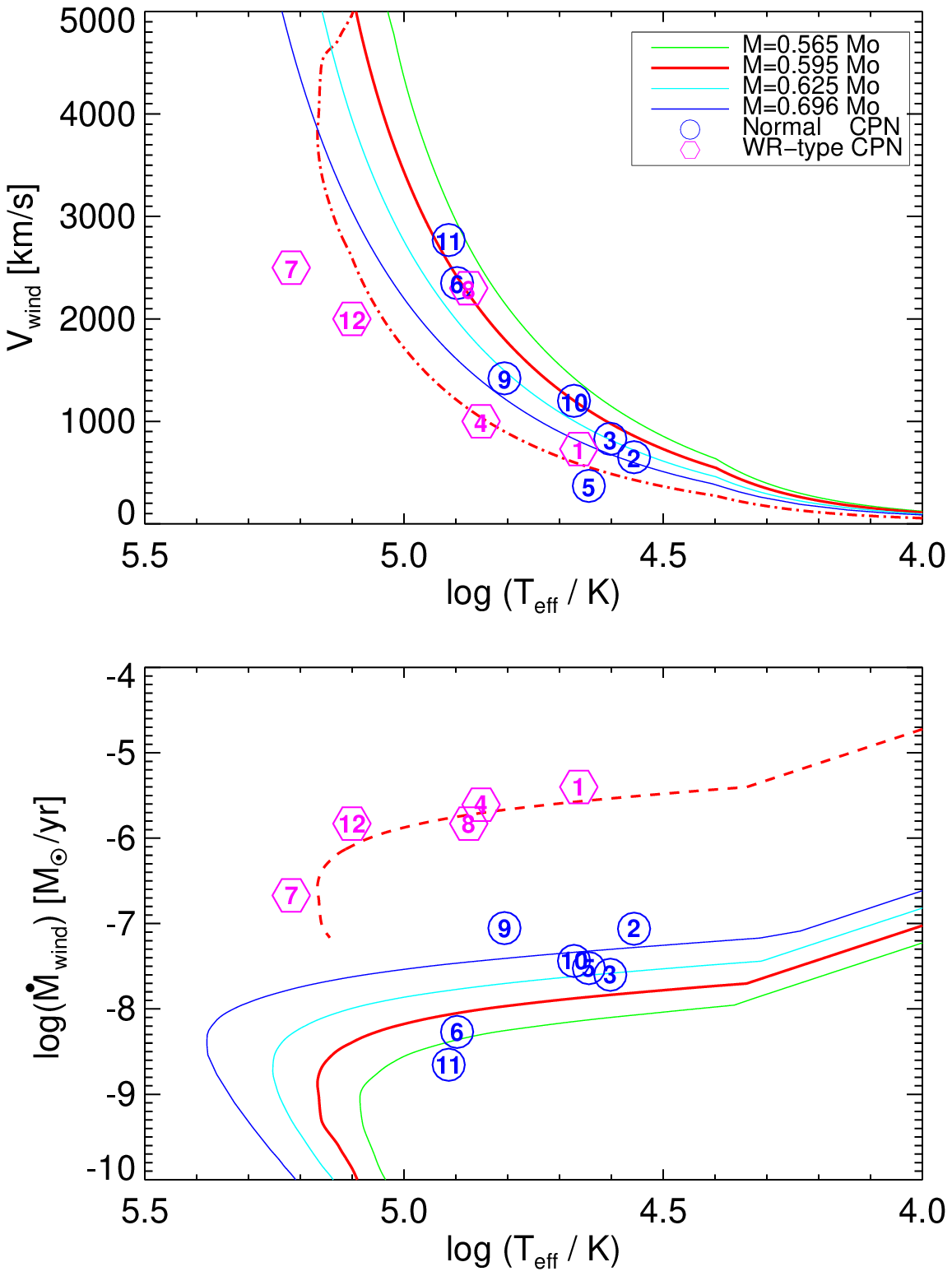}
\vskip -1mm                  
\caption{\label{fig:wind.model.2} \
  Stellar-wind velocities (\emph{top}) and mass-loss rates (\emph{bottom})
  versus stellar $ \teff $ for the same post-AGB tracks and sample objects as
  in Fig.~\ref{fig:wind.model}.  The dash-dotted line in the top panel
  represents the wind velocity of the 0.595~\Msun\ model halved, the dashed
  line in the bottom panel the mass-loss rate of the 0.595~\Msun\ model
  increased by a factor of 200.  The positions of the objects no.~6
  (NGC\,3242) and no.~8 (NGC\,5315) nearly coincide in the top panel, and
  those of no.~5 (NGC\,2392) and no.~10 (NGC\,6826) in the bottom panel.  }
\end{figure}
   
The classical Hertzsprung-Russell diagram in the top panel of
Fig.~\ref{fig:wind.model} shows that all the displayed sample objects are
still on or very close to the horizontal part of their post-AGB evolution and
are embraced by our 0.565 and 0.625~\Msun\ post-AGB tracks.
   
The mean luminosity of the seven hydrogen-rich, O-type objects is about
4900~\Lsun, close to the luminosity of our 0.595~\Msun\ model.  This mean
luminosity corresponds rather well with the mean luminosity of about
5000~\Lsun\ found for a sample of 15 nebulae with hydrogen-rich nuclei whose
distances have been derived by the expansion-parallax method by \citet{SBJ.18}.
The mean luminosity of the nine hydrogen-rich central stars in the
Magellanic Cloud sample of \citet{HB.04, HB.07} is 4200~\Lsun.  Considering
the rather small sample sizes and the uncertainties inherent to the luminosity
determination, the agreement between the mean stellar
luminosities of these three samples is rather satisfying.
     
The mean luminosity of the four displayed hydrogen-poor [WR] central stars is
about 6000~\Lsun.  This rather high mean luminosity is in contrast to the work
of \citet{HB.04, HB.07} on planetary nebulae in the Magellanic Clouds.  Their
MC sample contains three objects with hydrogen-deficient nuclei (the `odd'
object SMP LMC\,83 neglected) whose luminosities range from about 2400 to
4200~\Lsun, with a mean of about 3600~\Lsun.  However, both samples are really
too small for a meaningful comparison between the Milky Way and MC [WR]-type
central stars.  A significant difference between the mean stellar luminosities
of the O-type and [WR]-type sample objects is not evident.
   
However, a significant difference between O- and [WR]-type central stars i
obvious if one considers their wind luminosities,
$ \dot{M}_{\rm wind} \times V_{\rm wind}^{2} /2 $,
(Fig.~\ref{fig:wind.model}, bottom panel).  A clear segregation
between both samples is now visible. The observed wind luminosities of the
hydrogen-rich O-type objects cluster around our four post-AGB tracks,
but with a spread that is much wider than the spread of the model tracks (see
also below).  Nevertheless, the trend with stellar effective temperature is
compatible with the predictions by \citet{Pauletal.88}.
      
In contrast to the O-type central stars, the [WR]-type objects have
systematically much higher wind luminosities, varying between about
$ 10^2\ldots 10^3 $~\Lsun\ over the whole temperature range. If NGC~5189
(no.~7) is ignored, there appears to be a similar trend with $ \teff $, too.
A fair match to the observed values of the five [WR]-type sample objects can
be achieved if the wind power of our 0.595~\Msun\ post-AGB reference model is
increased by a factor of 50 (dashed line in the bottom panel).
    
Figure~\ref{fig:wind.model.2} disentangles how the wind power of our sample
objects is composed of wind velocity (top) and mass-loss rate (bottom).  The
wind velocities of the hydrogen-rich sample objects follow, on the average,
closely the predictions of our models, i.e.  of \citet{Pauletal.88}.  Only the
wind velocity of NGC~2392 (no.~5) is far too low.  In contrast, the wind
velocities of our [WR]-type objects are systematically lower (NGC~5315, no.~8,
excepted) by a factor of about two (cf. dash-dotted line in the top panel).
   
This result is in agreement with the findings in \citetalias{Sandin13}.  Based
on a sample of 13 [WR]-type central stars which contains only BD\,+30\degr
3639 and NGC~40 from our sample, the authors found that the
\citet{Pauletal.88} wind velocities derived for hydrogen-rich central stars
must be reduced by a factor of 0.583, on the average.  This result is at
variance with the computations of \citeauthor{Pauletal.88} who found, for
given $ \teff $, higher wind velocities for hydrogen-poor central stars.
However, these authors used simple hydrogen-poor model atmospheres without
enhanced carbon and oxygen abundances. Based on these findings, we
will consider only the simulations WR\#V05 as relevant for
interpreting the properties of the observed bubbles around hydrogen-poor
central stars.

Concerning the mass-loss rates (bottom panel of Fig.~\ref{fig:wind.model.2}),
the model predictions are consistent with the observations of our sample
O-type stars. The scatter is very high, probably indicating (i) that either
the observed mass-loss rates are more uncertain than generally believed or
(ii) that our understanding of radiation-driven mass loss needs revision.
This mass-loss scatter is obviously responsible for the scatter of the wind
power seen in Fig.~\ref{fig:wind.model} (bottom panel).

The bottom panel of Fig.~\ref{fig:wind.model.2} indicates that the mass-loss
rates of our [WR]-type central stars are significantly higher than for their
O-type counterparts. The hydrogen-poor wind models of \citet{Pauletal.88} are
unable to explain the high mass-loss rates of [WR]-type central stars.  
Nevertheless, they are astonishingly well matched by scaling up the standard
mass-loss rates of the \citet{Pauletal.88} wind of our 0.595~\Msun\ post-AGB model
by a factor of 200.  The combination of this mass-loss enhancement with the
\citeauthor{Pauletal.88} wind velocities halved leads then to wind powers of
our [WR]-sample objects which are about 50 times higher than for our
0.595~\Msun\ reference model (cf. bottom panel of Fig.~\ref{fig:wind.model}).

\subsection{Bubble sizes}
\label{subsec:bubble.sizes}
        
Another important aspect of post-AGB evolution is the speed of the central
star across the HRD which is heavily mass dependent and cannot be disentangled
in the Figs.~\ref{fig:wind.model} and \ref{fig:wind.model.2}. The post-AGB age
of a particular planetary nebula is difficult to determine from its size and
the nebular `expansion' velocity.  An in-depth discussion of the various
methods used in the literature to estimate post-AGB ages and their
systematic uncertainties can be found in \citet{schoenetal.14}.

Here we avoid ages at all and use instead the bubble size, $ R_2 $, defined by
the radial position of the bubble/nebula interface,
together with the stellar temperature as a marker of the bubble's evolution
across the HRD.
The use of $ \teff $ instead of the post-AGB age has two advantages: (i) the
\Teff\ range does not vary much with remnant mass, and (ii) \Teff\ is an
observable quantity.
   
The expansion of bubbles formed by stellar winds colliding with the
interstellar medium has been studied extensively by \citet{weaver.77}.  They
showed (their Eq.~21) for the then only interesting case of constant wind
power and constant ambient (interstellar) medium that the bubble's expansion
with time depends weakly on wind power or ambient matter density.  The
expansion velocity decreases with time and comes eventually to a halt when the
bubble's thermal pressure equals that of the ambient, snow-ploughed matter.
However, the results of \citet{weaver.77} cannot be applied to bubbles inside
expanding planetary nebulae.  Here we have an accelerating wind power combined
with a radially decreasing ambient matter density which is also strongly
modified by ionisation.
   
This deficiency was eliminated by the work of \citet{ZP.96}.  These authors
approximated the post-AGB evolution of Bl\"ocker's (\citeyear{B.95}) 0.605~\Msun\ model
including the mass-loss rates by analytical expressions.  The outer boundary
condition was described by a slow (AGB) wind with constant mass-loss rate and
velocity.  Solving the relevant equations, \citeauthor{ZP.96} found for the
expansion of an adiabatic bubble (their Eq.~2)

\begin{equation}  \label{eq:ZP}   
 R_2 \propto \Big({L}_{\rm wind}^{(0)}\Big)^{1/3}\times 
             \Big(V_{\rm agb}/\dot{M}_{\rm agb}\Big)^{1/3} \times t^{1.87} , 
\end{equation}     
where $ \Lwind^{(0)} $ is the central star's wind power at
$ t = t_0 = 1000 $~yr, $ \dot{M}_{\rm agb} $ and $ V_{\rm agb} $ the
(constant)\footnote{Valid only initially; once ionisation sets in, the density
ahead of $ R_2 $ will no longer follow a simple $\rho \propto r^{-2}$ law.}
AGB-wind mass-loss rate and (terminal) velocity, and $ t $ is the
bubble's age.  In contrast to the interstellar case, the bubble now expands
into a surrounding medium whose density decreases with distance from the star:
$ \rho(r) = \dot{M}_{\rm agb}/(4 \pi V_{\rm agb}\, r^{2} ) $.  Of course,
Eq.~(\ref{eq:ZP}) is only valid for the horizontal part of evolution across
the HRD.

The time dependence of $ R_2 $ in Eq.~(\ref{eq:ZP}) ensures that the bubble's
expansion is always accelerated during the evolution across the HRD, in
contrast to the \citet{weaver.77} case where $ R_2 \propto t^{0.6} $.  It is
noted that the acceleration of the expansion slowly decreases with time,
$ \ddot{R_2} \propto t^{-0.13} $.  According to Eq.~(\ref{eq:ZP}), the size
evolution of the bubble depends rather weakly on the wind power and the AGB
wind parameters, but such that the bubble expands faster for higher
central-star wind power and lower AGB wind densities, or vice versa.

\begin{figure}
\center
\includegraphics[trim= 1.3cm 0.0cm 0.7cm 1.6cm, width= 0.98\linewidth, clip]
                {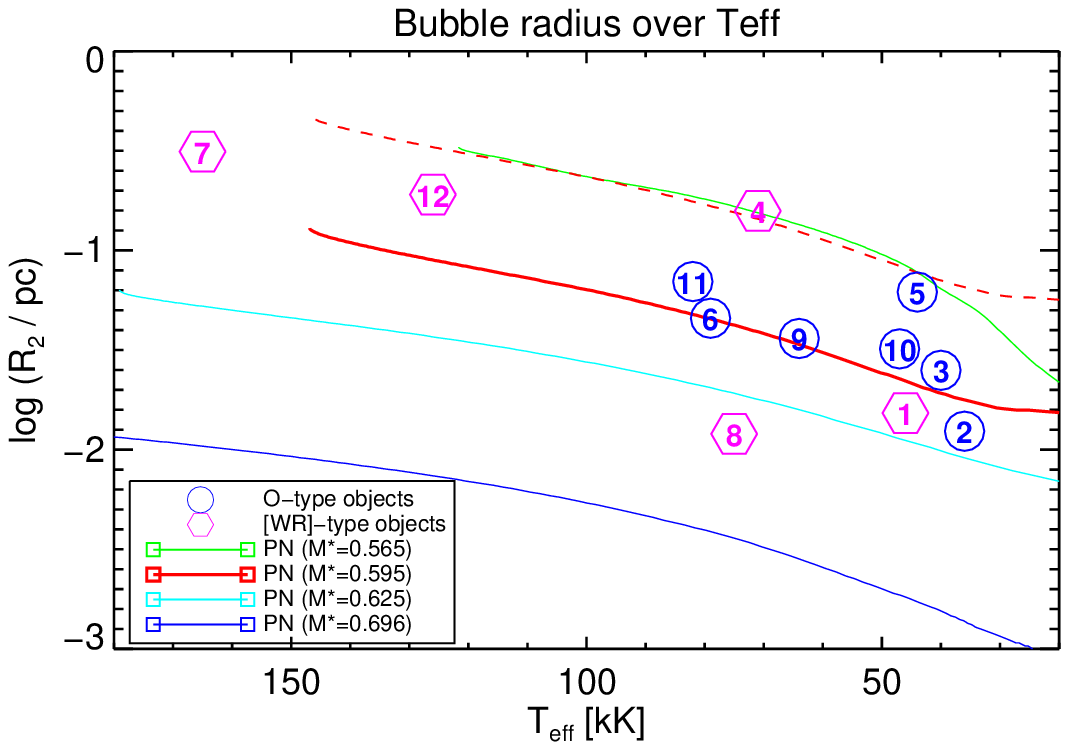}
\vskip -1mm 
\caption{\label{fig:r2.teff}      
         Bubble radius, $ \log(R_2) $ versus central-star effective temperature, \Teff, 
         as predicted by our hydrodynamical simulations of the 0.565, 0.595 (PN), 0.625, and
         0.696~\Msun\ sequences together with the sample objects. 
         The dashed line refers to that in Fig.~\ref{fig:wind.model} (bottom panel) and is an 
         estimate based on Eq.~(\ref{eq:ZP}) if the wind power of the 0.595~\Msun\ model is 
         increased by a factor of 50.  The simulations are plotted from the beginning of
         bubble formation till maximum stellar temperature.   
         }
\end{figure} 

The sensitivity of the $ R_2(\teff) $ relation on the central-star evolution
is demonstrated in Fig.~\ref{fig:r2.teff} for the four post-AGB simulations
used in the Figs.~\ref{fig:wind.model} and \ref{fig:wind.model.2}. The
curves of the 0.565, 0.595, 0.625, and 0.696~\Msun\ models
run nearly parallel and are well separated by about 1.7~dex in total.  The
main reason for this separation in the $ R_2 \mbox{-} \teff $ plane is the
strong dependence of the evolutionary speed across the HRD on stellar mass.
For instance, our post-AGB tracks have crossing times\footnote
   {The crossing time is the time needed by an AGB remnant to evolve from the tip of the AGB to
    maximum effective temperature (see, e.g. \citealt{B.95}).} 
  of about 800~yr (0.696~\Msun), 10\,000 (0.595~\Msun,
  Fig.~\ref{fig:mod.prop}) and 20\,000~yr (0.565~\Msun), i.e. a factor of
  about 25 between 0.696 and 0.565~\Msun, or 1.4~dex.  The corresponding
  increase of the wind power from the 0.565 to the 0.696~\Msun\ model is with
  a factor of about 2.5 only rather modest (cf. Fig.~\ref{fig:wind.model},
  bottom panel) and influences only little ($-0.13$ dex) the separation of the
  bubble sizes. Similarly, the influence of the density term in
  Eq.~(\ref{eq:ZP}) is quite small. For hydrogen-rich central stars, the
  bubble size $R_2$ at given $ \teff $ is therefore an indicator of the
  evolutionary timescale.

  However, if the wind power is increased to higher values typical for the
  [WR]-type central stars, the wind-power term in Eq.~(\ref{eq:ZP}) becomes
  important and must be considered.  For instance, an increase of the wind
  power of the reference simulation (0.595~\Msun) by a factor of 50 would
  increase the bubble size by a factor $3.7$, provided the evolutionary time
  scale remains the same (dashed line in Fig.~\ref{fig:r2.teff}).
      
  In principle, the HRD crossing time also depends on the mass-loss rate
  \citep{SB.93, B.95}.  The four simulations of the evolution of the bubbles
  around hydrogen-rich central stars displayed in Fig.~\ref{fig:r2.teff} are
  fully consistent with their respective mass-loss rates.  [WR]-type central
  stars have much higher mass-loss rates (cf. Table~\ref{tab:data}) which in
  turn would reduce their HRD crossing times.  However, since the formation
  and evolution of the hydrogen-poor central stars are unknown, we also do not
  know a priori (i) their HRD crossing times and (ii) how these depend on
  stellar mass.  We consider therefore the HRD crossing times of [WR]-type
  central stars as a free parameter, scaled to the crossing time of our
  0.595~\Msun\ reference model.

  The distribution of our sample objects in Fig.~\ref{fig:r2.teff} reveals
  evolutionary differences between the two central-star subsamples.  With the
  exception of NGC~2392 (no.~5) and possibly IC\,418 (no.~2), the bubble sizes
  of the O-type central stars are remarkably close to the predictions of our
  0.595~\Msun\ reference sequence. The spread is much less than one would
  expect from the luminosity spread in the top panel of
  Fig.~\ref{fig:wind.model}.  This confirms our finding that, thanks to \gaia,
  we are in a situation where the uncertainties of the luminosity
  determinations are now the limiting factor, and not the distances.
      
  We conclude, in conjunction with Figs.~\ref{fig:wind.model},
  \ref{fig:wind.model.2}, and \ref{fig:r2.teff}, that the hydrodynamics
  simulations of wind envelopes around post-AGB models as described in
  \citet{peretal.04} provide a fairly good description of the evolution of
  hydrogen-rich AGB remnants in terms of evolutionary timescale, mass-loss
  rate, wind velocity, and hot-bubble formation and expansion.
    
  Only NGC~2392 (no.~5) does not fit into this scheme. It has (i) a
  high-luminosity central star with (ii) a very low wind power due to the
  unusually low wind speed (Figs.~\ref{fig:wind.model} and
  \ref{fig:wind.model.2}, top panels), and (iii) a comparatively big bubble
  for the low central-star temperature (Fig~\ref{fig:r2.teff}).  The stellar
  luminosity is not consistent with the size of the bubble which corresponds
  better to our low-luminosity model of 0.565~\Msun. Needless to say that the
  low wind power cannot be responsible for the extraordinary bubble size.  We
  can only speculate that the wind power of NGC~2392's central star has been
  much higher in the past.  We note that individual luminosities which led to
  our mean value of NGC~2392 listed in Table~\ref{tab:data} show an
  unusually high spread, viz. a factor of 2.7 from the lowest to the highest
  value, where the lowest luminosity \citep{HB.11} is consistent
  with our 0.565~\Msun\ remnant.  More work is certainly needed for clarifying
  the case of NGC~2392.
   
  In contrast to our O-type sample, the distribution of the bubble sizes of
  the five hydrogen-poor [WR]-type objects seems to indicate more diverse
  evolutionary histories. Since all five objects have very similar wind powers
  (cf. bottom panel of Fig.~\ref{fig:wind.model}) and because the influence of
  the outer boundary conditions are small, their evolutionary timescales must
  vary considerably.  A detailed discussion of the properties of the [WR]-type
  objects in Fig.~\ref{fig:r2.teff} is postponed to
  Sect.~\ref{sec:WR-bubbles.observ} where more appropriate hydrodynamical
  simulations are used.
   
\subsection{X-ray properties}

\label{subsec:obs.xray}

The hydrodynamical modelling of \citetalias{SSW.08} showed that the bubble's
X-ray luminosity increases with time (or stellar temperature) because heat
conduction leads to evaporation of nebular mass, resulting in a steady
increase of the bubble mass with a corresponding increase of the X-ray
luminosity.  The increase of the X-ray luminosity with central-star mass is
caused by the corresponding increase of the wind power which drives the
evaporation, and the shorter evolutionary timescales resulting in a smaller
bubble radius. Altogether, the volume emission measure, and hence the X-ray
luminosity, of a hot bubble increases with stellar mass.  When the stellar
luminosity drops at the end of evolution, the wind power (and the
evaporation) ceases, but the bubble continues to expand.
Consequently, the bubble's X-ray luminosity must fade, too.  More details on
the X-ray emission of heat-conducting wind-blown bubbles can be found in
\citetalias{SSW.08}.
   
\begin{figure}
\vskip -2mm
\centering
\includegraphics[width=1.0\linewidth]{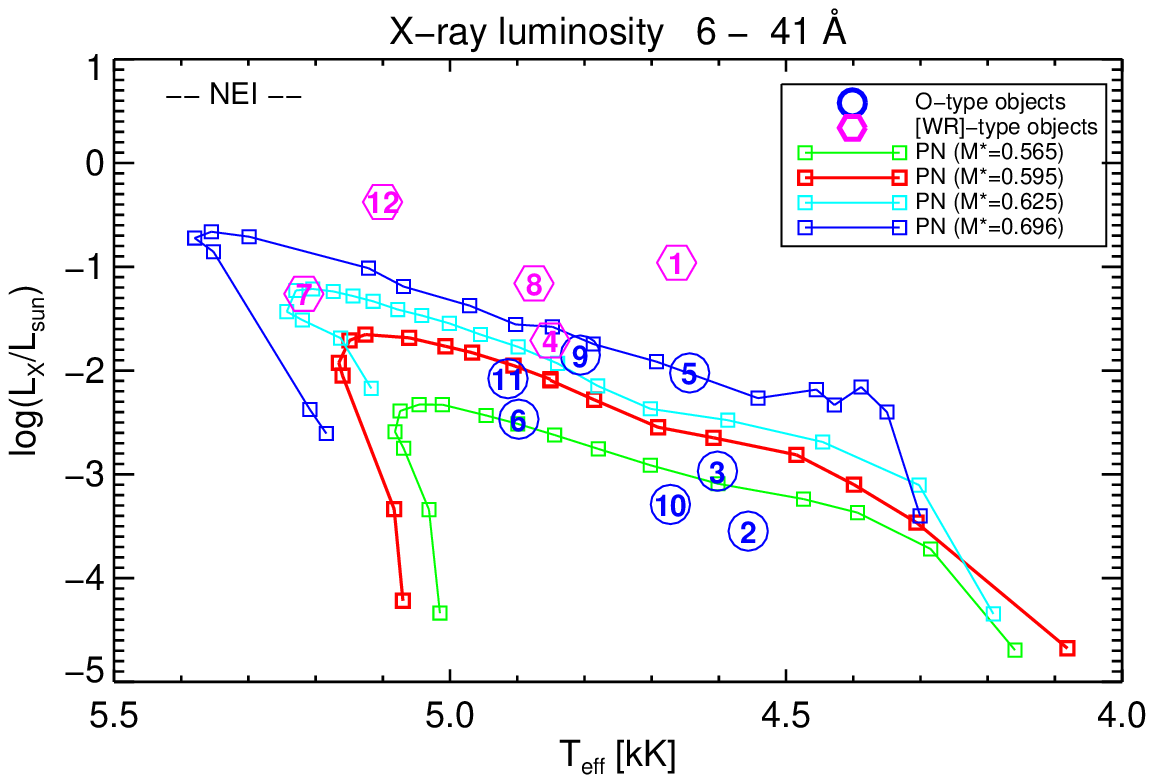}
\vskip 0mm
\includegraphics[width=1.0\linewidth]{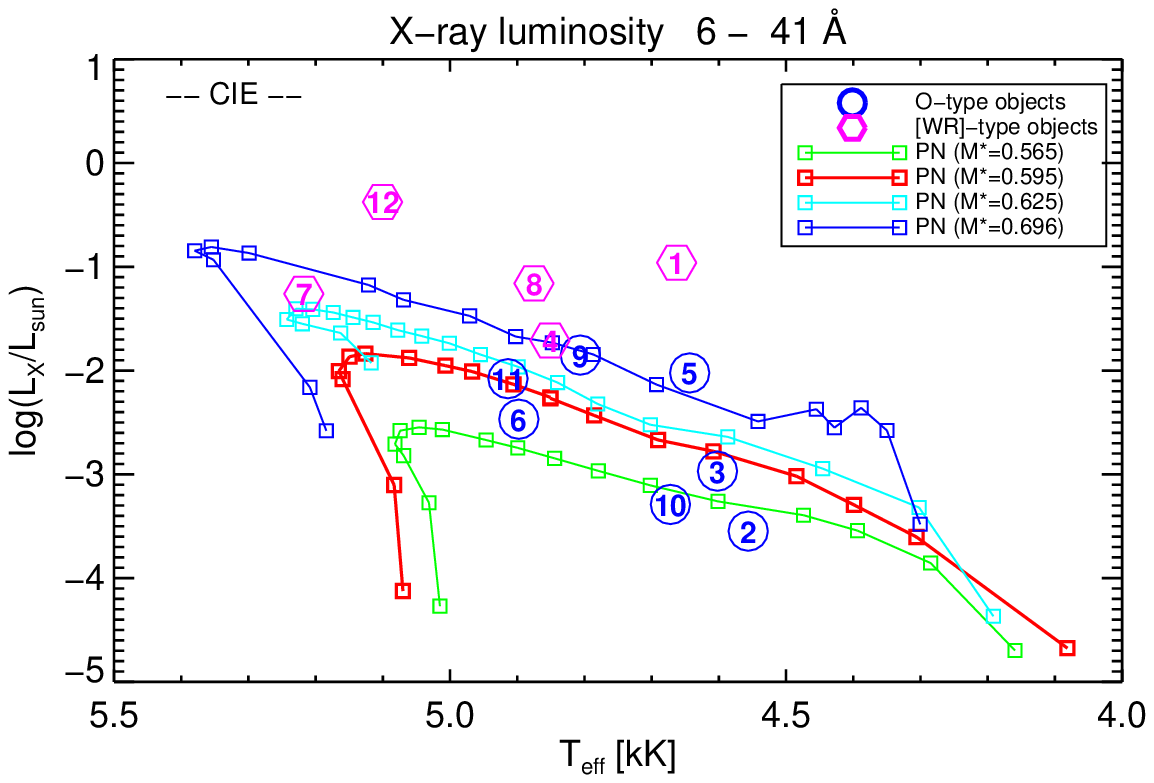}
\caption{\label{fig:xray.hrd}   
         X-ray HRD for our sample objects of Table~\ref{tab:data} (symbols as in previous 
         figures) together with the predictions of heat-conducting hydrodynamical bubble models 
         around the 0.565, 0.595, 0.625, and 0.696~\Msun\ sequences. 
         The symbols (squares) along the sequences denote the individual models for which the X-ray
         emission has been computed by means of the \texttt{CHIANTI} software package.  
         The \emph{top panel} shows the NEI, the \emph{bottom panel} the CIE case.
         }
\end{figure}

In Fig.~\ref{fig:xray.hrd}, the so-called X-ray HRD, the bubble
luminosities of all sample objects are plotted as a function of stellar
effective temperature and compared with the predictions of hydrodynamical
simulations based on the post-AGB evolutionary sequences displayed in
Fig.~\ref{fig:wind.model} (top panel). As expected from the above
discussion, we see a considerable luminosity spread between the 0.565 and
0.696~\Msun\ sequences of about 1~dex.  We also see that
the differences between NEI (top panel) and CIE (bottom panel) are
comparatively small: the X-ray luminosities are in general higher in the NEI
case by only about 0.1\ldots 0.2~dex.  These differences are too small to
have any impact on the interpretation of the observations.

\begin{figure}
\centering
\vskip -2mm
\includegraphics[width=1.0\linewidth]{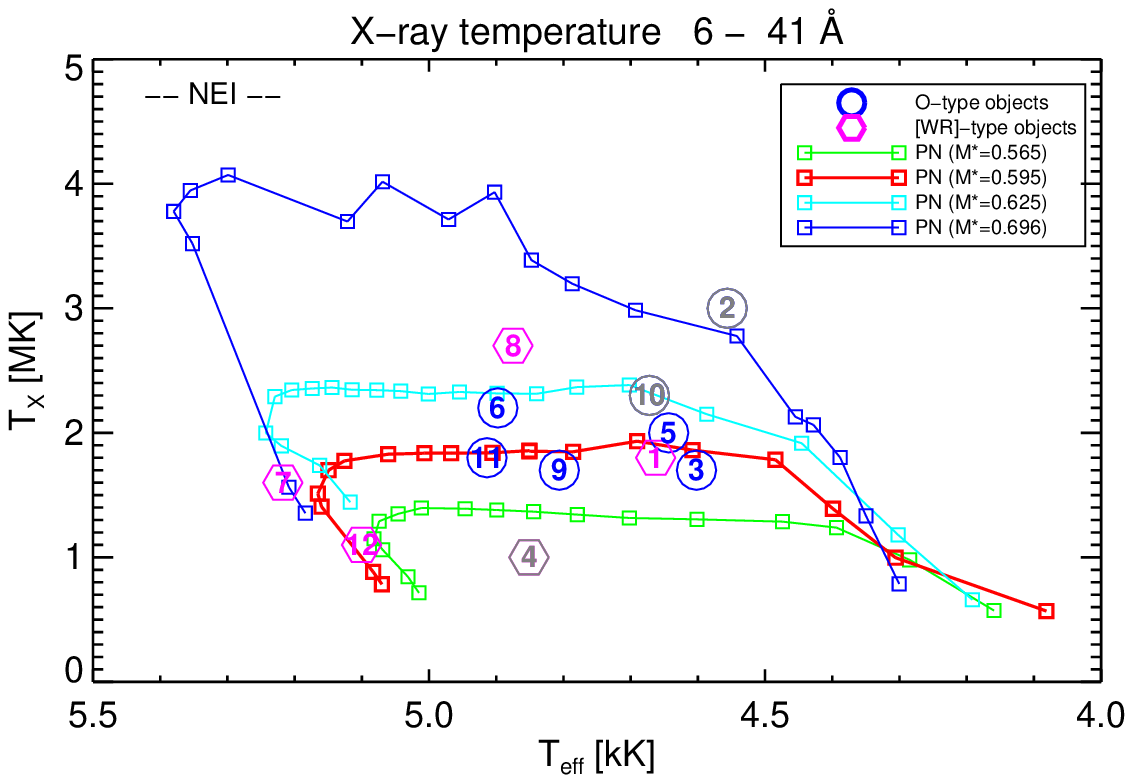}
\vskip 0mm
\includegraphics[width=1.0\linewidth]{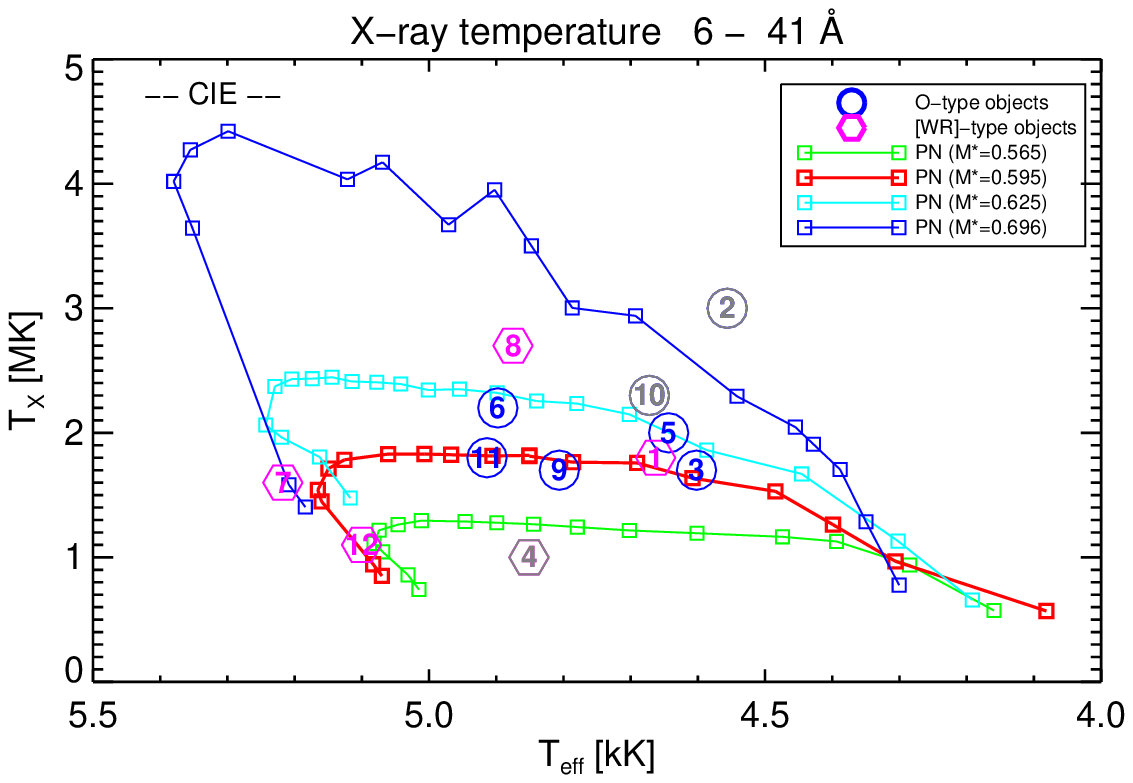}
\caption{\label{fig:tx.teff}
         Same as in Fig.~\ref{fig:xray.hrd}, but for the characteristic
         bubble temperature \Tx. The objects with very uncertain $ \tx $
         values are shown in grey (\Tx\ entries of Table~\ref{tab:data} in italics). }
\end{figure}

The mean X-ray temperatures behave differently and remain (0.696~\Msun\
excepted) nearly constant for most of the post-AGB evolution (see
Fig.\,\ref{fig:tx.teff}).  This temperature behaviour can be explained by
Eq.~(\ref{eq:tx.wind.size.2}) introduced in Sect.~\ref{sub:mean.tx} and the
wind model used in our hydrodynamics simulations: The increasing wind power,
(Fig.~\ref{fig:wind.model}, bottom) drives the expansion of the bubble such
that the ratio of wind power over bubble size remains fairly constant or
increases moderately (0.696~\Msun) during the horizontal evolution across the
HRD.  Both \Tx\ and $ \lx $ drop at the end of our simulations as the wind
power declines sharply while bubble expansion continues
(cf. Eq.~\ref{eq:tx.wind.size.2}). The general trend of increasing \Tx-values
with stellar mass is also explained by this equation: faster stellar evolution
leads to smaller bubbles and, combined with a stronger wind, to higher bubble
temperatures.
Figure~\ref{fig:tx.teff} also demonstrates that mean bubble temperatures
do not differ significantly between the NEI and CIE cases.  In all sequences
(the 0.696~\Msun\ sequence excepted), the NEI temperatures are a bit higher
than the CIE temperatures, but by no more than about 0.2~MK.

From Figs.~\ref{fig:xray.hrd} and \ref{fig:tx.teff}, we draw the following conclusions:
 
\begin{itemize}
\item
The observed X-ray luminosities and mean X-ray temperatures of the
bubbles around O-type central stars and their run with effective temperature
are well covered by the \citetalias{SSW.08} hydrodynamical bubble models
around AGB remnants between about $\la$\,0.57 and 0.70~\Msun.
        
\item
The bubbles around the [WR]-type central stars are systematically more
luminous and would technically correspond to models with central stars of at
least 0.63~\Msun, while their mean \Tx\ values are comparable to the
bubbles around O-type stars, albeit with a larger scatter.
\end{itemize}     

   The close correspondence between the X-ray and the classical HRD of the O-type objects
   is very gratifying since in Fig.~\ref{fig:xray.hrd} two different entities of the stellar-nebula
   systems are linked together: the star and its wind-blown bubble.
   For the [WR]-type sample objects, the stellar luminosities are very similar
   to those of the O-type objects, but the X-ray luminosities of their bubbles
   are definitively higher, obviously due to their much higher wind powers.
   
 As the HRD does not contain any information on individual evolutionary
 timescales, it is instructive to use again the bubble radius instead of the
 stellar temperature as an independent indicator of the evolution of the X-ray
 properties, counting the time since bubble formation, which may vary from
 object to object because of different stellar masses, different wind
 densities, and radiation-cooling properties. As we have seen in
 Fig.~\ref{fig:r2.teff}, the size of the bubble, $ R_2 $, is very sensitive to
 the stellar evolutionary timescale, and hence to the stellar mass.

\begin{figure}
\centering
\vskip -2mm
\includegraphics[width=1.0\linewidth]{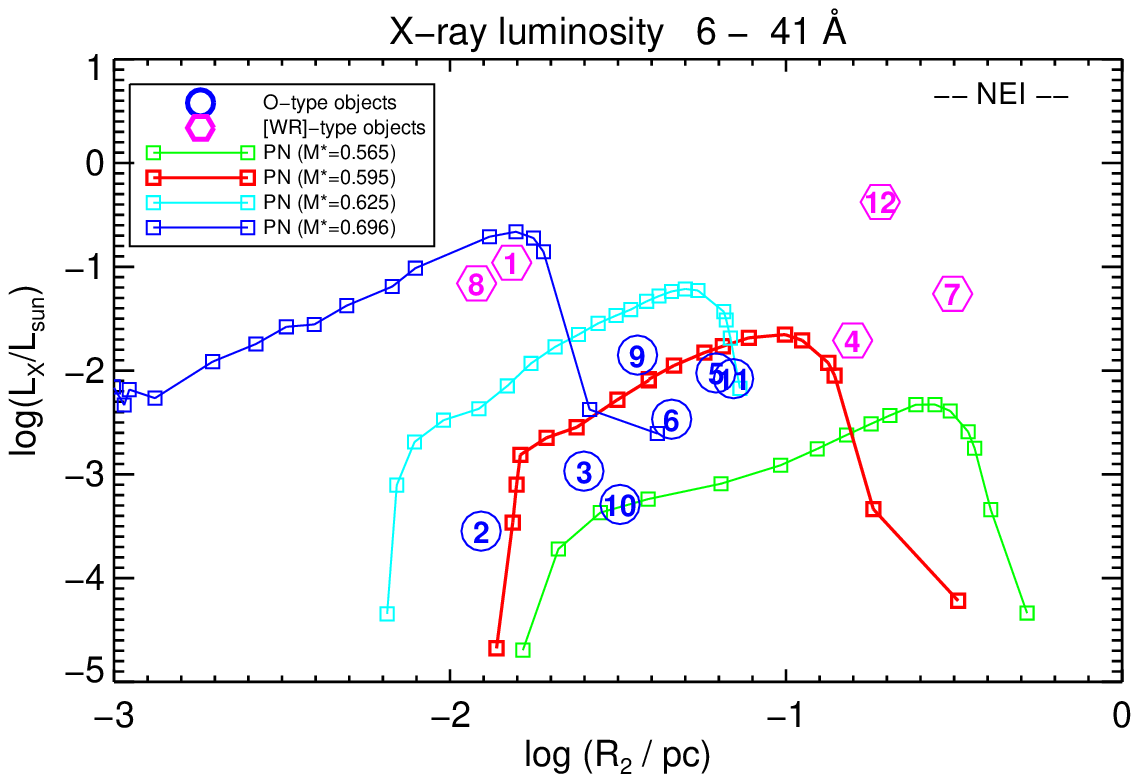}
\vskip 0mm
\includegraphics[width=1.0\linewidth]{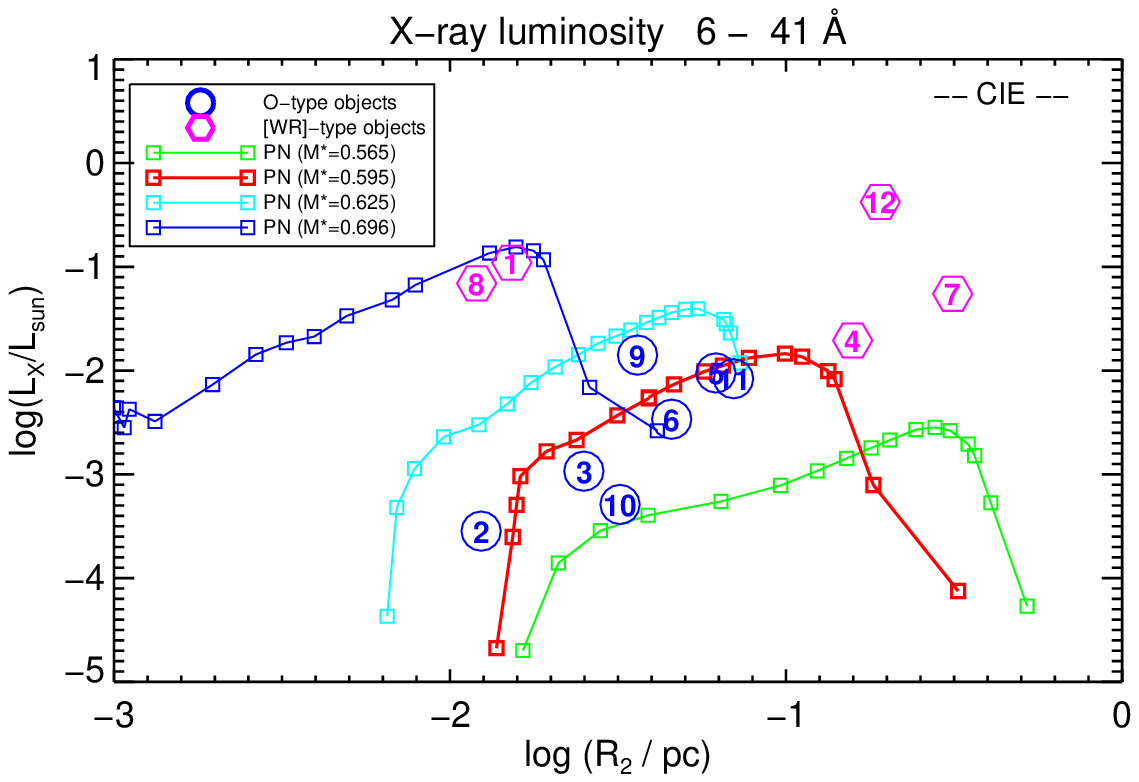}
\caption{\label{fig:lx.r2.pn}   
         X-ray luminosity versus bubble size of the sample objects listed in Table~\ref{tab:data} 
         (numbered symbols) together with the predictions of the four hydrogen-rich bubble sequences 
         of the previous figures. Again, the NEI and CIE case is shown at \emph{top} and
         \emph{bottom}, respectively.
         } 
\end{figure}                       

\begin{figure}
\centering
\vskip -2mm
\includegraphics[width=1.0\linewidth]{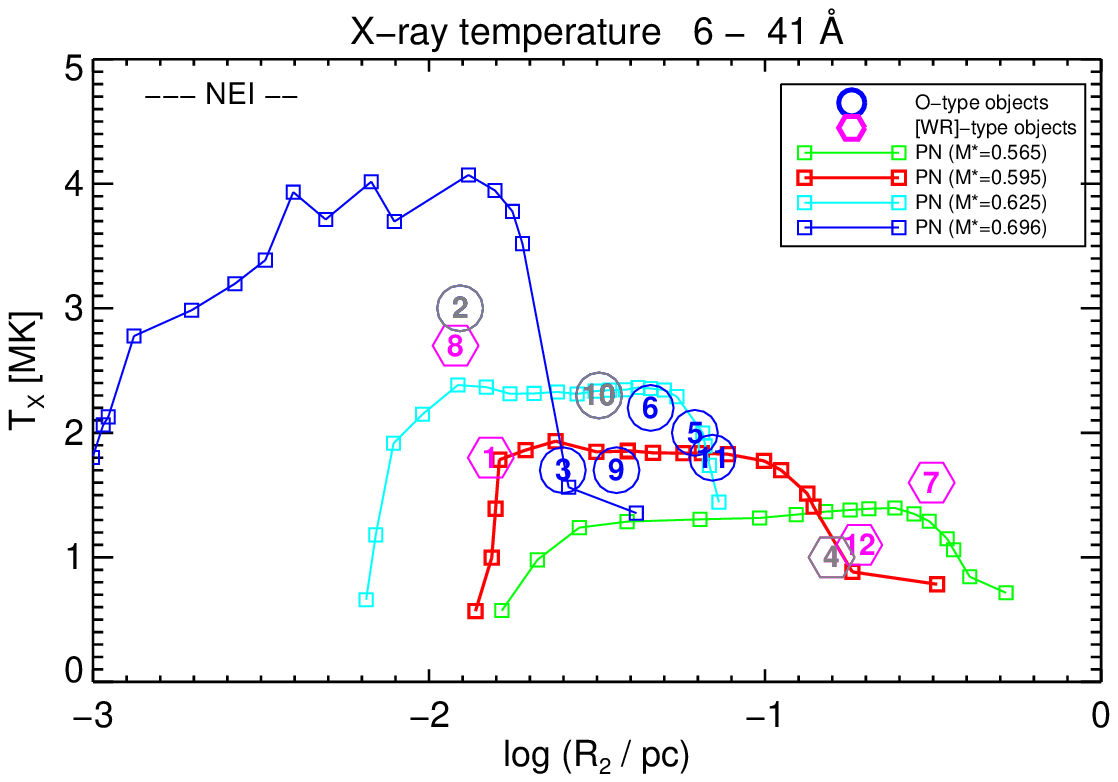}
\vskip 0mm
\includegraphics[width=1.0\linewidth]{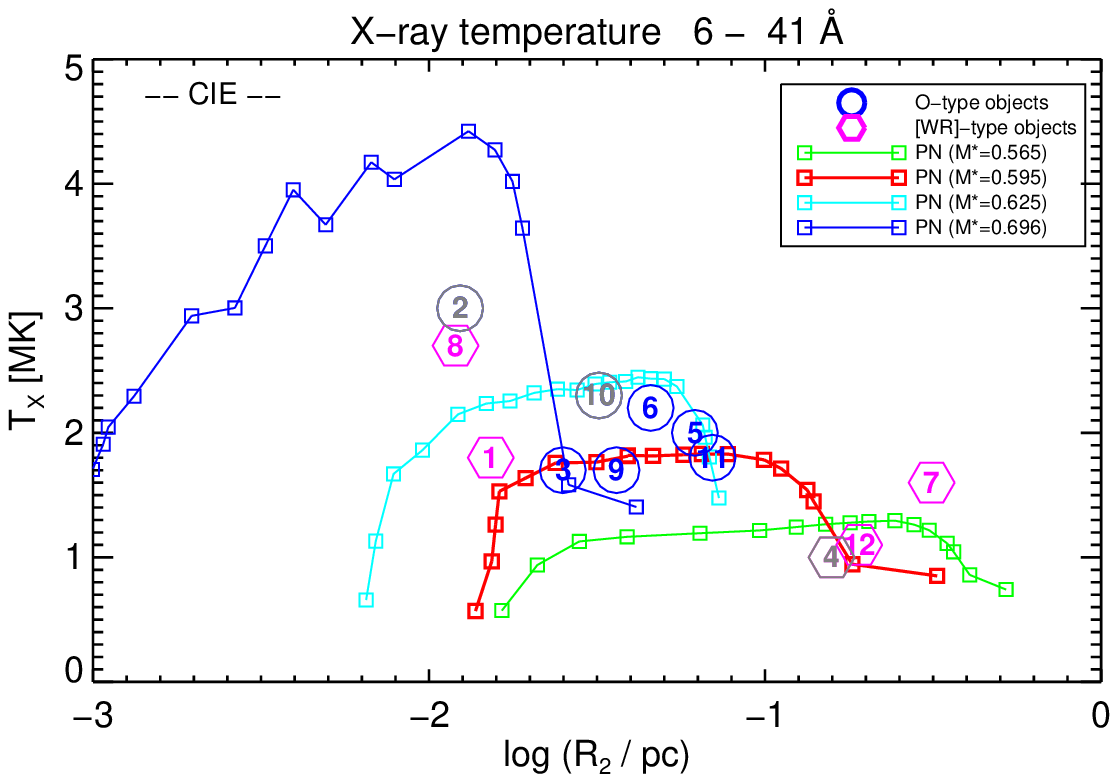}
\caption{\label{fig:tx.r2.pn}
         Same as in Fig.~\ref{fig:lx.r2.pn}, but for the characteristic
         bubble temperature \Tx. The objects with very uncertain $ \tx $
         values are shown in grey (\Tx\ entries of Table~\ref{tab:data} in italics). }
\end{figure}

Figures~\ref{fig:lx.r2.pn} and \ref{fig:tx.r2.pn} show the X-ray
luminosity and characteristic (mean) X-ray temperature $ \tx $, respectively,
versus bubble radius $ R_2 $ for our sample objects and the (hydrogen-rich)
\citetalias{SSW.08} bubble simulations. As already seen in
Fig.~\ref{fig:r2.teff}, the dependence of the crossing timescale on
remnant mass leads to a substantial horizontal spread of the individual simulations
by more than an order of magnitude.
            
Thanks to the mass spread of the different bubble sequences seen in Figs.~\ref{fig:lx.r2.pn} and
\ref{fig:tx.r2.pn} we come to an even more detailed conclusion concerning the hydrogen-rich objects:

\begin{itemize}
  
\item  The X-ray luminosities of all seven bubbles around the O-type nuclei are constrained by our 
    0.565 and 0.625~\Msun\ simulations, where six objects are closely reproduced by the
    0.595~\Msun\ model bubbles.

\item  Virtually the same holds for the characteristic bubble temperatures if the very uncertain 
       \Tx\ value of IC~418's bubble (no.~2) is ignored. 
\end{itemize}
 
   Altogether, the observed evolution of $ \lx $ and $ \tx $ of the bubbles around the O-type sample
   objects is fully in line with the existence of heat conduction with evaporation as introduced by
   \citetalias{SSW.08}. Moreover, 
   the observed increase of $ \lx $ with bubble radius by about two orders of magnitude and the 
   observed moderate evolution of \Tx\ seems to validate our models of hot bubbles around
   hydrogen-rich AGB remnants of around 0.6~\Msun, in particular their evolutionary timescales and the 
   chosen mass-loss prescription.  
   Nevertheless, any firm assignment of a particular object to one of the post-AGB sequences shown in
   Figs.~\ref{fig:wind.model} through \ref{fig:r2.teff} is problematic because of the uncertainties
   of the stellar luminosities ($ L_{\rm star} $ and $ \lx) $, and especially of the mass-loss rates.
      
    In contrast, a consistent evolutionary picture is not evident for our [WR]-type sample objects.
\begin{itemize}
          
\item    There is no clear indication of an evolution of $ \lx $ with bubble size. The smallest two
         bubbles, nos.~1 (BD\,+30\degr 3639) and 8 (NGC 5315), have X-ray luminosities that correspond
         well to the mean value of the three big bubbles, nos.~4 (NGC~40), 7 (NGC 5189), and 12 
         (NGC~7026), viz. $ \log (\lx/L_\odot) \simeq -1.0 $.  
          
\item    Concerning the mean bubble temperatures we can only state that the two young and
         small bubbles are relatively hot whereas the temperatures of the big bubbles
         are quite low, more comparable to those of the 0.565~\Msun\ bubbles
         close to the end of our simulation.
          
\end{itemize}

   Although the two samples are rather small, the obviously different X-ray properties of the bubbles
   around O-type and [WR]-type central stars suggest a diverse evolutionary history of these two
   central-star spectral types in general, and between the young and old [WR]-type objects in
   particular. Their different chemical compositions may certainly play a decisive role,
   too. For instance, the comparably low X-ray luminosity of the hydrogen-poor bubbles is in contrast
   to the prediction of \citetalias{helleretal.16}, according to which bubbles with WR composition
   should have a 70--100\,times higher X-ray luminosity than their hydrogen-rich counterparts.
   In order to tackle the problem of the comparatively low observed X-ray luminosities of 
   hydrogen-poor bubbles, hydrodynamical models which allow for evaporation are mandatory.   
   In the following section we therefore discuss our new simulations with 
   hydrogen-poor winds in terms of their X-ray luminosities and temperatures.

\section{Our hydrogen-poor bubble models and the observations}
\label{sec:WR-bubbles.observ}

Evaluating the global X-ray properties of the simulations with [WR]-type winds
introduced in Sect.~\ref{sec:param.study}, it turns out that most of our
simulations are of no relevance for interpreting the so far existing
observations of the hydrogen-poor bubbles. Given the observed values of
mass-loss rate and wind speed of the [WR]-type objects
displayed in the Figs.~\ref{fig:wind.model}, \ref{fig:wind.model.2}, and
\ref{fig:r2.teff}, we found that only the two simulations, WR100V05 and
WR100V05x5.5, should suffice for a comparison with the observations.  Our
choice of a 100\ times higher mass-loss rate only instead of 200\ times as
suggested by the observations if no wind clumping is assumed (Fig.~\ref{fig:wind.model.2},
bottom) may be justified by assuming a moderately clumpy wind\footnote{The
true (clumpy) mass loss rate is relevant for heating the bubble, even
though our models assume a homogeneous wind (but see also
Sect.\,\ref{sec:clumps})}
characterised by a volume filling factor of $ f = 0.25 $.
   
\subsection{Evolutionary stages}
\label{subsec:evol.stage}

\begin{figure}
\center
\includegraphics[trim= 0.4cm 0.0cm 0.3cm 1.1cm, width= 0.98\linewidth, clip]
                {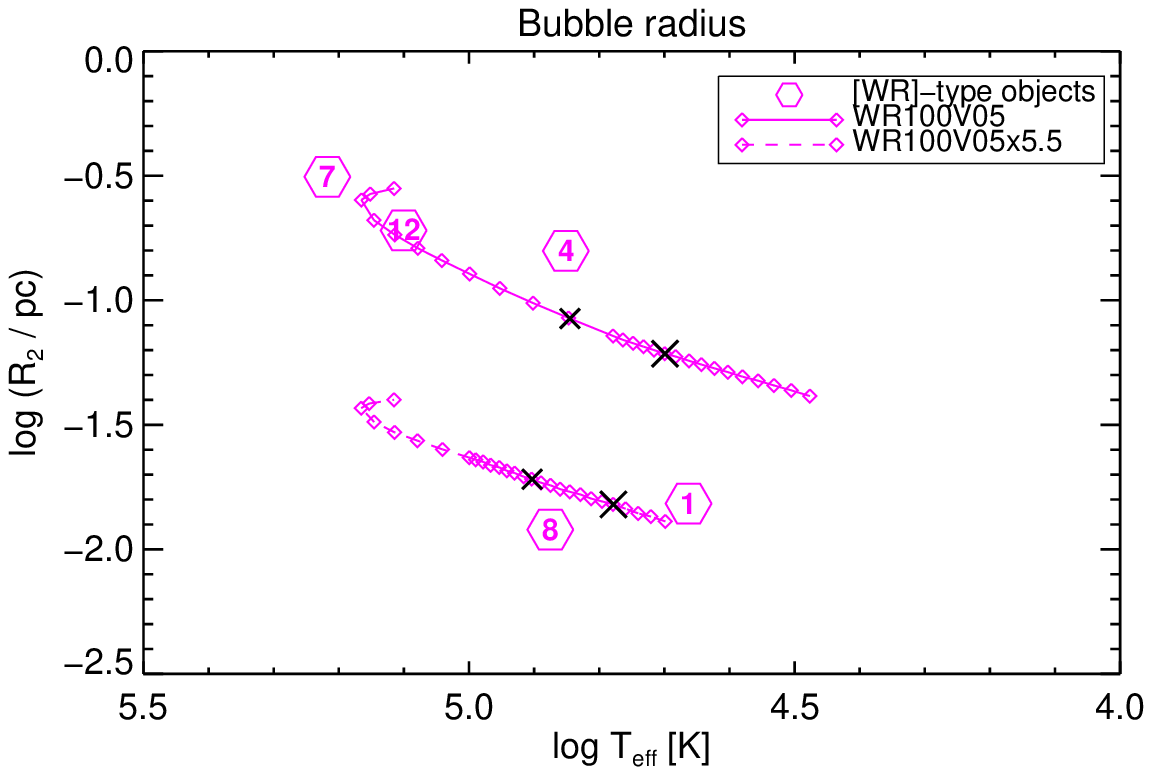}
\caption{\label{fig:r2.teff.wr}
         Same as in Fig.~\ref{fig:r2.teff} but now only for the new hydrogen-poor bubble sequences 
         WR100V05 (upper) and WR100V05x5.5 (lower) together with the bubbles of the [WR] objects in
         Table~\ref{tab:data}. The black crosses indicate the moments of hot-bubble formation 
         (big crosses) and the start of evaporation (small crosses).                   
         }      
\end{figure}

  Figure~\ref{fig:r2.teff.wr} is the same as Fig.~\ref{fig:r2.teff} but contains only the
  two selected WR simulations and the bubble positions of our [WR]-type sample objects. 
  The conclusions concerning the evolution of [WR]-type central stars can be
  refined as follows:
  
\begin{itemize}
\item  The central stars of BD\,+30\degr 3639 (no.~1) and NGC~5315 (no.~8) evolve fast and
       are fairly well represented by our accelerated WR100V05x5.5 simulation. 

\item  NGC~40 (no.~4) appears as having a central star which evolves more slowly than our WR100V05
       sequence, viz. by a factor of about two (--0.30~dex). 

\item  The evolution of the bubbles of NGC~5189 (no.~7) and NGC~7026 (no.~12) is well matched by our
       WR100V05 simulation. 
            
\end{itemize}    

   These conclusions remain valid even if one allows for the considerable uncertainty
   of the stellar mass-loss rates, 
   mainly caused by the filling factor $ f $.  For instance, an uncertainty of 
   the mass-loss rate by a factor of three translates into an $ R_2 $ change of $ \pm 0.16 $~dex
   only, provided the other two terms in Eq.~(\ref{eq:ZP}) remain virtually unchanged.   
   
   An important question still remains: are our models with WR-type winds and
   their delayed hot-bubble formation able to explain the existence of bubbles
   around the youngest objects of our [WR] sample, viz. of BD\,+30\degr 3639
   (no.~1) and NGC 5315 (no.~8)? Figure~\ref{fig:r2.teff.wr} contains the
   answer: we have marked the positions where the hot bubble forms (X) and
   where evaporation begins (x) for both simulations.
   Considering the uncertainties of individual stellar mass-loss rates (wind powers) and the
   corresponding changes of the times of hot-bubble formation and onset of
   evaporation, we can state the following: 
\begin{itemize}
\item   The bubble of BD\,+30\degr 3639 (no.~1) is close to the stage where our WR100V05x5.5 simulation
        predicts hot-bubble formation, while the bubble of NGC~5315 (no.~8) appears a bit more
        evolved, right before the beginning of evaporation.   
        
        Both bubbles can therefore be considered to be chemically homogeneous with a hydrogen-poor 
        WR composition. 
\item   The bubble of NGC~40 (no.~4) is still close to the evaporation stage 
        and may already contain a small amount of evaporated hydrogen-rich matter.

\item   The two most evolved objects, NGC 5189 (no.~7) and NGC~7026 (no. 12), should obviously have
        bubbles that are far into their evaporating stage. However, these
        objects with very extended bubbles show rather smooth X-ray intensity
        distributions which are not consistent with the model prediction
        of a sudden radial intensity drop towards an outer region of fainter X-ray
        emission next to the nebular rim as seen in Fig.~\ref{fig:bubble.structure}.
\end{itemize} 
   
   Altogether, the observed parameter combination bubble size and stellar effective temperature  
   found for our [WR]-type subsample can be matched by our simulations of hydrogen-poor bubbles 
   if the wind power is selected appropriately and the evolutionary speed adjusted if necessary.

\subsection{X-ray luminosities}
\label{subsec:xray.lum}

\begin{figure}
\center
\includegraphics[trim= 0.3cm 0.0cm 0.6cm 1.2cm, width=0.99\linewidth, clip]
                 {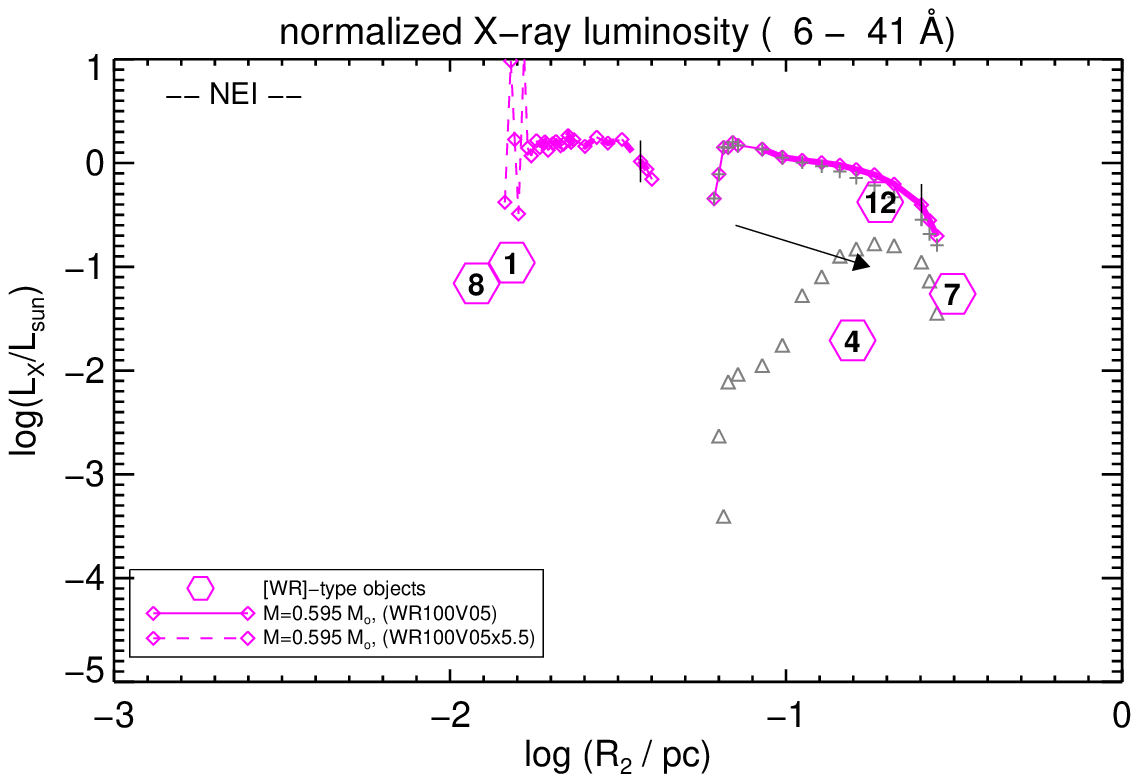}
\vskip 2mm
\includegraphics[trim= 0.3cm 0.0cm 0.6cm 1.2cm, width=0.99\linewidth, clip]
                 {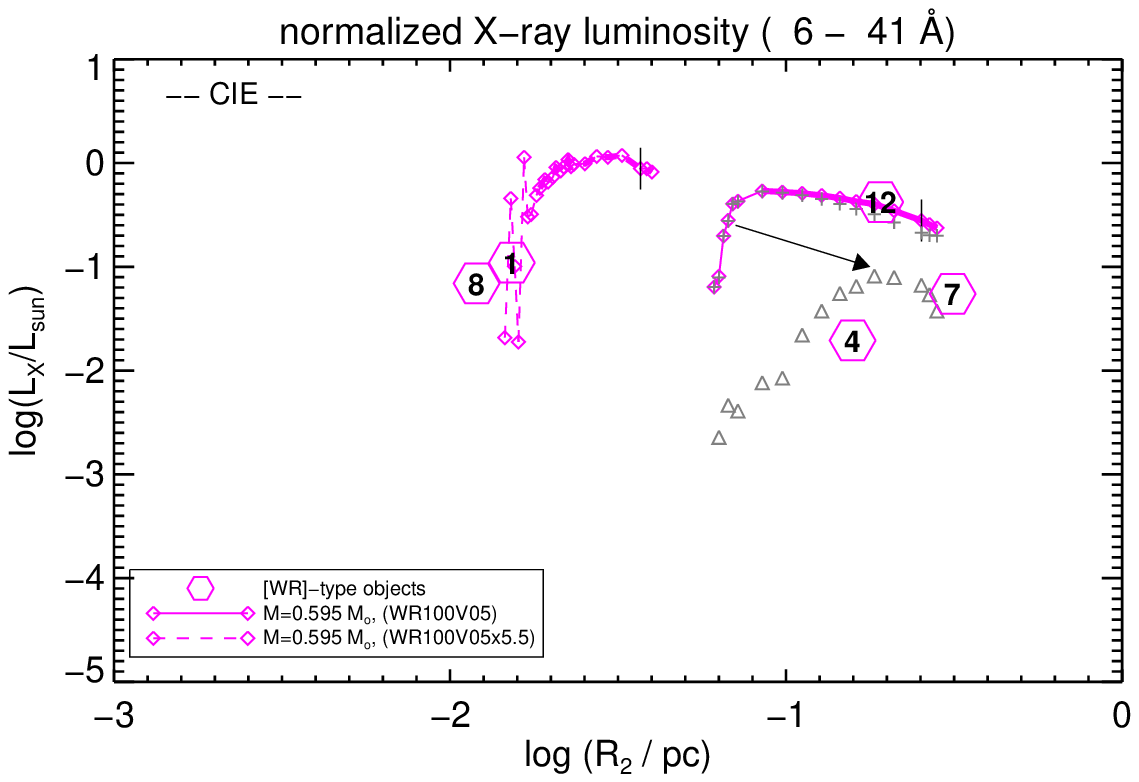}
\vskip -1mm
\caption{\label{fig:lx.r2.wr}  
  Bubble X-ray luminosities $ \lx $ versus bubble radius $ R_2 $ of the sequences WR100V05 
  (right, solid) and WR100V05x5.5 (left, dashed) and the bubbles of the [WR]-type sample objects of
  Table~\ref{tab:data}.  
  Evolution occurs from left to right, and maximum stellar temperatures are indicated by a
  vertical mark.   Evaporating bubble models are connected by a thick solid or dashed line. 
  The WR100V05 bubbles also contain the information on the luminosity contributions of the
  hydrogen-poor WR part (crosses) and the evaporated hydrogen-rich PN part (triangles). 
  The arrow indicates the (estimated) shift of the WR100V05 bubbles if the evolutionary 
  timescale is increased by a factor of 2.5. The \emph{top panel} displays the NEI,
  the \emph{bottom panel} the CIE case.
}
\end{figure}
  
Figure~\ref{fig:lx.r2.wr} shows the evolution of the X-ray luminosity with
bubble size for the WR100V05 and WR100V05x5.5 simulations, again for both
the NEI and CIE case. The NEI bubbles are more luminous than their
CIE counterparts, and the differences are a bit higher than seen
above for the hydrogen-rich bubbles.  The general trend is that the
`overluminosity' of the NEI bubbles is (i) higher during the earlier phases of
evolution, and (ii) less pronounced for the `accelerated' WR100V05x5.5
sequence.  The higher bubble densities of the latter sequence
obviously favour a bubble ionisation closer to equilibrium.
   
   In general, the bubble evolution can be described as follows:       
   When the hot bubble starts to form, its X-ray luminosity is already rather high, thanks to the high
   emissivity of the WR matter with ${ \langle Z \rangle \simeq 4 }$. 
   After this initial increase, the $ \lx $ evolution differs from the hydrogen-rich case.
   Instead of a continuous increase while the central star crosses the HRD, $ \lx $ remains fairly
   constant or even decreases with evolution.  The reason lies in the fact that the bubble's main mass
   input is due to evaporation of hydrogen-rich nebular matter which, however, contributes
   little to the total X-ray luminosity. For illustration, the individual luminosity contributions of the 
   WR wind matter and the evaporated PN matter are separately shown for the
   WR100V05 sequence in Fig.~\ref{fig:lx.r2.wr}. The input of hydrogen-poor
   wind matter is comparatively small and can hardly compensate for the decrease of the
   emission measure by expansion. We interpret Fig.~\ref{fig:lx.r2.wr} as follows: 
\begin{itemize}
        
\item   The young and small bubbles around the nuclei of BD\,+30\degr 3639 (no.~1) and NGC~5315
        (no.~8) are again reasonably well represented by the young and still chemically homogeneous
        hydrogen-poor bubbles of our accelerated WR100V05x5.5 simulation (cf. Fig.~\ref{fig:r2.teff.wr}).
        
\item   The same holds for the bubble of NGC~40 (no.~4) if we shift the WR100V05
        simulation by a factor of about 2.5 (0.4~dex) to the right, thereby approximating a slowed-down  
        stellar evolution by a factor of 2.5 (arrow in Fig.~\ref{fig:lx.r2.wr}).   
        NGC~40 can then be explained as an object on the ascending part of the WR100V05 bubble sequence 
        where bubbles are still (nearly) chemically homogeneous.

\item   The X-ray luminosities of NGC~5189 (no.~7) and NGC~7026 (no.~12) are, within   
        the uncertainties, fairly well represented by our WR100V05 bubbles during their
        evaporating stage.   
        
\end{itemize}  

It is very gratifying that both the Figs.~\ref{fig:r2.teff.wr} and
\ref{fig:lx.r2.wr} allow consistent interpretations concerning the present
evolutionary stage of our [WR]-type sample objects with respect to bubble size
and X-ray luminosity. It appears that our hydrogen-poor bubble sequences
WR100V05 and WR100V05x5.5 somewhat overestimate the X-ray luminosities of our
sample objects, suggesting that their real wind powers (mass-loss rates) are a
bit lower than those assumed for the models. Nonetheless, our simulations do
not indicate any evolutionary connection between the young and old [WR]-type sample objects,
contrary to what one might have assumed from Fig.~\ref{fig:xray.hrd} alone.

\subsection{The characteristic bubble temperatures $ \tx $}
\label{subsec:bubble.temp}         
   
\begin{figure} 
\center 
\vskip 1mm
\includegraphics[trim= 0.3cm 0cm 0.6cm 1.15cm, width=0.99\linewidth, clip]
               {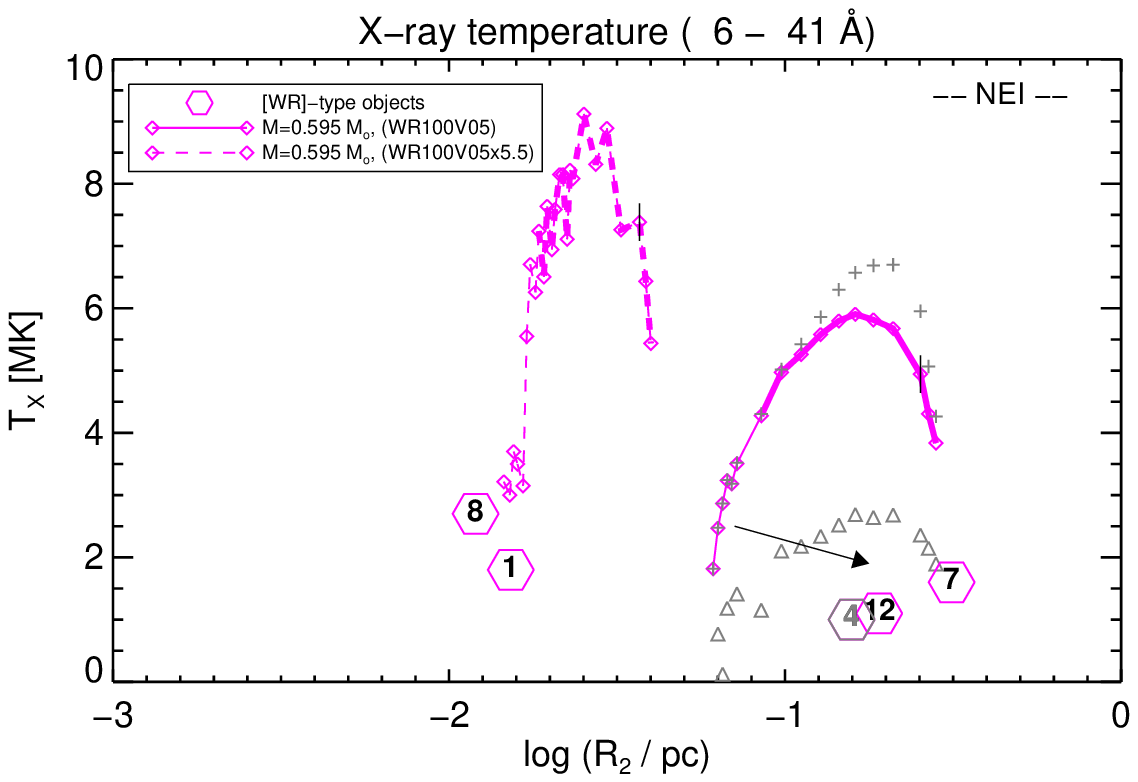}
\vskip 2mm                
\includegraphics[trim= 0.3cm 0cm 0.6cm 1.15cm, width=0.99\linewidth, clip]
               {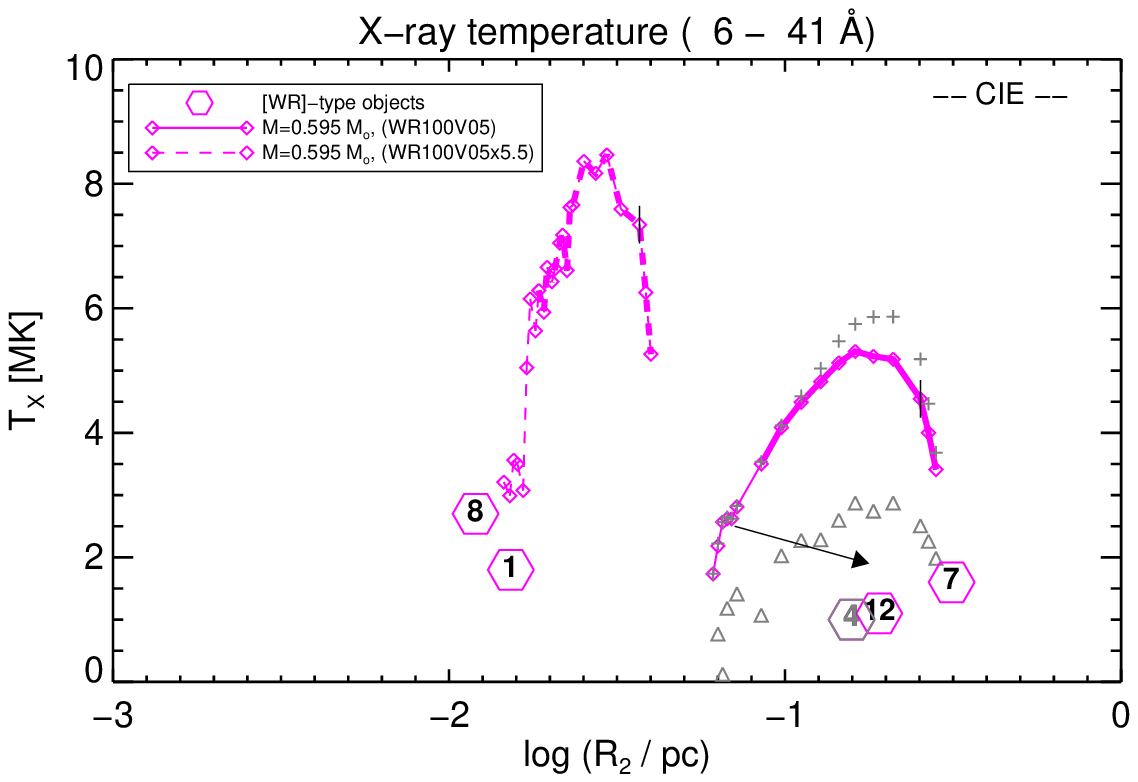}
\vskip -1mm 
\caption{\label{fig:tx.r2.wr}  
         Characteristic X-ray temperature $ \tx $ versus bubble radius $ R_2 $
         for the sequences WR100V05 (right, solid) and WR100V05x5.5 (left,
         dashed) and the bubbles of our sample (Table~\ref{tab:data}).  Again,
         bubble evolution occurs from left to right, and maximum stellar
         temperatures are indicated by a vertical mark.  Evaporating bubbles
         with inhomogeneous composition are connected by thick solid or dashed
         lines. The mean temperatures of the hydrogen-poor part,
         $ \tx(\rm WR) $, and of the hydrogen-rich part, $ \tx(\rm PN) $, are shown
         as crosses and triangles, respectively, for the WR100V05 bubbles. As
         in the previous figure, the arrow indicates the (estimated) shift of
         the WR100V05 bubbles if the evolutionary timescale is increased by a
         factor of 2.5.  The symbol of NGC~40 (no. 4) is shown in grey because
         of the large uncertainty of the derived \Tx\ value. The \emph{top}
         and \emph{bottom} panel displays the NEI and the CIE case, respectively.
         }
\end{figure}        

In Fig.~\ref{fig:tx.r2.wr}, the comparison between the X-ray temperature
predictions of our WR100V05 and WR100V05x5.5 bubble sequences and the
observation is presented, again separately for the NEI and CIE case. The
mean bubble temperatures are systematically higher in the NEI case by
about 0.5~MK at most.
     
The temperature evolution predicted by the models follows an inverted `U'
shape.  First, the bubble temperature increases rapidly with bubble size, in
pace with their X-ray luminosities (Fig.~\ref{fig:lx.r2.wr}), goes through a
maximum, and finally declines towards the end of our simulations.  A longer
lasting phase of roughly constant bubble temperature as found for the
hydrogen-rich case is missing, obviously because the bubbles form much later
at higher stellar temperatures and hence with bigger bubble
radii. Furthermore, the achieved bubble temperatures are significantly higher
than those of the hydrogen-rich bubble models (cf. Fig.~\ref{fig:tx.r2.pn}).
The reason is the very high stellar wind power at comparable bubble sizes
(cf. Eq.~\ref{eq:tx.wind.size.2}) which cannot be compensated fully by the
higher bubble size $ R_2 $ and its faster evolution.
   
   Although the wind model is the same in both displayed simulations, the achieved bubble temperatures
   are quite different, i.e. 6.0 vs. $ \simeq $9.0~MK at maximum (NEI case). 
   The reason is the faster stellar
   evolution of the WR100V05x5.5 simulation where the bubbles are smaller by about a factor of 5.5 and 
   correspondingly hotter.  An estimate using Eq.~(\ref{eq:tx.wind.size.2}) yields a factor of
   ${ 5.5^{2/7} = 1.6 }$, and therefore the maximum mean bubble temperature reached by the accelerated 
   simulation should be $ 6.0\times 1.6 = 9.6 $, in fair agreement with the simulation. 
   
   For completeness, we also show in Fig.~\ref{fig:tx.r2.wr} the mean
   temperatures of the hydrogen-poor and hydrogen-rich bubble parts of the
   WR100V05 sequence separately.  The differences are considerable, simply
   because the evaporated hydrogen-rich matter resides in the outer, cooler
   parts of the bubble while the wind matter occupies the inner, very hot part
   of the bubble (cf.~Figs.~\ref{fig:WR100V05.suite.bubbles} and
   \ref{fig:bubble.structure}).  However, the hydrogen-poor WR matter always
   dominates the X-ray emission and the overall mean bubble temperature \Tx\
   is always below but close to \Tx(WR). We conclude from Fig.~\ref{fig:tx.r2.wr}:

\begin{itemize} 

\item  The mean bubble temperatures of the young objects BD\,+30\degr 3639 (no.~1) and NGC~5315 (no.~8)
       are consistent with the prediction of the WR100V05x5.5 models with purely hydrogen-deficient 
       composition and an accelerated stellar evolution.

\item  The mean temperatures of the two old bubbles of NGC~5189 (no.~7) and NGC~7026 (no.~12) completely
       disagree with the predictions of the WR100V05 bubbles.  They range between about 1.1 and 1.6~MK
       while the models predict temperatures of up to 5\ldots 6~MK in the relevant $ \teff $ regime.  
       Even the evaporated, hydrogen-rich bubble regions have mean temperatures as high as about 2.6~MK. 
                 
\end{itemize}    
   Concerning NGC~40 (no.~4), we repeat that any temperature estimate of bubble plasmas from
   spectral X-ray distributions requires observations of sufficiently high quality.
   In our opinion, this is not the case for NGC~40 where the total number
   of useful X-ray photons is around 60 only.  

   However, taking the observed low mean bubble temperature as granted, the following interpretation
   can be made. Assuming a slower central-star evolution as discussed above for interpreting the   
   X-ray luminosity and size of NGC~40's bubble, its low mean temperature
   would be a natural consequence (cf. arrow in Fig.~\ref{fig:tx.r2.wr}).
   The hot bubble of NGC~40 would therefore be rather young and still 
   chemically homogeneous but already quite big because of the slow central-star evolution.
 
    The obvious mismatch between theory and observation found for the bubble temperatures of 
    NGC~5189 and NGC~7026 is surprising in view of the rather good agreement concerning their 
    bubble sizes and X-ray luminosities. It is certainly real because the spectral appearance
    of a 5~MK hot plasma with WR composition can by no means be mistaken with that of a plasma as
    cold as about 1.6~MK or less.  
    Apparently, our 1D hydrodynamical \texttt{NEBEL} simulations are missing some physical ingredients 
    that are necessary to provide an adequate description of evaporating hydrogen-poor wind-blown
    hot bubbles.   

\section{Discussion}
\label{sec:discussion}

The present study, a follow-up of the \citetalias{SSW.08} and
\citet{ruizetal.13} works, shows again the usefulness of our 1D
hydrodynamical code \texttt{NEBEL/CORONA} in conjunction with the heat conduction
paradigm for interpreting the soft X-ray emission from the wind-blown
bubbles of planetary nebulae. For computing synthetic \mbox{X-ray} spectra, we have
for the first time allowed for deviations from collisional ionisation
equilibrium for nine key chemical elements that dominate the emission in the
considered energy range. The comparison of the NEI results with those based
on the usual assumption of CEI revealed considerable differences of the
local ionisation fractions inside the bubbles. Concerning the global bubble
parameters like X-ray luminosity and characteristic X-ray temperature, the
differences between the NEI and CIE cases are less dramatic, and therefore
the previous conclusions based on the CIE alone remain valid.
In contrast, the determination of abundance ratios from spectral features
will depend critically on the details of the ionisation balance.
    
While the diffuse X-ray emission of wind-blown bubbles around hydrogen-rich
central stars could be very well explained in all aspects by our simulations,
open questions still remain for bubbles around [WR] central stars where
the models predict the extended bubbles to become chemically inhomogeneous
due to evaporation of hydrogen-rich nebular matter.
We will discuss some important issues in the following.

\subsection{The importance of mixing by dynamical instabilities}
For the hot bubbles around hydrogen-rich central stars, we can compare the
observed X-ray temperatures with the predictions of the pure mixing models of
\citet{TA.16}.  For central-star masses around 0.6~\Msun\ relevant here, the
pure mixing models reach rather high mean bubble temperatures from about 2.6
up to about 4~MK in the stellar temperature range from about 40\,000 to
100\,000~K (cf. Fig.~8 in \citealt{TA.16}). Although this stellar temperature
range is almost covered by our sample objects (cf. Table~\ref{tab:data}), such
high bubble temperatures are not observed: With the exception of IC~418 whose
\Tx\ value is uncertain and which also has most probably a more massive
central star, the \Tx\ values of the six remaining bubbles do not indicate any
evolution with bubble size or stellar temperature (Fig.~\ref{fig:tx.r2.pn}).
Instead, they cluster around the mean value of ${ 1.96\pm 0.25 }$~MK, very
close to the mean bubble temperatures of our 0.595~\Msun\ sequence, viz.
${ \simeq\! 1.8 }$~MK.
     
Regarding the question about the role of dynamical mixing, this mismatch
suggests that mixing processes as considered by \citet{TA.16} cannot alone be
responsible for the observed low mean bubble temperatures.  The reason is that
mixing is a slow process; the time needed for producing enough relatively cool
bubble matter seems to be comparable to the HRD crossing time. In contrast,
thermal conduction works on an extremely short timescale and imposes the
typical temperature profile nearly `instantaneously' \citep{ZP.96}.  Indeed,
if \citet{TA.16} also allow for thermal conduction in their simulations, the
mean bubble temperature of their 0.597~\Msun\ sequence does not exceed 1.6~MK.

\subsection{Low mean temperatures of hydrogen-poor bubbles}

   Our bubble models with thermal conduction are able to explain the 
   rather moderate mean temperatures of the young bubbles of BD\,+30\degr 3639 and NGC~5315 but fail
   in the case of large, old bubbles (e.g. of NGC~5189 and NGC~7026).   The reason for this failure
   is evaporation of hydrogen-rich nebular matter during later stages which leads in our models 
   to a chemical stratification where the evaporated matter encloses the hydrogen-poor wind matter.
   These chemically stratified bubbles have 
    very high mean plasma temperatures since the X-ray spectrum is always dominated by the
   emission of the hydrogen-poor but hot inner bubble region.   
   
   Such high temperatures have so far not been observed, suggesting that in
   nature an additional physical process is at work that somehow prevents a clear 
   separation of the two chemistries within a bubble.  
   Dynamical mixing across the chemical discontinuity 
   by hydrodynamical instabilities is certainly conceivable.
   Complete mixing would result in 
   (i)   a lowering of the mean ionic charge,   
   (ii)  a surface-brightness distribution without a jump, and therefore
   (iii) a lower characteristic X-ray temperature. 
   However, a more thorough investigation of this scenario is beyond the scope of the present work.  

  Another possibility is a sudden considerable decrease of the wind power 
  (cf. Eq.~\ref{eq:tx.wind.size.2}) during the late stages of evolution.  However, this is not observed:
  the present-day wind-powers of NGC~5189 and NGC~7026 correspond closely to those used in our models
  (Fig.~\ref{fig:wind.model}, bottom panel, dashed line).       

Finally, we would like to emphasise that the mean X-ray temperature
derived from the observation and from the models, respectively, are based
on different methods. The usual method used to interpret the observations
is to fit spectral features assuming a single-temperature plasma, while
the X-ray temperature of the models is defined as the emissivity-weighted
temperature of the heat-conducting bubble structure. Hence the
characteristic mean temperatures derived from the models are not directly
compatible with those derived from the observations.  However, it was
shown in \citetalias{helleretal.16} (Fig.~6 therein, albeit assuming CIE)
that the differences are quite moderate and do not exceed 0.4~MK in the
relevant temperature range, in the sense that the bubble models predict
lower temperatures for given line ratios. Therefore, this inconsistency
cannot be held responsible for the temperature discrepancies discussed
here.

  \subsection{The problem of wind clumping}
  \label{sec:clumps}
    As mentioned previously, especially the dense winds of the [WR]-type
    objects may not be homogeneous but rather `clumpy'. However, we neither
    can model clumpy winds nor can we estimate the consequences of wind
    clumping. These uncertainties of the mass loss from [WR]-type central
    stars are the main reason why we did not perform simulations
    which provide individual matches to our sample objects.    
   Fortunately, any uncertainties of the mass-loss rate, and hence wind power, 
   by factors of 2\ldots 3 only have a rather modest
   influence on the bubble's size, X-ray luminosity, and mean temperature, and
   would not change our interpretations and conclusions. 
   
   It is possible that the hot bubble has density inhomogeneities (`clumps')
   as well because a clumpy wind flow becomes simply compressed and
   decelerated while passing through the shock.  High-density regions within a
   bubble would be sites of very efficient radiation cooling, reducing thereby
   considerably the part of the wind power capable of heating the bubble. However, it
   is unclear how such density fluctuations could survive under the
   conditions of constant pressure and thermal conduction. Based on 2D
   hydrodynamical models, \citet{Dwar.23} recently put forward the idea of
   additional bubble cooling by density inhomogeneities in order to explain
   the low temperatures of wind-blown bubbles around massive stars.
   He concluded that the inclusion of thermal conduction is not necessary
   to reproduce the X-ray temperatures and spectra. On the other hand,
   \citet{zhekov.11} and \citet{zhekov.14} had previously demonstrated that
   heat conduction is an efficient physical mechanism controlling the
   temperature structure in 1D numerical models of this type of hot bubbles,
   and therefore a viable explanation of the low plasma temperatures deduced from
   the observed X-ray spectra of both massive WR-type stars, as is the case
   for Planetary Nebulae.
   
\subsection{The chemical composition of hydrogen-poor bubbles}

 By using the WR abundances of Table~\ref{tab:abundances} in our simulations with hydrogen-poor 
 stellar winds it is guaranteed that, at least in the beginning, the bubbles
 are truly hydrogen-poor, too.  
 The hydrogen-poor character of our WR mixture is evident from the numbers in Table~\ref{tab:abundances}
 where helium and carbon are the most abundant species.
    
 This is obviously not the case in Table~A1 of \citet{TA.18} where the abundance values of 
 BD\,+30\degr 3639 indicate that hydrogen and helium are still the most abundant elements 
 (${ \epsilon_{\rm H} = 12 }$, ${ \epsilon_{\rm He} = 10.99, \epsilon_{\rm C} = 10.59,
 \epsilon_{\rm O} = 9.53 }$).   The abundance of 
 carbon, oxygen are scaled relative to helium according to the ratios found by 
 \citet{marco.07} and \citet{yuetal.09}.  This procedure ensures that helium, carbon, and oxygen have 
 the correct proportions, but the chemical mixture is still hydrogen-rich, at variance with 
 existing spectral analyses which proved that BD\,+30\degr 3639's surface is definitively hydrogen-poor
 (e.g. \citealt{leuetal.96, marco.07}). 
 {Such an element distribution ignores the fact that the surface
  layers of [WR]-type central stars consist of matter that has gone through hydrogen and     
  helium burning in the past.}
 
  Since hydrogen and helium have no X-ray line signatures, such a `pseudo' hydrogen-poor composition 
  is able to provide correct X-ray spectra once the proportions of the main X-ray emitters, carbon, 
  oxygen, and neon have been adjusted accordingly.  However, the `pseudo'
  hydrogen-poor composition has a lower mean electronic charge than a true WR mixture
  (4.4) and is therefore not suited to compute realistic X-ray luminosities and cooling rates of 
  hydrogen-poor hot bubbles.  
 
\subsection{Formation scenarios of hydrogen-poor central stars}
We have assumed here, based on single-star simulations, that the change to a
hydrogen-free (-poor) stellar surface occurs immediately when the star leaves
the AGB, possibly by a thermal pulse right at the tip of the AGB.
A thermal pulse occurring later while the AGB remnant evolves across the HRD, is
no real option for the following reasons:

\begin{itemize}
\item Such a thermal pulse (late or very late) forces the star to expand back
  to the tip of the AGB for some time until it shrinks again towards
  hotter regions of the HRD. During this second giant phase, a deep envelope
  mixing is responsible for the hydrogen depletion of the surface layers.
  Possible examples of such a scenario are, e.g. A\,30 \citep{guerrero.12}
  and A\,78 \citep{toala.20}.  However, none of the three evolved objects in
  our [WR]-type sample show indications of nested nebular shells with
  different chemistries.  Also, the existence of [WR]-type central stars as cold as
  ${ \simeq\! 20\,000 }$~K \citep[Fig.~2 in][]{hamann.97}, seem to support our
  assumptions.
        
\item There remains the possibility of binary-star interactions which may lead
  to the loss of the remnant's hydrogen-rich envelope.  However, this should
  occur only while the object in question is big, i.e. still on or close
  to the tip of the AGB.              
\end{itemize}  
We therefore believe that the hydrogen-poor stellar wind starts right at the beginning
of the post-AGB evolution. We admit that such simple
scenario cannot provide any explanation why our sample of [WR]-type objects
seems to contain two distinct subgroups with obviously rather different
evolutionary histories. Our models can only suggest a combination of wind
power and evolutionary timescale which is able to reproduce the observations.
         
\subsection{The Miller Bertolami \& Althaus  post-AGB models}

Before closing the discussion, we would like to comment on the modern
evolutionary calculations of \citet{MMA.06} which include updated radiative
opacities and a better treatment of the boundaries between radiative and
convective stellar layers by considering convective overshoot. The
differences concerning the relevant values of post-AGB luminosities and
crossing times between these and our calculations are discussed in more
detail in Appendix~\ref{app:post.AGB}.  The important result is that there
exist a scaling relation between the \citet{MMA.06} masses and the older
ones of \citet{S.79, S.83} and \citet{B.95} that is valid for both the
luminosity and crossing time.
   
Since stellar mass is not entering the hydrodynamical simulations, and since the stellar 
(bolometric) luminosity is not important for the bubble evolution, the use of the older post-AGB
evolutionary tracks is of no concern for the present work because our remnant masses can easily be
converted to the Miller Bertolami mass scale (Fig.~\ref{appfig.comp}, top).
A very similar relation exists for the conversion of the \citet{VW.94} post-AGB masses to the
Miller Bertolami mass scale (Fig.~\ref{appfig.comp}, bottom).
  
We note that post-AGB evolutionary tracks of WR-type central stars with a fully consistent
treatment of the enhanced mass loss rate and evolutionary timescale are not
available to date.

\section{Summary and conclusion}
\label{sec:conclusion}  

   We conducted a parameter study by means of 1D radiation-hydrodynamic simulations with our
   \texttt{NEBEL/CORONA} code and followed the evolution of all circumstellar structures (wind, hot bubble,
   nebula proper) from the tip of the AGB across the HRD into the white-dwarf domain.   
   The hydrodynamical evolution of these structures,
   especially of the wind-blown bubbles, is primarily determined by the \citet{Pauletal.88} wind
   model. Consistently using the same post-AGB remnant of
   0.595~\Msun\ and the same initial circumstellar envelope, we varied 
   (i) the stellar wind power via the mass-loss rate and wind velocity, 
   (ii) the chemical composition of the stellar wind ---that is, hydrogen-rich vs. hydrogen-poor--- and 
   (iii) the post-AGB evolutionary timescale, guided by observational evidence. 
   
   We employed, for the first time, a hybrid method for the computation of
   the X-ray fluxes emerging from the hot bubbles; that is, we forced the \texttt{CHIANTI}
   code to use the non-equilibrium ionisation (NEI) fractions of nine
   important elements provided by our \texttt{NEBEL/CORONA} code while
   collisional equilibrium ionisation (CIE) was kept for the remaining (trace)
   elements. We find that the NEI fractions inside the bubbles differ
   considerably from the equilibrium values, but the effects on the global
   bubble parameters, such as X-ray luminosity ($ \lx $) and X-ray temperature
   (\Tx), are comparably moderate.
   We conclude that the results of previous studies that exclusively rely on
   the assumption of CIE remain valid, at least when analysing the global
   parameters $ \lx $ and \Tx.

 Our main focus is on the scenario where a hydrogen-poor wind interacts with
 the hydrogen-rich nebular matter. In particular, we studied the formation and
 evolution of bubbles fed by a hydrogen-poor wind typical for
 [WR]-type central stars and computed the X-ray emission of these bubbles at
 selected evolutionary phases. We find that chemically inhomogeneous wind-blown
 bubbles can develop at advanced evolutionary stages by evaporation of nebular
 matter across the bubble--nebula interface.
      
   Our model computations were used to interpret the diffuse X-ray emission of hot bubbles inside
   planetary nebulae observed by the \xmm\ or \chan\ satellite.  Altogether, a set 
   of seven nebulae with hydrogen-rich O-type central stars and five nebulae with hydrogen-poor 
   [WR]-type central stars have been collected. By means of the \gaia\ DR3 parallaxes, we determined
   all the relevant (optical and X-ray) parameters available from the literature as accurately as
   possible.  Our detailed comparisons of the observations with the corresponding predictions of our
   hydrodynamical bubble models lead us to the following conclusions, in part
   addressing the four open questions posed in Sect. 1:
     
\begin{description}
\item[\sl Hydrogen-rich objects.]   \hspace*{0cm}

 \begin{itemize}

\item[]
       Our seven sample objects form a relatively homogeneous group (with the possible exceptions of
       IC~418 and NGC~2392),  whose properties with respect to bubble radius, X-ray temperature, and
       X-ray luminosity can be well described by a hydrodynamical simulation of the circumstellar
       structures around a 0.595~\Msun\ AGB remnant, and by imposing the post-AGB wind model of
       \citet{Pauletal.88}.  Thermal conduction is mandatory while dynamical mixing does not seem 
       to play a significant role. 
       
 \end{itemize}

\item[\sl Hydrogen-poor objects.]   \hspace*{0cm}

  \begin{itemize} 
   
  \item  Our sample of five objects seems to be split into two subgroups: one consisting of two very
         young objects (BD\,+30\degr 3639, NGC~5315) with central stars that evolve considerably
         faster than our original 0.595~\Msun\ model, and the other consisting of three comparatively old objects
         (NGC~40, NGC~5189, NGC~7026) with central stars that evolve on a timescale similar 
         to or slightly longer (NGC~40) than our 0.595~\Msun\ model.

  \item  Due to the very high radiation-cooling efficiency 
         of hydrogen-poor but carbon- and oxygen-rich matter, the hot-bubble formation is delayed
         to higher stellar wind powers (or effective temperatures).  
         Furthermore, we find that heat conduction does not necessarily lead
         to efficient evaporation of hydrogen-rich nebular matter at all
         times. Under certain conditions, even condensation of matter out of
         the bubble may occur for some time, such that evaporation may be
         postponed to later evolutionary stages.
         Therefore, the existence of hot bubbles with homogeneous WR composition is no argument against
         the presence of heat conduction. 
         
         Guided by our simulations, we identified the bubbles of the two young objects BD\,+30\degr 3639
         and NGC~5315 together with the older object NGC~40 as being in a stage where evaporation is
         still absent or very weak.
         The bubbles of NGC~5189 and NGC~7026 are more evolved and should be
         in the evaporating stage, but verification would require much
         better X-ray observations. 

        \item The evaporating models appear very hot because the hydrogen-poor
          wind matter of high emissivity resides in the inner, very hot bubble
          region and dominates the X-ray emission.  However, the predicted
          high characteristic bubble temperatures are not observed. Apparently,
          our models with WR-type central star winds are lacking some physical
          ingredients.
         
  \item  The present study shows ---albeit based on a   
         small sample--- that at least two formation scenarios for hydrogen-poor central stars must
         exist.  One scenario leads to very fast evolving AGB remnants, and the other 
         to more slowly evolving remnants.  These two different timescales of post-AGB evolution
         most likely reflect primarily different remnant masses, but an individual mass 
         assignment is not possible as long as dedicated post-AGB evolutionary
         tracks for WR-type central stars are unavailable. 

\end{itemize}

\end{description}   

  The present study confirms that heat conduction is necessary to explain
   the low temperatures found in wind-blown bubbles inside planetary nebulae.
   This implies that magnetic fields are unlikely to be responsible for shaping the nebulae proper,
   at least for the objects showing diffuse X-ray emission.  

\begin{acknowledgements}
   We are grateful to the (unknown) referee who pointed out that the
   ionisation within a wind-blown bubble cannot be expected to be in equilibrium.
   D. S. thanks the Leibniz-Institut f\"ur Astrophysik Potsdam for hospitality and support.
\end{acknowledgements}



\begin{appendix}
\section{Influence of the X-ray band width}
\label{appsec:calib.bandwidth}          
          
\begin{figure}[hb]
\vskip -1mm
\centering
\includegraphics[width= 0.72\linewidth, angle= -90]{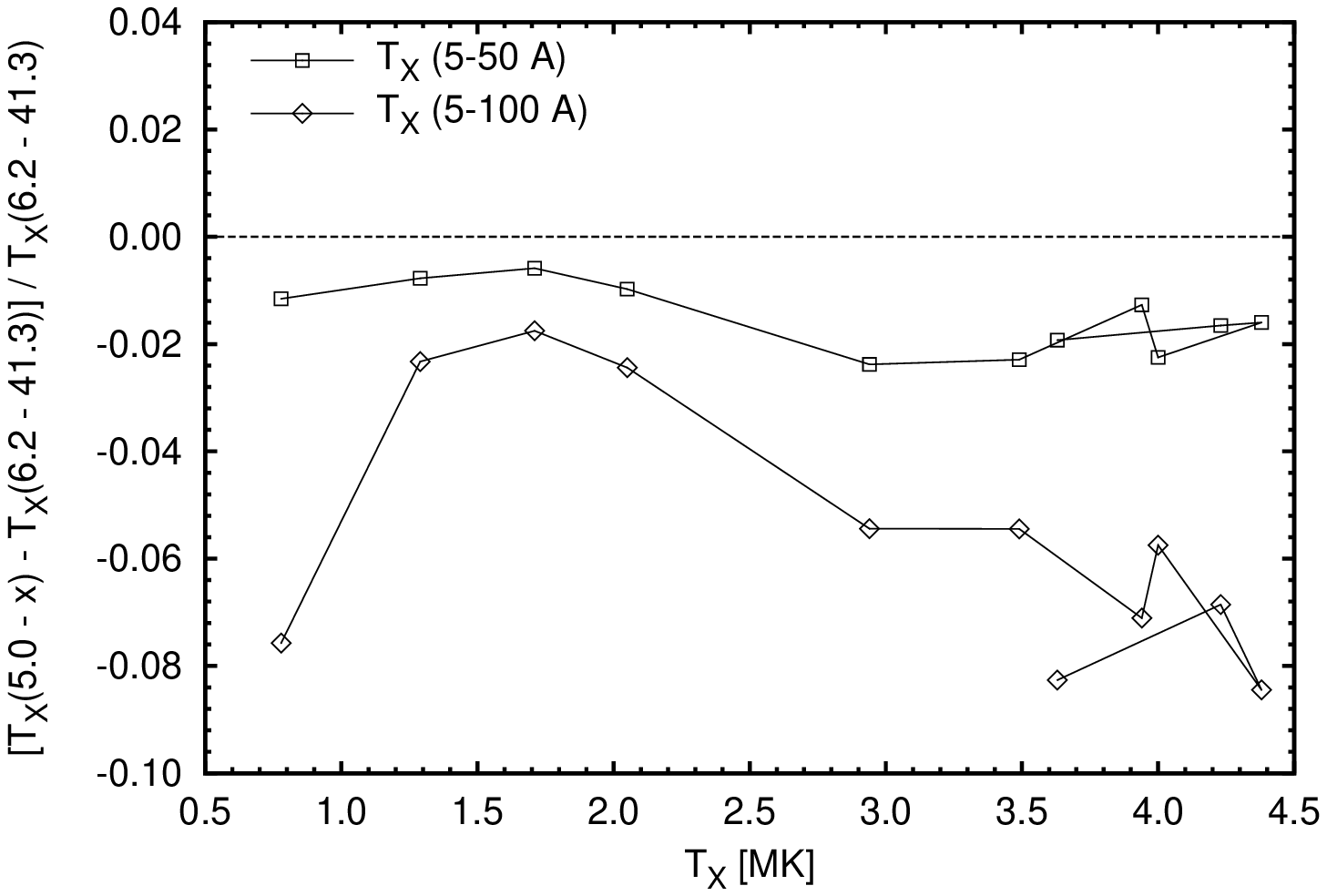}
\vskip -2mm
\caption{\label{appfig:tx.band}
  The differences of the \Tx\ values as computed by Eq.~(\ref{eq:tx}) for the
  two selected band widths (legend) relative to the one used in this work
  (6.2--41.3~\AA), assuming CIE. Symbols belong to the PN bubbles of the
  0.696~\Msun\ simulation. The evolution of \Tx\ occurs from low to high
  values and back when the star fades (cf. Fig.~\ref{fig:tx.r2.pn}).
  }
\end{figure}

    The definition of \Tx\ in Sect.~\ref{sub:mean.tx} implies that the temperature values
    depend on the chosen X-ray band width. This band-width effect is illustrated in
    Fig.~\ref{appfig:tx.band} where the bubbles of the (hydrogen-rich) 0.696~\Msun\ simulation
    have been used as an example because they cover a wide range of \Tx\ values.
     
   Figure~\ref{appfig:tx.band} shows that
   the differences to the band width used in the present work
   6.2--41.3~\AA\ (0.3--2~keV) depend primarily on the low-energy limit.  
   This is so since below 5~\AA\ (2.5~keV) there
   is hardly any X-ray emission, even for the hotter bubbles.   
   The closest agreements between the various \Tx\ values of the three band widths appear
   around ${ \tx \approx 1.7 }$~MK. At these temperatures most of the bubble's emission occurs
   just within 10--50~\AA\ (0.25--1.26~keV).  In general, the band-width dependence of \Tx\
   is quite modest and is not exceeding 10\% in the worst case.     
   The X-ray band used in the literature (which is dictated by the existing X-ray space
   observatories) is obviously sufficient to provide a useful temperature estimate of the X-ray 
   emitting bubbles of planetary nebulae.  
     
\begin{figure}[hb]
\vskip -1mm
\centering
\includegraphics[width= 0.72\linewidth, angle= -90]{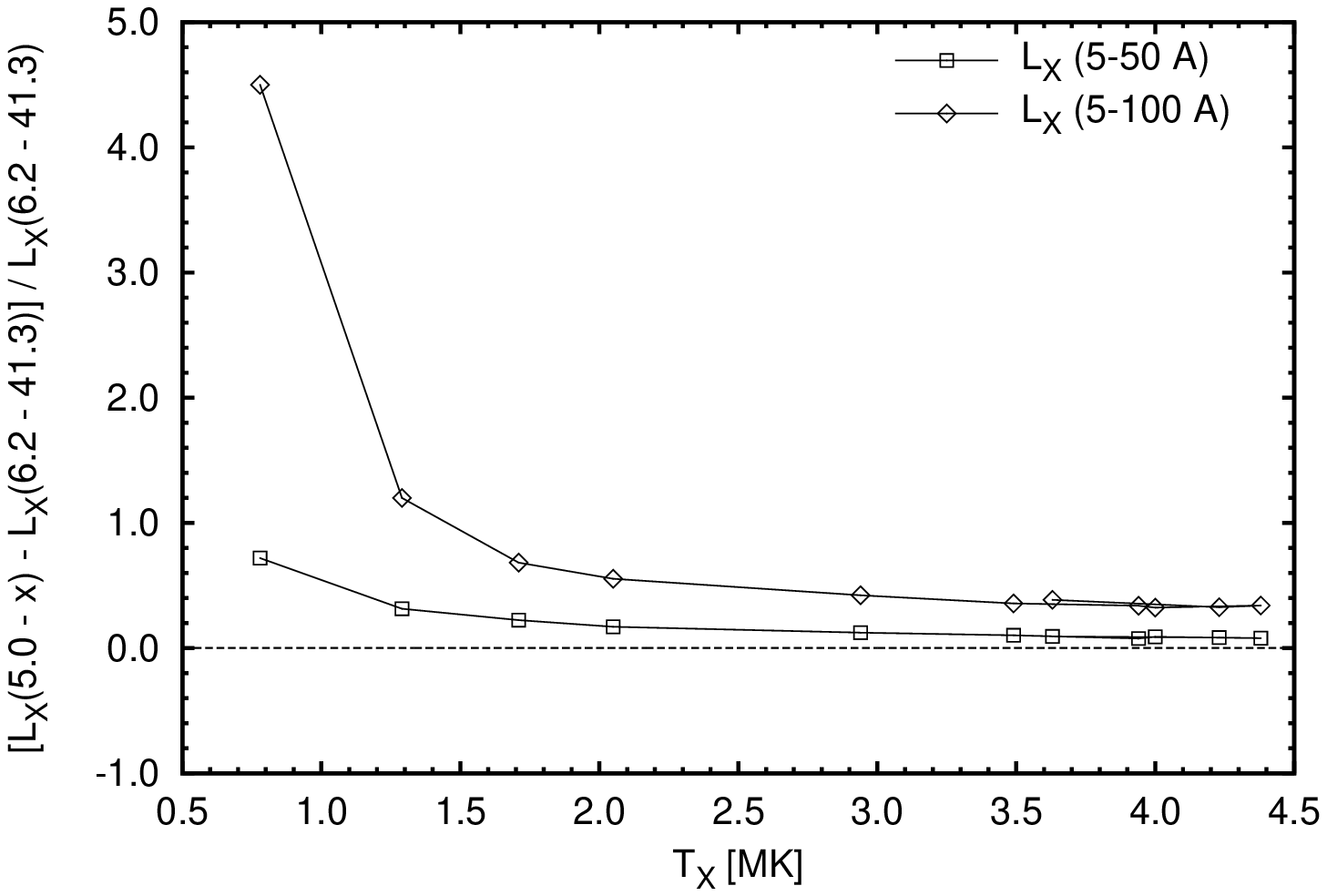}
\vskip -2mm
\caption{\label{appfig:lx.band}
  The difference of the $ \lx $ values based on the two selected band widths
  (legend) relative to the one used in this work (6.2--41.3~\AA), assuming
  CIE.  Symbols belong to the same PN bubbles as Fig.~\ref{appfig:tx.band}.
  }
\end{figure}
         
   Contrary to the bubble temperature, the corresponding X-ray luminosity is severely dependent on
   the chosen band width.  This is demonstrated in Fig.~\ref{appfig:lx.band} where the $ \lx $
   differences of the same bubbles as in Fig.~\ref{appfig:tx.band} are displayed.   
   As expected, the luminosity differences increase with decreasing low-energy limit to very high
   values for cool bubbles.   Any observed $ \lx $ value without indication of the used
   energy band is therefore useless. 
    
\section{Comparison of the properties of different post-AGB evolutionary sequences}
\label{app:post.AGB}

   The existing hydrodynamical simulations of planetary-nebula evolution employ
   the evolutionary models available at the time, i.e. either those of
   \citet{S.79, S.83} and \citet{B.95} (used in the present work) or those of
   \citet{VW.93,VW.94} \citep[used by, e.g. ][]{TA.14,TA.16}. Both sets of models
   are based on the same or very similar physical assumptions with respect to
   radiative opacities and treatment of convection without overshoot.
   
   The question now arises whether these hydrodynamical simulations are still of relevance in view of
   more recent post-AGB evolutionary calculations like those of \citet{MMA.06} with the latest
   radiative opacities and inclusion of convective overshoot across radiative-convective boundary
   layers.   Similar calculations have been performed by \citet{WF.09}, but they are not useful 
   for our purposes because of the limited mass coverage.  
   \citet{MMA.19} presented an extensive overview of the differences between the old and new 
   evolutionary calculation with respect to the post-AGB phase.  We concentrate here on the
   parameters that are of relevance for the hydrodynamical simulations, i.e. the luminosities of
   the AGB remnants and their timescales for crossing the HRD.  

\begin{figure}[hb]
\vskip -4mm
\center
\includegraphics[width= 0.71\linewidth, angle= -90]{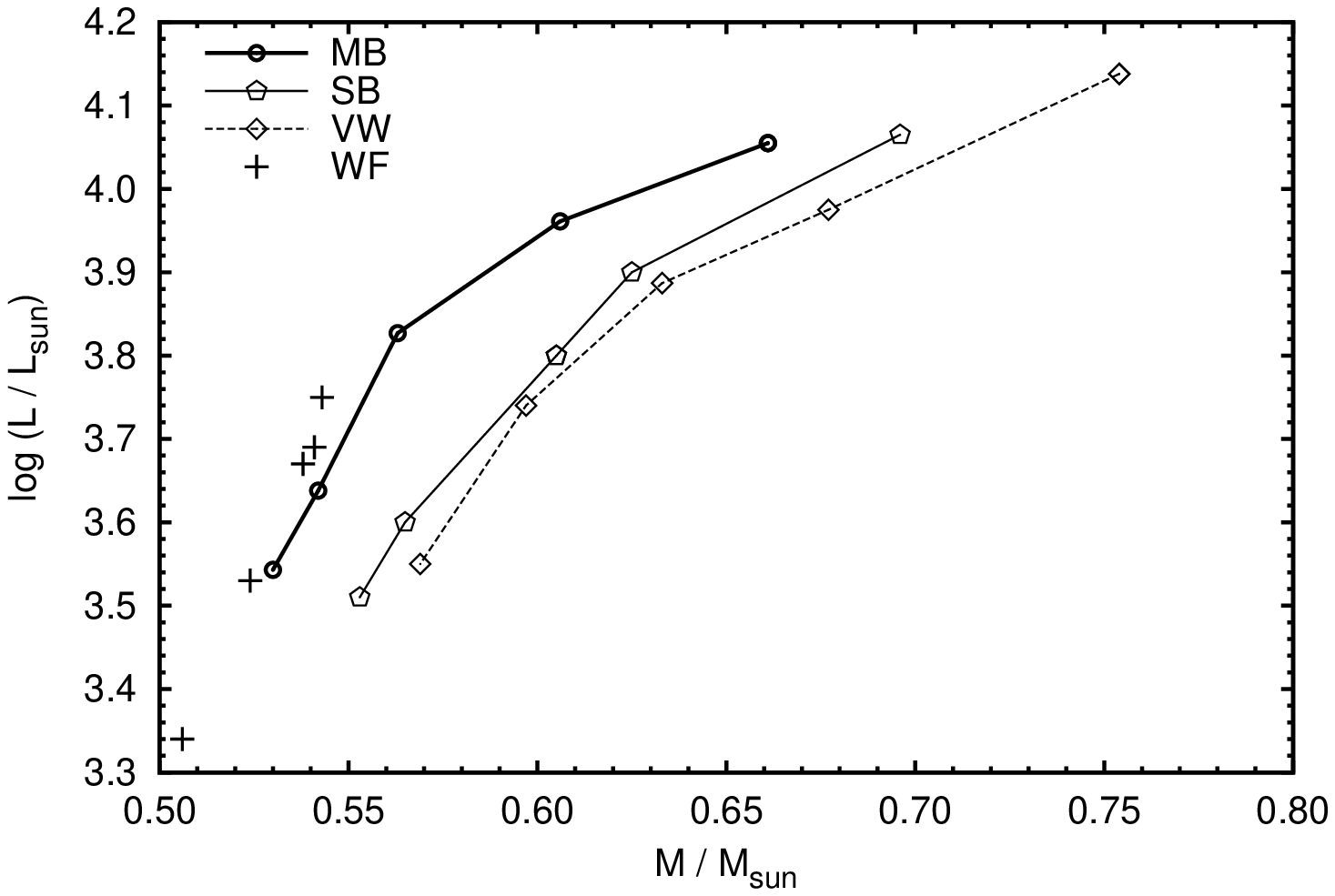}
\vskip -2mm                
\caption{\label{appfig:core.lum}   
         Luminosity of AGB remnants at  $ \teff \approx 30\,000 $~K over remnant
         mass for the evolutionary calculations considered here:
         \citeauthor{MMA.06} (MB), Sch\"onberner/Bl\"ocker (SB), Vassiliadis \& Wood
         (VW), and Weiss \& Ferguson (WF).
         }
\end{figure}
 
Figure~\ref{appfig:core.lum} shows a comparison of the luminosities of the AGB
remnants when they enter the planetary nebulae region, i.e. at
$ \teff \approx 30\,000 $~K (core-mass luminosity relation) from four
different sets of evolutionary calculations. We see a large discrepancy between the older
and newer calculations, in the sense that the new models have higher
luminosities for given remnant masses.\footnote{We note that the remnant
luminosity depends not only on the remnant mass but also somewhat on the
thermal-pulse cycle phase when the star leaves the tip of the AGB, which
can differ from sequence to sequence.}
    
The rather high dependence of the post-AGB luminosity on the physics employed is of paramount 
importance because the bolometric luminosity is the only possibility for estimating an AGB remnant's
mass via evolutionary calculations.  For given luminosity, we find from Fig.~\ref{appfig:core.lum}
a mass difference of 0.04~\dots 0.05~\Msun\ between the Sch\"onberner/Bl\"ocker and Miller Bertolami
mass scale.  Therefore, it is absolutely mandatory to indicate the source of the post-AGB tracks used for 
any mass estimation.  
   
\begin{figure}[hb]
\vskip -4mm
\center
\includegraphics[width= 0.71\linewidth, angle= -90]{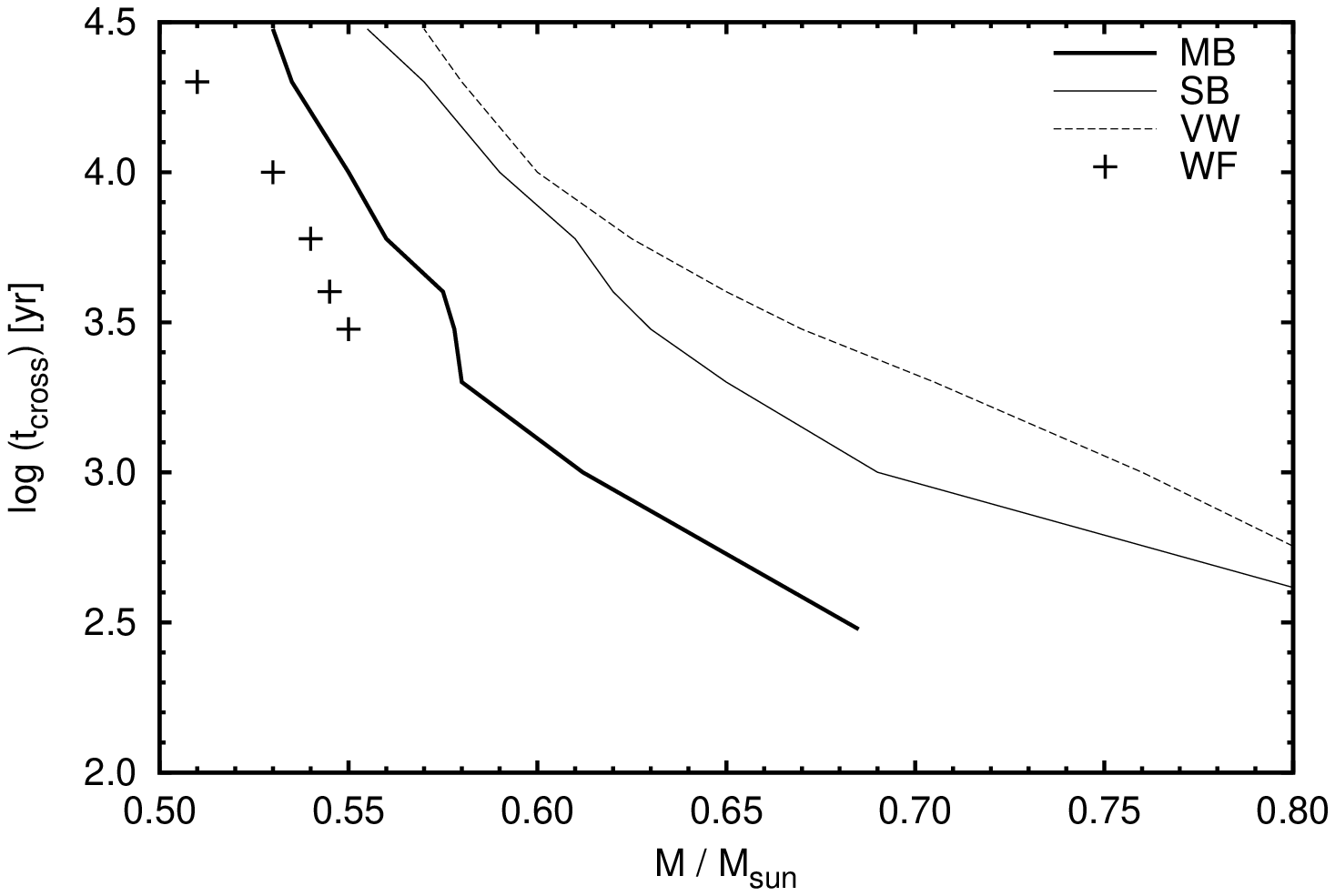}
\vskip -2mm                
\caption{\label{appfig.tcross}
         Crossing time, $ t_{\rm cross} $, of AGB remnants from ${ \teff = 10\,000 }$~K to maximum
         stellar temperature for the same models as in Fig.~\ref{appfig:core.lum}, adapted from 
         Fig.~1 of \citet{MMA.19}. Differences caused by the metallicity are small and not relevant
         here.
         }
\end{figure}

   The differences of the post-AGB luminosities translate directly into the differences of the
   crossing times from ${ \teff = 10\,000 }$~K to maximum stellar temperature, as one can clearly see
   in Fig.~\ref{appfig.tcross}.  This difference is considerable: the Miller Bertolami models 
   evolve faster than the Sch\"onberner/Bl\"ocker models by factors between four and six.    
   
    The smooth decrease of $ t_{\rm cross} $ with mass is distorted around 0.58~\Msun\
     \citep{MMA.06} or 0.62~\Msun\ \citep{B.95}.  This distortion is the signature of the
     transition from initially `low mass' to `intermediate-mass' stars, i.e. from stars which ignite
     helium via a core flash to stars which ignite helium quiescently.  Hence, the stars of these two
     groups start their AGB evolution with a different internal structure which in turn leads to an
     adjustment of the respective crossing times.
     A thorough discussion of the impact of the internal structure of an AGB star on the HRD crossing
     time can be found in \citet{B.95}.

\begin{figure}
\vskip -4mm
\center
\includegraphics[width= 0.71\linewidth, angle= -90]{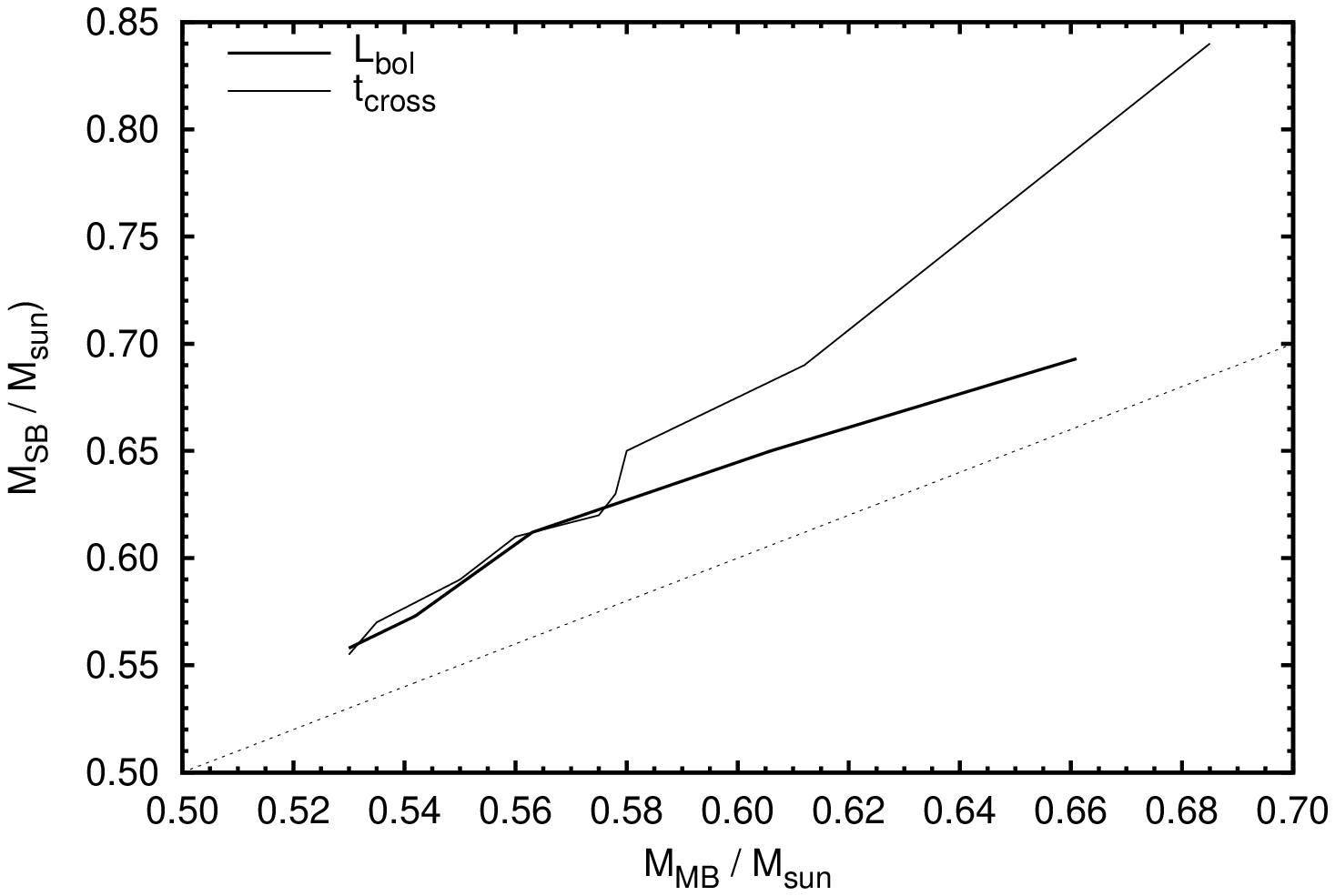}
\vskip -2mm                
\includegraphics[width= 0.71\linewidth, angle= -90]{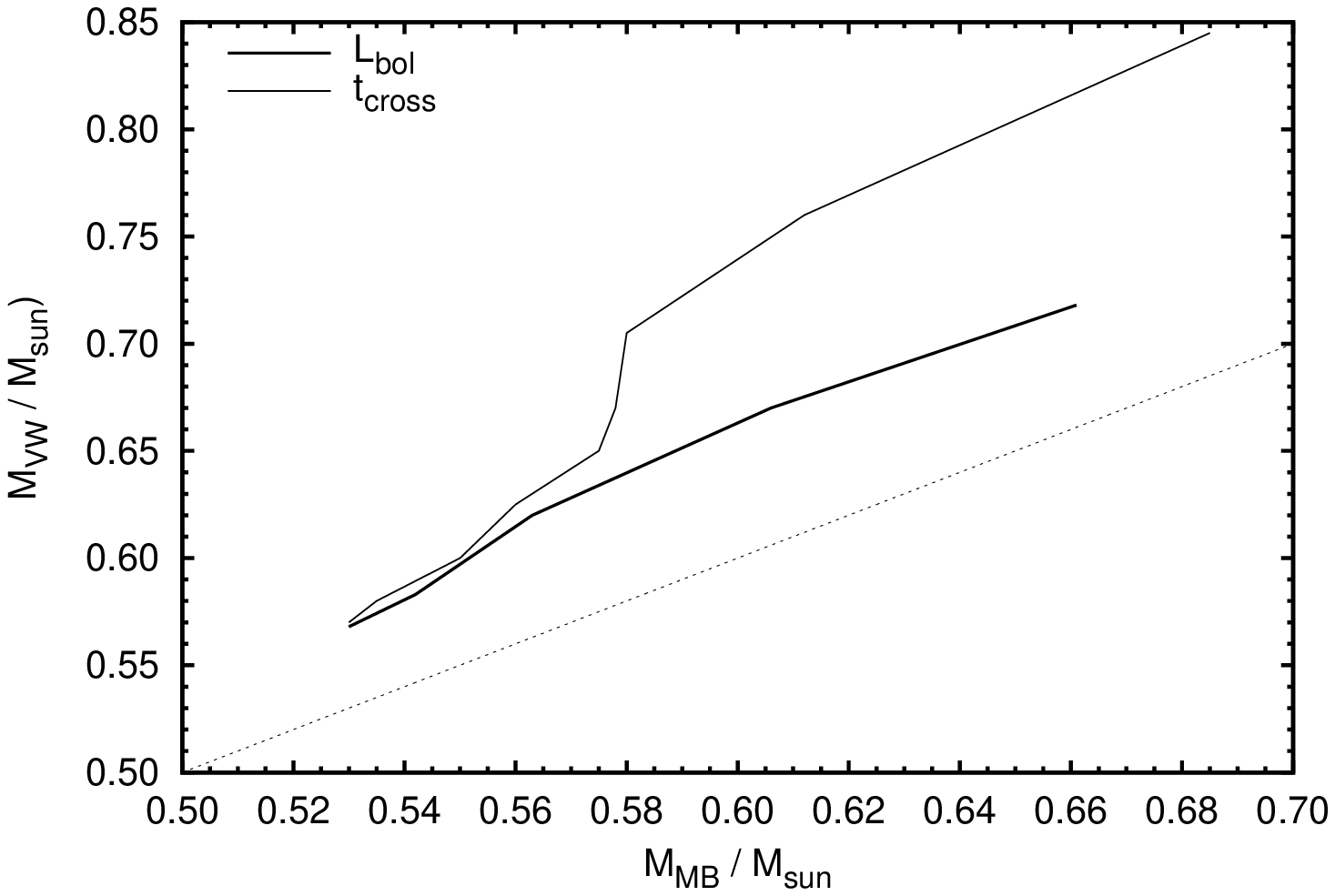}                
\vskip -2mm                
\caption{\label{appfig.comp}
         Relations between the masses of the Sch\"onberner/Bl\"ocker (\emph{top}) and the 
         Vassiliadis \& Wood (\emph{bottom}) post-AGB models against the Miller Bertolami post-AGB
         models along a common sequence of luminosity ($ L_{\rm bol}) $ and
         crossing time ($ t_{\rm cross} $), respectively.
         The dotted line is the 1:1 relation.
         }
\end{figure}

   Figure~\ref{appfig.comp} is a combination of the Figs.~\ref{appfig:core.lum} and \ref{appfig.tcross}
   and shows the relations between post-AGB masses of Sch\"onberner/Bl\"ocker (top) and 
   Vassiliadis \& Wood (bottom) to the Miller Bertolami ones, such that the two curves trace
   a sequence of common luminosity and crossing time, respectively. First of all,
   it is very gratifying that both the crossing time and the bolometric luminosity lead to virtually 
   the same relation as long as $ M_{\rm MB} \la 0.58 $~\Msun\ (or $ M_{\rm SB} \la 0.62 $~\Msun).
   Up to these mass limits, virtually constant offsets between the three post-AGB evolutionary 
   calculations can be used. 

   This bevaviour is a consequence of 
   (i)  the fact that low-mass stars start their helium burning (by a core flash) at virtually the 
        same core mass, independently of the total mass, and      
   (ii) that a homologous core-envelope structures develop during the AGB phase of evolution.
   
   This is not the case for intermediate-mass stars which begin helium burning quiescently in cores whose
   mass increases with total mass.  Hence homologous structures cannot develop, and consequently a common
   scaling of masses based on luminosity and crossing time is not possible:
   above about 0.58~\Msun\ (MB) both the luminosity and the
   crossing time give different relations between the Miller Bertolami and the Bl\"ocker/Sch\"onberner or
   Vassiliadis \& Wood post-AGB masses (Fig.~\ref{appfig.comp}).     
   However, AGB remnants with such relatively high masses are not really
   important in the context of the present study because of their high
   luminosities ($ \ga\! 10^{4} $~\Lsun) and short crossing times ($ \la\! 10^{3} $~yr).
  
   Numerically, the unique part of the relation between
   $ M_{\rm  MB} \leftrightarrow M_{\rm SB} $ shown in Fig.~\ref{appfig.comp} (top) gives 
   $ 0.535 \leftrightarrow 0.565 $~\Msun, $ 0.555 \leftrightarrow 0.595 $~\Msun, and 
   $ 0.575 \leftrightarrow 0.625 $~\Msun. For higher remnant masses, the
   evolution of the two sets of models is truly different. While a MB model
   with 0.665~\Msun has the same post-AGB luminosity as a SB model with
   0.696~\Msun, its HRD crossing time a about a factor of 5 shorter.
   
\end{appendix}
    

\begin{thebibliography}{}

\bibitem[Bl\"ocker(1995)]{B.95}
   Bl\"ocker, T. 1995, \aap, 299, 755
   
\bibitem[Borkowski et al.(1990)]{BBF.90}
   Borkowski, K. J., Balbus, S. A., \& Fristrom, C. C. 1990, \apj, 355, 501  
   
\bibitem[Cahn et al.(1992)]{cahn.92}
   Cahn, J. B., Kaler, J. B., \& Stanghellini, L. 1992, \aaps, 94, 399    
  
\bibitem[Cox \& Tucker(1969)]{CT.69}   
   Cox, D. P., \& Tucker, W. H. 1969, \apj, 157, 1157  
   
\bibitem[Crowther et al.(2006)]{CMS.06}
   Crowther, P. A., Morris, P. W., \& Smith, J. D. 2006, \apj, 636, 1033  
   
\bibitem[de Freitas Pacheco et al.(1993)]{FCAP.93}
   de Freitas Pacheco, J. A., Costa, R. D. D., de Ara\'ujo, F. X., \& Petrini, D.
        1993, \mnras, 260, 401    

\bibitem[{{Dere} {et~al.}(1997){Dere}, {Landi}, {Mason}, {Monsignori Fossi}, \&
  {Young}}]{dere.97}
{Dere}, K.~P., {Landi}, E., {Mason}, H.~E., {Monsignori Fossi}, B.~C., \&
  {Young}, P.~R. 1997, \aaps, 125, 149

\bibitem[{{Dere} {et~al.}(2009){Dere}, {Landi}, {Young}, {Del Zanna},
  {Landini}, \& {Mason}}]{dere.09}
  {Dere}, K.~P., {Landi}, E., {Young}, P.~R., {et~al.} 2009, \aap, 498, 915  
  
\bibitem[Dwarkadas(2023)]{Dwar.23}
   Dwarkadas, V. V. 2023, Galaxy, 11, 78  
  
\bibitem[Freeman \& Kastner(2016)]{FK.16}
    Freeman, M. J., \& Kastner, J. H. 2016, \apjs, 226, 15     

\bibitem[Freeman et al.(2014)]{freeman.14}
   Freeman, M., Montez, R., Jr., Kastner, J. H., et al. 2014, \apj, 794, 99
   
\bibitem[Frew et al.(2016)]{frew.16}
   Frew, D.J., Parker, Q. A., \& Boji\u{c}i\'c, I. S. 2016, \mnras, 455, 1459   
   
\bibitem[Georgiev et al.(2008)]{georgiev.08}
   Georgiev, L. N., Peimbert, M., Hillier, D. J., et al. 2008, \apj, 681, 333   
   
\bibitem[Girard et al.(2007)]{GKA.07}
   Girard, P., K\"oppen, J., \& Acker, A. 2007, \aap, 463, 265
   
\bibitem[Gonz\'ales-Santamar\'ia et al.(2019)]{gon.19} 
   Gonz\'ales-Santamar\'ia, I., Manteiga, M., Manchado, A., Ulla, A., \& Dafonte, C. 2019,
        \aap, 630, A150   
   
\bibitem[G\'orny \& Stasi\'nska(1995)]{GS.95}   
   G\'orny, S. K., \& Stasi\'nska, G. 1995, \aap, 303, 893    
   
\bibitem[Gruendl et al.(2006)]{gruendl.06}
   Gruendl, R. A., Guerrero, M. A., Chu, Y.-H., \& Williams, R. M. 2006, \apj, 653, 339  
   
\bibitem[Guerrero et al.(2005)]{guerrero.05}  
   Guerrero, M. A., Chu, Y.-H., Gruendl, R. A., \& Meixner, M. 2005, \aap, 430, L69 
   
\bibitem[Guerrero et al.(2012)]{guerrero.12}   
   Guerrero, M. A., Ruiz, N., Hamann, W.-R., et al. 2012, \apj, 755, 129
   
\bibitem[Hamann(1997)]{hamann.97}
   Hamann, W.-R. 1997, in Planetary Nebulae, eds. H. J. Habing \& H. J. G. L. M. Lamers, 
          IAU Symp. No. 180, 91         
   
\bibitem[Heller et al.(2018)]{helleretal.16}
   Heller, R., Jacob, R., Steffen, M., Sch\"onberner, D., \& Sandin, C. 2018, \aap, 620, A98
         (Paper II)   
         
\bibitem[Herald \& Bianchi(2004)]{HB.04}
   Herald, J. E., \& Bianchi, L. 2004, \apj, 611, 294
   
\bibitem[Herald \& Bianchi(2007)]{HB.07}
   Herald, J. E., \& Bianchi, L. 2007, \apj, 661, 845   
   
\bibitem[Herald \& Bianchi(2011)]{HB.11}
   Herald, J. E., \& Bianchi, L. 2011, \mnras, 417, 2440      
   
\bibitem[Hillier \& Miller(1999)]{HM.99}
   Hillier, D. J., \& Miller, D. L. 1999, \apj, 519, 354  
   
\bibitem[Kahn \& Breitschwerdt(1990)]{KB.90}
   Kahn, F. D., \& Breitschwerdt, D. 1990, \mnras, 242, 505           

\bibitem[Kastner et al.(2008)]{kastner.08} 
   Kastner, J. H., Montez, R., Jr., Balick, B., De Marco, O. 2008, \apj, 672, 957  

\bibitem[Kastner et al.(2012)]{kastner.13}        
   Kastner, J. H., Montez, R., Jr., Balick, B., et al. 2012, \aj, 144, 58 
   
\bibitem[Keller et al.(2014)]{keller.14}
   Keller, G. R., Bianchi, L., \& Maciel, W. J. 2014, \mnras, 442, 1379   
   
\bibitem[Koo \& McKee(1992a)]{KMcK.92a}
   Koo, B.-C., \& McKee, C. F. 1992a, \apj, 338, 93    
   
\bibitem[Koo \& McKee(1992b)]{KMcK.92}
   Koo, B.-C., \& McKee, C. F. 1992b, \apj, 338, 103    
   
   
\bibitem[Kudritzki et al.(2006)]{KUP.06}
   Kudritzki, R.P., Urbaneja, M.A., \& Puls, J. 2006 in Planetary Nebulae in our Galaxy 
         and Beyond, eds. M. J. Barlow \& R. H. M\'endez, IAU Symp. No. 234, 119    
         
\bibitem[Leuenhagen et al.(1996)]{leuetal.96}
   Leuenhagen, U., Hamann, W.-R., \& Jefferey, C. S. 1996, \aap, 312, 167  
   
\bibitem[Liedahl(1999)]{liedahl.99}
   Liedahl, D. A. 1999, Lect. Notes of Phys., 520, 189         
   
\bibitem[Lindegren et al.(2018)]{Lin.18}
   Lindegren, L., Hern\'andez, Bornbrun, A, et al. 2018, \aap, 616, 2    
   
\bibitem[{Marcolino} {et~al.}(2007)]{marco.07}
   {Marcolino}, W.~L.~F., {Hillier}, D.~J., {de Araujo}, F.~X., \& {Pereira}, C.~B. 
        2007, \apj, 654, 1068   
        
\bibitem[Marten \& Szczerba(1997)]{MS.97}
   Marten, H., \& Szczerba, R. 1997, \aap, 325, 1132    
   
\bibitem[Mellema(1998)]{mellema.98}
   Mellema, G. 1998, \apss, 260, 203       
         
\bibitem[Mellema \& Lundqvist(2002)]{ML.02}
   Mellema, G., \& Lundqvist, P. 2002, \aap, 394, 901  
   
\bibitem[M\'endez et al.(1992)]{MKH.92}
   M\'endez, R. H., Kudritzki, R. P., \& Herrero, A. 1992, \aap, 260, 329 
   
\bibitem[Mewe(1999)]{mewe.99}
   Mewe, R. 1999, Lect. Notes of Phys., 520, 109      
   
\bibitem[Miller Bertolami(2019)]{MMA.19}   
   Miller Bertolami, M. M. 2019, in Why Galaxies Care About AGB stars IV, F. Kerschbaum, 
       M. Gronewegen, H. Olofsson, \& V. Baumgartner, Proc. IAU Sympos. 343, 36           
  
\bibitem[Miller Bertolami \& Althaus(2006)]{MMA.06}
   Miller Bertolami, M. M., \& Althaus, L. G. 2006, \aap, 454, 845   
   
\bibitem[Morisset \& Georgiev(2009)]{MG.09}
   Morisset, C., \& Georgiev, L. 2009, \aap, 507, 1517    
           
\bibitem[Pauldrach et al.(2004)]{PHM.04}
   Pauldrach, A.W.A., Hoffmann, T.L., \& M\'endez, R.H. 2004, \aap, 419, 1111            
           
\bibitem[Pauldrach et al.(1988)]{Pauletal.88}
   Pauldrach, A., Puls, J., Kudritzki, R.-P., M\'endez, R. H., \& Heap, S. R. 1988, 
         \aap, 207, 123          
   
\bibitem[Perinotto et al.(1998)]{peretal.98}        
   Perinotto, M., Kifonidis, K., Sch\"onberner, D., \& Marten, H. 1998, \aap, 332, 1044       
        
\bibitem[Perinotto et al.(2004)]{peretal.04}
   Perinotto, M., Sch\"onberner, D., Steffen, M., \& Calonaci, C. 2004, \aap, 414, 993  
   
\bibitem[Reimers(1975)]{Reim.75} 
   Reimers, D. 1975, in Problems in Stellar Atmospheres and Envelopes, eds. 
          B. Baschek, W. H. Kegel, \& G. Traving (Berlin: Springer), 229    
   
\bibitem[Ruiz et al.(2013)]{ruizetal.13}   
   Ruiz, N., Chu, Y.-H., Gruendl, R. A., et al. 2013, \apj, 767, 35    
   
\bibitem[{{Sandin} et~al.(2016)}]{Sandin13}
   Sandin, C., Steffen, M., Sch\"onberner, D., \& R\"uhling, U.  2016, \aap, 586, A57
          (Paper I)   
          
\bibitem[Sandin etal.(2011)]{Sandin11}
   Sandin, C., Steffen, M.; Sch\"onberner, D., R\"uhling, U., \& Hamann, W. R. 2011,
       Asymmetric Planetary Nebulae 5 Conf., eds. A. A. Zijlstra, F. Lykou, I. McDonald, 
       \& E. Lagadec,  Jodrell Bank Centre for Astrophysics, A53  
       
\bibitem[Sch\"onberner(1979)]{S.79}
   Sch\"onberner, D. 1979, \aap, 79, 108
   
\bibitem[Sch\"onberner(1983)]{S.83}
   Sch\"onberner, D. 1983, \apj, 272, 708                   
          
\bibitem[Sch\"onberner \& Bl\"ocker(1993)]{SB.93}
   Sch\"onberner, D., \& Bl\"ocker, T. 1993, in Luminous High-Latitude Stars, 
          ed. D. D. Sasselov, ASP Conf. Ser., 45, 337  
          

\bibitem[Sch\"onberner et al.(2018)]{SBJ.18}          
   Sch\"onberner, D., Balick, B. \& Jacob, R. 2018, \aap, 609, A126                 
          
\bibitem[Sch\"onberner et al.(2014)]{schoenetal.14} 
   Sch\"onberner, D., Jacob, R., Lehmann, H., et al. 2014, AN, 335, 378                
   
\bibitem[Soker(1994)]{soker}
   Soker, N. 1994, \aj, 107, 276    
   
\bibitem[Solf \& Weinberger(1984)]{SW.84}
   Solf, J., \& Weinberger, R. 1984, \aap, 130, 269    
   
   
\bibitem[Steffen et al.(2014)]{steffenetal.14}   
    Steffen, M., Hubrig, S., Todt, H., et al. 2014, \aap, 570, A88
   
\bibitem[Steffen et al.(2012)]{steffenetal.12}
   Steffen, M., Sandin, C., Jacob, R., \& Sch{\"o}nberner, D. 2012, in
         Planetary Nebulae: An Eye to the Future, eds. A. Manchado, L. Stanghellini, \&
         D. Sch{\"o}nberner, IAU Symp. No. 283, 215      
   
\bibitem[{{Steffen} {et~al.}(2008){Steffen}, {Sch{\"o}nberner}, \& {Warmuth}}]{SSW.08}
   {Steffen}, M., {Sch{\"o}nberner}, D., \& {Warmuth}, A. 2008, \aap, 489, 173 (SSW)  
   
\bibitem[Stute \& Sahai(2006)]{SS.06}
   Stute, M., \& Sahai, R. 2006, \apj, 651, 882   
   
\bibitem[Toal\'a \& Arthur(2014)]{TA.14}
   Toal\'a, J. A., \& Arthur, S. J. 2014, \mnras, 443, 3486
   
\bibitem[Toal\'a \& Arthur(2016)]{TA.16}  
   Toal\'a, J. A., \& Arthur, S. J. 2016, \mnras, 463, 4438 
   
\bibitem[Toal\'a \& Arthur(2017)]{TA.16b}  
   Toal\'a, J. A., \& Arthur, S. J. 2017, \mnras, 464, 178   
   
\bibitem[Toal\'a \& Arthur(2018)]{TA.18}  
   Toal\'a, J. A., \& Arthur, S. J. 2018, \mnras, 478, 1218  
   
\bibitem[Toal\'a et al.(2020)]{toala.20}  
   Toal\'a, J. A., Guerrero, M. A., Bianchi, L., Chu, Y.-H., \& De Marco, O, 2020, 
        \mnras, 494, 3784        
   
\bibitem[Toala et al.(2015)]{toala.15}   
   Toal\'a, J. A., Guerrero, M. A., Todt, H., et al. 2015, \apj, 799, 67    
   
\bibitem[Toal\'a et al.(2019a)]{toala.19}
   Toal\'a, J. A., Montez Jr. R., \& Karovska, M. 2019a, \apj, 886, 30
   
\bibitem[Toal\'a et al.(2019b)]{toalaetal.19}   
   Toal\'a, J. A., Ramos-Larios, G., Guerrero, M. A., \& Todt, H. 2019b, \mnras, 485, 3360  
   
\bibitem[Todt et al.(2015)]{todt.15}
   Todt, H., Kniazev, A. Y., Gvaramadze, V. V., et al. 2015, ASP Conf. S., 493, 539 
   
\bibitem[Vassiliadis \& Wood(1993)]{VW.93}
   Vassiliadis, E., \& Wood, P. R. 1993, \apj, 413, 641    
   
\bibitem[Vassiliadis \& Wood(1994)]{VW.94}
   Vassiliadis, E., \& Wood, P. R. 1994, \apjs, 92, 125      
   
\bibitem[Villaver et al.(2002)]{villa.02}
   Villaver, E., Manchado, A., \& Garc\'ia-Segura, G. 2020, \apj, 581, 1204    
   
\bibitem[{{Weaver} {et~al.}(1977){Weaver}, {McCray}, {Castor}, {Shapiro}, \&
  {Moore}}]{weaver.77}
{Weaver}, R., {McCray}, R., {Castor}, J., {Shapiro}, P., \& {Moore}, R. 1977, \apj, 218, 377 

\bibitem[Weiss \& Ferguson(2009)]{WF.09}
   Weiss, A, \& Ferguson, J. W. 2009, \aap. 508, 1343       
   
\bibitem[{{Yu} {et~al.}(2009){Yu}, {Nordon}, {Kastner}, {Houck}, {Behar}, \&
          {Soker}}]{yuetal.09}
    {Yu}, Y.~S., {Nordon}, R., {Kastner}, J.~H., {et~al.} 2009, \apj, 690, 440 

\bibitem[{{Zhekov} \& {Perinotto}(1996)}]{ZP.96}
   {Zhekov}, S.~A. \& {Perinotto}, M. 1996, \aap, 309, 648   

\bibitem[Zhekov \& Park(2011)]{zhekov.11}
    Zhekov, S.~A. \& Park, S.\ 2011, \apj, 728, 135

\bibitem[Zhekov(2014)]{zhekov.14}
    Zhekov, S.~A.\ 2014, \mnras, 443, 12

\end{thebibliography}
\end{document}